\documentclass[a4paper,twoside,12pt]{book}

\usepackage{amsmath}
\usepackage{graphicx}

\newcommand{\cE}{\mathcal{E}}
\newcommand{\cH}{\mathcal{H}}
\newcommand{\cJ}{\mathcal{J}}
\newcommand{\cL}{\mathcal{L}}
\newcommand{\cR}{\mathcal{R}}
\renewcommand{\vec}{\boldsymbol}
\newcommand{\ten}{\overleftrightarrow}
\newcommand{\re}{\mathrm{Re~}}
\newcommand{\im}{\mathrm{Im~}}
\newcommand{\pr}[1]{{\sc{\lowercase{#1}}}}

\begin{document}

\pagenumbering{roman}

\thispagestyle{empty}

\setlength{\baselineskip}{8mm}

\begin{flushright}
IFT preprint 19/2004
\end{flushright}
\vspace{0.5cm}

\begin{center}
{\LARGE\bf Chiral and magnetic rotation \\ in atomic nuclei studied within \\ self-consistent mean-field methods}
\end{center}
\vspace{1cm}

\begin{center}
{\Large Przemys\l{}aw Olbratowski}
\end{center}
\vspace{0.5cm}

\hspace{0.5cm}
\begin{minipage}{14cm}
Instytut Fizyki Teoretycznej, Warszawa, Uniwersytet Warszawski \\ \\
Institut de Recherches Subatomiques, Strasbourg, \\
Universit\'e Louis Pasteur de Strasbourg I, \\
Institut National de Physique Nucl\'eaire et de Physique de Particules
\end{minipage}
\vspace{2cm}

\begin{minipage}{15cm}
A Ph.D. thesis prepared in collaboration between the Warsaw University, Poland, and the Louis Pasteur University, France, thanks to a scholarship from the French Government. This work was supported in part by the Polish Committee for
Scientific Research (KBN) and by the Foundation for Polish Science
(FNP).
\end{minipage}
\vspace{1cm}

\begin{tabular}{ll}
{\bf Supervisors:} & \\
Jacek Dobaczewski  & Instytut Fizyki Teoretycznej, Warszawa \\
Jerzy Dudek        & Institut de Recherches Subatomiques, Strasbourg \\
                   & \\
{\bf Referees:}    & \\
Wojciech Satu\l{}a & Instytut Fizyki Teoretycznej, Warszawa \\
Jan Stycze\'n      & Instytut Fizyki J\c{a}drowej \\
                   & im. H. Niewodnicza\'nskiego, Krak\'ow\\
Johann Bartel      & Institut de Recherches Subatomiques, Strasbourg \\
Hubert Flocard     & Centre de Spectrom\'etrie Nucl\'eaire \\
                   & et Spectrom\'etrie de Masse, Orsay
\end{tabular}
\vspace{2cm}

\begin{flushright}
Warsaw/Strasbourg, July 12, 2004
\end{flushright}

\newpage

\thispagestyle{empty}

\centerline{}

\newpage

\thispagestyle{empty}

\setlength{\baselineskip}{5.1mm}

\begin{flushright}
{\large Wierz tym, kt\'orzy szukaj\c{a} prawdy, \\ i nie dowierzaj tym, kt\'orzy j\c{a} znale\'zli.} \\
~ \\
Andr\'e Gide
\end{flushright}
\vspace{14cm}

\begin{flushright}
{\LARGE\it Moim Rodzicom}
\end{flushright}

\newpage

\thispagestyle{empty}

\centerline{}

\newpage

\thispagestyle{empty}

{\bf Podzi\c{e}kowania} Pragn\c{e} podzi\c{e}kowa\'c mojemu promotorowi, prof. Jackowi Do\-ba\-czew\-skie\-mu, za przyj\c{e}cie mnie pod sw\c{a} opiek\c{e} w Instytucie Fizyki Teoretycznej w Warszawie, propozycj\c{e} tematyki pracy naukowej, bezcenn\c{a} pomoc w rozwijaniu narz\c{e}dzi numerycznych, kt\'orymi si\c{e} pos\l{}ugiwa\l{}em, za po\'swi\c{e}can\c{a} mi ci\c{a}gle uwag\c{e}, oraz za wnikliw\c{a} korekt\c{e} moich publikacji i niniejszej pracy.
\vspace{1cm}

{\bf Remerciements} Je tiens \`a remercier mon co-directeur de th\`ese, prof. Jerzy Dudek, de m'avoir accueilli \`a l'Institut de Recherches Subatomiques \`a Strasbourg, d'avoir surveill\'e continuellement mon travail scientifique, et d'avoir attentivement lu et corrig\'e mes publications et ma th\`ese de doctorat.

\newpage

\thispagestyle{empty}

\centerline{}

\newpage

\thispagestyle{empty}

\setlength{\baselineskip}{4.2mm}

{\bf Streszczenie} {\small Jednym z zastosowa\'n metod \'sredniego pola w fizyce j\c{a}drowej jest obecnie badanie egzotycznych symetrii j\c{a}der atomowych. Wi\c{a}\.ze si\c{e} to w szczeg\'olno\'sci z analiz\c{a} rotacji j\c{a}der wok\'o\l{} osi pochylonej wzgl\c{e}dem osi g\l{}\'ownych rozk\l{}adu masy w modelu Tilted-Axis Cranking (TAC). Niniejsza praca przedstawia jedne z pierwszych oblicze\'n TAC wykonanych metodami w pe\l{}ni samozgodnymi. Zastosowano w niej metod\c{e} Hartree'ego-Focka z dwucia\l{}owym od\-dzia\-\l{}y\-wa\-niem efektywnym Skyrme'a. Stworzono program komputerowy pozwalaj\c{a}cy na \l{}amanie wszystkich symetrii przestrzennych rozwi\c{a}zania. Jako pierwsze zastosowanie przeprowadzono obliczenia dla pasm {\it magnetycznych} w $^{142}$Gd oraz {\it chiralnych} w $^{130}$Cs, $^{132}$La, $^{134}$Pr i $^{136}$Pm. Ich wyst\c{e}powanie zwi\c{a}zane jest odpowiednio z nowym mechanizmem \l{}amania symetrii sfe\-rycznej oraz ze spontanicznym \l{}amaniem symetrii {\it chiralnej}. Samozgodne rozwi\c{a}zania w $^{142}$Gd potwierdzaj\c{a} istotn\c{a} rol\c{e} mechanizmu {\it shears} w tworzeniu ca\l{}kowitego momentu p\c{e}du. Zgodno\'s\'c z danymi do\'swiadczalnymi nie jest jednak zadowalaj\c{a}ca, prawdopodobnie wskutek nie\-u\-wzgl\c{e}d\-nie\-nia korelacji par lub przeszacowania deformacji. Wyniki w $^{132}$La stanowi\c{a} pierwszy w pe\l{}ni samozgodny dow\'od, \.ze rotacja j\c{a}drowa mo\.ze przybiera\'c charakter chiralny. Wykazano, \.ze rotacja chiralna mo\.ze wyst\c{e}powa\'c tylko powy\.zej pewnej krytycznej cz\c{e}sto\'sci obrotu. Spraw\-dzo\-no te\.z, \.ze cz\l{}ony pola \'sredniego Skyrme'a nieparzyste wzgl\c{e}dem odwr\'ocenia czasu nie maj\c{a} jako\'sciowego wp\l{}ywu na wyniki.}
\vspace{2mm}

{\bf R\'esum\'e} {\small Une des applications r\'ecentes des m\'ethodes de champ moyen en physique nucl\'eaire est l'\'etude des sym\'etries exotiques du noyau. Cette probl\'ematique est reli\'ee, en particulier, \`a l'\'analyse de la rotation nucl\'eaire autour d'un axe inclin\'e par rapport aux axes principaux de la distribution de masse dans le mod\`ele dit de Tilted-Axis Cranking (TAC). Cette th\`ese pr\'esente l'un des premiers calculs TAC effectu\'es dans le cadre de m\'ethodes enti\`erement auto-coh\'erentes. La m\'ethode Hartree-Fock avec l'interaction effective \`a deux corps de Skyrme a \'et\'e utilis\'ee. Un code num\'erique a \'et\'e \'ecrit qui permet de briser toutes les sym\'etries spatiales des solutions. Comme premi\`ere application, des calculs pour les bandes {\it magn\'etiques} dans $^{142}$Gd et pour les bandes {\it chirales} dans $^{130}$Cs, $^{132}$La, $^{134}$Pr et $^{136}$Pm ont \'et\'e effectu\'es. L'existence de ces bandes est d\^ue \`a un nouveau m\'ecanisme de brisure de la sym\'etrie sph\'erique, et de brisure spontan\'ee de la sym\'etrie {\it chirale}, respectivement. Les solutions auto-coh\'erentes dans $^{142}$Gd confirment le r\^ole important du m\'ecanisme {\it shears} dans la g\'en\'eration du moment angulaire. Pourtant, l'accord avec les donn\'ees exp\'erimentales n'est pas satisfaisant, probablement \'a cause de l'omission des corr\'elations d'appariement dans les calculs ou de la possible surestimation de la d\'eformation. Les r\'esultats obtenus dans $^{132}$La constituent la premi\`ere preuve enti\`erement auto-coh\'erente que la rotation nucl\'eaire peut acqu\'erir un caract\`ere chiral. Il a \'et\'e d\'emontr\'e que la rotation chirale ne peut avoir lieu qu'au-dessus d'une certaine fr\'equence angulaire critique. Il a \'et\'e \'egalement v\'erifi\'e que les termes du champ moyen de Skyrme impair par rapport au renversement du temps n'ont pas d'influence qualitative sur les resultats.}
\vspace{2mm}

{\bf Abstract} {\small Currently, one application of the mean-field methods in nuclear physics is the investigation of exotic nuclear symmetries. This is related, in particular, to the study of nuclear rotation about an axis tilted with respect to the principal axes of the mass distribution in the Tilted-Axis Cranking (TAC) model. The present work presents one of the first TAC calculations performed within fully self-consistent methods. The Hartree-Fock method with the Skyrme effective two-body interaction has been used. A computer code has been developed that allows for the breaking of all spatial symmetries of the solution. As a first application, calculations for the {\it magnetic} bands in $^{142}$Gd and for the {\it chiral} bands in $^{130}$Cs, $^{132}$La, $^{134}$Pr, and $^{136}$Pm have been carried out. The appearance of those bands is due to a new mechanism of breaking the spherical symmetry and to the spontaneous breaking of the {\it chiral} symmetry, respectively. The self-consistent solutions for $^{142}$Gd confirm the important role of the {\it shears} mechanism in generating the total angular momentum. However, the agreement with experimental data is not satisfactory, probably due to the lack of the pairing correlations in the calculations or to the possibly overestimated deformation. The results obtained for $^{132}$La constitute the first fully self-consistent proof that the nuclear rotation can attain a chiral character. It has been shown that the chiral rotation can only exist above a certain critical angular frequency. It has also been checked that the terms of the Skyrme mean field odd under the time reversal have no qualitative influence on the results.}

\newpage

\thispagestyle{empty}

\centerline{}

\newpage

\pagenumbering{arabic}

\pagestyle{headings}

\setlength{\baselineskip}{5.1mm}

\tableofcontents

\listoffigures

\listoftables

\chapter{Introduction}

The phenomenon of {\it spontaneous symmetry breaking} is a leading theme of a multitude of quantum effects. For instance, it is due to the spontaneous breaking of the particle-number symmetry that superconducting condensates appear in metals. Spontaneous breaking occurs if a system in its endeavor to attain the minimal energy chooses a symmetry-violating state even though the underlying interactions are invariant under the concerned symmetry. Nevertheless, it is the nature of the interactions that determines which symmetries are broken and under which conditions. Therefore, study of symmetry-violating states brings one closer to understanding the interactions - the most fundamental goal in physics.

In the nuclear structure physics, one important example of spontaneous symmetry breaking is the existence of deformed nuclei. Like for molecules, violation of the spherical invariance leads to the appearance of rotational excitations, which manifest themselves in specific sequences of levels, the rotational bands. Depending on the conservation or violation of other symmetries, like the plane reflection, those bands can have different structures. Recently, two novel types of bands have been observed experimentally, that have challenged the theoretical models. They are called {\it magnetic} and {\it chiral} bands.

The peculiarity of the magnetic structures is that, contrary to all the previously known rotational bands, they appear in nuclei whose charge distribution is nearly spherical, which can be inferred from the very weak electric quadrupole transitions. On the other hand, magnetic dipole transitions, to which the bands owe their name, are strong. Thus, the bands constitute the first evidence for the breaking of the spherical symmetry by a large dipole moment, which, in turn, is produced by highly asymmetric current distributions. In the absence of charge deformation, no collective rotation is possible, and the magnetic bands also entail a new mechanism of generating the angular momentum, in which valence nucleons play a crucial role. The angular momenta of the valence particles align coherently along one direction, and those of the valence holes along a perpendicular direction. The spin of the levels in the band is produced by gradual alignment of the two angular momenta vectors. This has been dubbed {\it shears mechanism}, because it resembles the closing of a pair of shears used for cutting the sheep wool. The magnetic bands were first observed in early 1990s, and more that 130 such structures have been found so far.

The chiral bands have the form of doublets of bands lying very close in energy. Since other hypotheses failed to reproduce the vanishing energy splitting between them, it has been suggested that the doublets are due to the possible existence of two enantiomeric (left- or right-handed) forms of the nucleus. The chiral bands are observed mainly in well-deformed nuclei, in which there is one valence proton particle and one valence neutron hole in an orbital of high angular momentum. The former drives the nucleus towards elongated shapes, while the latter towards oblate ones. The interplay of these opposite tendencies may result in a shape resembling a triaxial ellipsoid. In the triaxially deformed nucleus, the particle and hole align their angular momenta along the short and long axes of the density distribution, respectively. Moreover, the moment of inertia with respect to the medium axis is the largest, which favors the collective rotation around that axis. Thus, the total angular momentum vector has non-zero components on all the three axes, and those component vectors can form either a left-handed or a right-handed system. This means, according to the original Kelvin's definition, that the system is chiral, only that its chirality is related to the handedness of angular momenta rather than of position vectors. First chiral bands were identified about the year of 2000, and ten odd occurrences are known now. 

The chiral rotation has been successfully examined within the phe\-nom\-e\-no\-logical Par\-ti\-cle-Rotor Model, in which the nucleus is represented by a rigid core with a few valence nucleons coupled to it. However, the very concept of both the shears mechanism and rotational chirality came from considerations within the mean-field cranking approach, which is one of the most fundamental methods in the nuclear structure physics. In this approximation, each nucleon is moving independently in a rotating potential that represents an averaged interaction with other nucleons. The main point about the mean-field description of the two considered phenomena is that it must allow for rotation about an axis that does not coincide with any principal axis of the mass distribution; such a variant of the cranking model is called Tilted-Axis Cranking (TAC). Before the interest in magnetic and chiral bands arose, only rotations about principal axes were considered, so addressing the new effects required an upgrade of the existing numerical tools. A first TAC code was written by Frauendorf \cite{Fra93a}; it used a phenomenological mean potential to describe the properties of the valence particles, and the nuclear liquid-drop model to account for some bulk properties of the nucleus. A great deal of experimental data, mainly on magnetic bands, has been analyzed by using that code, and generally a good agreement was obtained. As far as more fundamental methods are concerned, only one magnetic band in $^{84}$Rb has been studied by Madokoro {\it et al.} \cite{Mad00a} within the Relativistic-Mean-Field method.

Although the phenomenological model of Frauendorf has led to a remarkable success, a more fundamental description requires self-consistent methods, in which the mean potential is indeed generated from averaging an effective two-body interaction with the nucleonic density. First of all, such an approach is capable to provide a strong test of the stability of the proposed shears and chiral configurations with respect to the core degrees of freedom. It is also necessary to take into account several important effects, like all kinds of polarization of the core by the valence particles and full minimization of the underlying energies with respect to all deformation variables, including the deformation of the current and spin distributions. Application of self-consistent methods to the description of the magnetic and chiral rotation is the subject of this PhD dissertation.

The author developed a code that can perform TAC calculations within the full Hartree-Fock (HF) method with the Skyrme effective interaction \cite{Rin00a}. It is one of the first existing tools of this kind, and allows for a study of many other effects; for instance, it has already been used to calculate the time-reversal violating Schiff moment of $^{225}$Ra \cite{Eng03a} in connection with the search for fundamental breaking of the $T$ or $CP$ symmetry beyond the Standard Model. In the present work, the first Skyrme-HF TAC solutions were found. For the study of the shears bands, the nucleus of $^{142}$Gd was chosen. The HF results corroborated that an important portion of the angular momentum is generated by the shears mechanism, although the shape deformation is non-negligible, and the collective rotation of the core is also present. The chiral solutions were looked for in four $N=75$ isotones, $^{130}$Cs, $^{132}$La, $^{134}$Pr, $^{136}$Pm, and found in the second one. It was established that the chiral rotation appears only above a certain critical value of the angular frequency. The origin of the critical frequency was explained in terms of a very simple classical model, that provides a surprising agreement with the HF results.

In Chapter~\ref{magchi_cha}, an introduction to the physics of the magnetic and chiral bands is given, together with a review of the literature. Chapter~\ref{thetol_cha} describes the theoretical tool used in this work - the Skyrme-HF TAC method. Particular emphasis is put on the issue of spontaneous symmetry breaking in rotational bands and on technical differences between the self-consistent and phenomenological approaches. One aspect that can be studied uniquely within the self-consistent approach is the influence of the HF time-odd densities and fields on the shears and chiral solutions. The way this is investigated in the present work is also described in that Chapter. Finally, the code \pr{HFODD} is presented. The rotational behavior of the nuclei here under study is determined, to a big extent, by the alignment properties of a few valence $h_{11/2}$ nucleons. In Chapter~\ref{valenc_cha}, their basic features are established from standard (non-TAC) cranking calculations and from pure symmetry considerations. The Skyrme-HF results for the shears band in $^{142}$Gd are presented in Chapter~\ref{magnet_cha}. Chapter~\ref{chiral_cha} gives the planar and chiral solutions for the $N=75$ isotones. There, the classical model is formulated, the expression for the critical frequency is derived, and its values are analyzed. Since the present calculations constitute the first application of the Skyrme-HF TAC method, several technical details on how the solutions were obtained are also given in that Chapter. Chapter~\ref{conout_cha} summarizes the main conclusions from the present work, and outlines prospects for further research.

\chapter{Magnetic and chiral rotations}
\label{magchi_cha}

A classical rigid body can rotate uniformly only about the principal axes of its inertia tensor. However, already Riemann showed \cite{Rie60a} that in liquids rotation about a tilted axis may take place if there is an intrinsic vortical motion. In a drop of nuclear matter, the nucleus, vorticity may arise from the presence of unpaired nucleons near the Fermi level. The swirling vortices produce the single-particle (s.p.) angular momenta that take various directions and add to the collective rotation drawing the total spin away from the principal axes. It is the rotation about a tilted axis that breaks the signature and chiral symmetries, and gives rise to the magnetic and chiral bands. The two phenomena show different ways of coupling the s.p.\ spins to the collective rotation.

\section{Magnetic rotation}
\label{magnet_sec}

In early 1990's, very regular rotational bands were observed in several light lead isotopes \cite{Bal92a,Cla92a,Kuh92a}, and were initially mistaken for superdeformed structures. But detailed measurements showed that they had an unusual feature: weak $E2$ and strong $M1$ transitions. To date, many more similar structures have been identified. They are called {\it magnetic dipole} or {\it shears} bands. This Section recapitulates their main characteristics and presents the mechanisms that are believed to lie at their origin.

\begin{figure}[t]
\begin{center}
\includegraphics{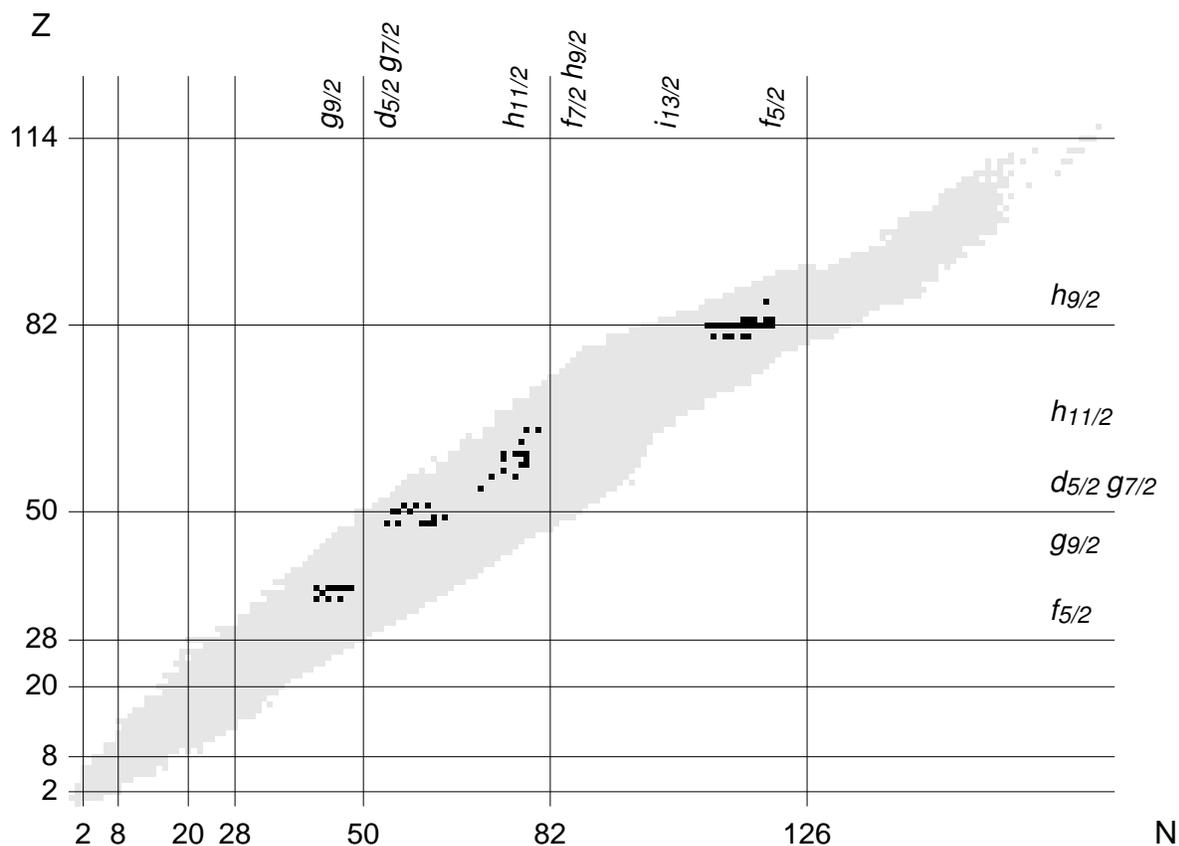}
\end{center}
\caption[Magnetic bands throughout the nuclear chart.]{Experimental evidence for the magnetic bands throughout the nuclear chart. Nuclei, in which magnetic bands have been found  \cite{Ami01a} (May 2001), are marked with black squares, and the gray background represents the known isotopes \cite{Ant02a} (July 2002). Principal high-$j$ orbitals available for the valence particles and holes are marked.}
\label{nuchart_fig}
\end{figure}

Taking into account the vast experimental evidence, one can enumerate the following properties that can be considered as a definition of the magnetic dipole bands:
\begin{itemize}
\item Levels in the band form the $I^+,(I+1)^+,(I+2)^+$ or $I^-,(I+1)^-,(I+2)^-$ spin-parity sequence.
\item Bands never start at spins less than about $10\hbar$, except for the lightest nuclides, and approximately follow the regular rotational dependence of $E\sim I(I+1)$.
\item The $E2$ transitions within the band are very weak, with reduced probabilities $B(E2)$ typically not exceeding $\sim0.1\,\mathrm{e}^2\mathrm{b}^2$. In contrast, the values of $B(M1)$ are exceptionally large, in the range of $\sim2-10\mu_N$. They exhibit a characteristic fall with increasing spin.
\item The ratio $\cJ^{(2)}/B(E2)$ assume the values of $\sim100\,\hbar^2\mathrm{MeV}^{-1}\mathrm{e}^{-2}\mathrm{b}^{-2}$ that is roughly an order of magnitude larger than for normal or superdeformed bands.
\end{itemize}
So far, more than 130 bands possessing such properties have been observed in more that 60 nuclides that are marked in Fig.~\ref{nuchart_fig}. They group in four islands situated not far from magic numbers.

If a nucleus exhibits rotational excitations, its spherical symmetry must be spontaneously broken. Before 1990, there was only evidence for breaking the rotational invariance by a deformation of the charge distribution. But in the shears bands, the very weak $E2$ transitions unequivocally point to an almost spherical shape. Therefore, the main question about the newly found bands is what violates the spherical symmetry, if not the shape. On the other hand, sequences of levels linked by relatively strong $M1$ and weak $E2$ transitions appear quite often in nearly spherical nuclei. They originate from recouplings of the angular momenta of valence nucleons, and are well understood within the spherical shell model. However, these sequences constitute complicated multiplets with energy changing abruptly with spin. From this point of view, the peculiarity of the magnetic bands consists in their surprisingly regular level spacing. 

Origins of that regularity were sought in a series of shell-model studies. Frauendorf {\it et al.} investigated the bands observed in light Pb isotopes \cite{Fra96a} and in odd In isotopes \cite{Fra97b} using the surface-delta interaction. They found that a substantial portion of the angular momentum originates from coupling of high-$j$ valence particles in one kind of nucleons with high-$j$ valence holes in the other kind. However, regular structures were obtained only if low-$j$ orbitals were included in the configuration space. These orbitals provide a certain polarizability of the core, which introduces the long-range, mainly quadrupole correlations. These correlations smooth out the $E(I)$ dependence, because they involve many nucleons and average their influence. Therefore, the magnetic rotation is a collective phenomenon, despite that mainly the valence high-$j$ nucleons contribute to the angular momentum\footnote{However, to avoid confusion, in the following the term {\it collective rotation} is reserved for the standard rotation of the deformed mass distribution.}. The core polarization plays a crucial role, even though it is weak as opposed to that in well deformed nuclei. These features are supported by several experimental observations of transitions from irregular to regular structures, as the $Z$ and $N$ numbers get away from magic shells, increasing the polarizability. Frauendorf \cite{Fra97b} found such a transition in $_{49}$In$_{58,60,62,64}$ isotopes with increasing $N$, in good correlation with the shell-model results. Similar behavior was reported for isotopes of Cd and Sn \cite{Fra01a}. In the region of $N=80$, it was found in $_{64}$Gd$_{78,79,80}$ \cite{Rza01a,Lie02a}. Inspecting the table of magnetic bands \cite{Ami01a}, one easily notices that the effect is also well pronounced when passing from the isotopes of $_{80}$Hg to the isotopes of $_{82}$Pb. In fact, the magnetic bands have been observed in close, but never immediate vicinity of doubly magic nuclei, see Fig.~\ref{nuchart_fig}.

\begin{figure}[t]
\begin{center}
\includegraphics{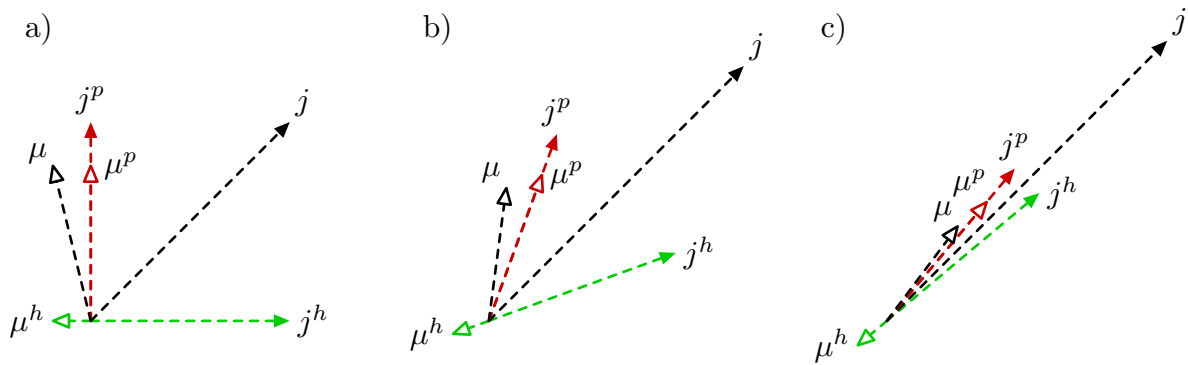}
\end{center}
\caption[The shears mechanism.]{The shears mechanism; from the bandhead a) to the maximum alignment c). The angular momentum, $j$, is generated by gradual alignment of spins of valence particles and holes, $j^p$ and $j^h$. The corresponding contributions, $\mu^p$ and $\mu^h$, to the magnetic dipole moment, $\mu$, are drawn schematically for the case in which the valence particles are protons, and the valence holes are neutrons.}
\label{shears_fig}
\end{figure}

If the low-$j$ orbitals are excluded from the configuration space in the discussed shell-model study, the valence high-$j$ particles and holes rearrange their angular momenta in a complicated way from one level to another. The appearance of quadrupole correlations with the inclusion of those orbitals causes the valence high-$j$ particles to orient their angular momenta coherently in the same direction and the holes to behave similarly. Such a {\it stretched} alignment of several high-$j$ contributions results in long particle and hole angular momenta vectors, $\vec{j}^p$ and $\vec{j}^h$, which are called {\it blades}. In general, both kinds of nucleons can contribute to each blade, but in most cases the particles are protons and the holes are neutrons. At the bandhead, the two blades turn out to form an angle of nearly $90^\circ$ (Fig.~\ref{shears_fig}a). Since the proton and neutron effective $g$-factors, $g_\pi$ and $g_\nu$, have opposite signs, the perpendicular coupling also gives rise to a large magnetic dipole moment, $\vec{\mu}$, with a significant component, $\mu_\perp$, perpendicular to the total spin, $\vec{j}$, (Fig.~\ref{shears_fig}a). This is the source of the strong $M1$ transitions, because the reduced probability $B(M1)$ is proportional to $\mu_\perp^2$. The perpendicular coupling of $\vec{j}^p$ and $\vec{j}^h$ has been confirmed by an experimental determination of the effective $g$-factor in the bandhead of a shears band in $^{193}$Pb \cite{Chm97a}. In the band, $\vec{j}^p$ and $\vec{j}^h$ align towards each other, retaining their constant lengths. In this way, the total spin increases, while $\mu_\perp$ decreases, causing the observed drop-off in the $M1$ strength (Fig.~\ref{shears_fig}b). Assuming the geometry shown in Fig.~\ref{shears_fig}, Macchiavelli and Clark \cite{Mac98a} calculated the $B(M1)$ values for the magnetic bands in $^{198}$Pb and $^{199}$Pb as proportional to $\mu_\perp^2$, and obtained a good overall agreement with experiment. Obviously, the maximum spin that can be generated in such a way approximately equals the sum of $j^p$ and $j^h$, whereas $\mu_\perp$ and $B(M1)$ vanish when the maximum alignment is achieved (Fig.~\ref{shears_fig}c). This mode of generating the angular momentum and the magnetic moment was originally proposed by Frauendorf \cite{Fra93a}, and was then dubbed {\it shears mechanism} \cite{Bal94a}, because it resembles the closing of a pair of shears used for cutting the sheep wool.

In view of the important role of the quadrupole polarization, the magnetic bands have been investigated within approaches that take that polarization into account in a model way. In \cite{Mac98b} and also in \cite{Mac98a,Mac98c,Mac99a}, Macchiavelli and Clark used the particle-vibration theory \cite{Boh75a}, considering the coupling of the valence high-$j$ particles and holes to the quadrupole vibrations of the nuclear shape about a spherical equilibrium. They postulated the stretched coupling of the constituent angular momenta in each blade and treated all the valence particles as one particle with spin $j^p$ and all the valence holes as one hole with spin $j^h$. In the second order of the perturbation series, the particle-vibration coupling gives rise to an effective interaction between the valence nucleons. It takes the form of the second-order Legendre polynomial, $P_2(\cos\vartheta)$, where $\vartheta$ is the angle between the position vectors of the valence particle and hole. It was shown in \cite{Boh53a} that in the limit of high angular momenta the expectation value of the interaction $P_2(\cos\vartheta)$ in the state $|j^pj^hj\rangle$ becomes proportional to $P_2(\cos\theta)$, where, this time, $\theta$ is the angle between $j^p$ and $j^h$, the {\it shears angle}. Actually, these results were obtained in such a regime of the particle-vibration theory, in which there already exist a static quadrupole polarization of the core. Therefore, it is not surprising that they could also be derived within the mean-field approximation. Frauendorf considered \cite{Fra01a} the quadrupole-quadrupole model with some simplifications similar to that made by Macchiavelli and Clark. He presumed that in both protons and neutrons the ensemble of the valence nucleons can be treated as one object with constant quadrupole tensor and mean angular momentum pointing along one of its principal axes. Under these assumptions and with the aid of the addition theorem for spherical harmonics, the quadrupole-quadrupole interaction between the valence particles and holes transforms directly into the form of $P_2(\cos\theta)$. He also postulated that the valence nucleons are the only source of polarization of the otherwise spherical core and that the core energy depends quadratically on the induced quadrupole moment. The core defined in this way only modifies the strength of the $P_2(\cos\theta)$ force.

It comes both from the particle-vibration theory and from the mean-field quadrupole-quadrupole model that for the relevant particle-hole case the coupling constant of the $P_2(\cos\theta)$ interaction is positive. Since the polynomial $P_2(\cos\theta)$ itself has a minimum at $\theta=90^\circ$, the perpendicular coupling at the bandhead is reproduced. These facts lie at the origin of the following popular picture that can be found in the literature. States belonging to high-$j$ orbitals have doughnut-like shapes with mean angular momenta pointing along the symmetry axis of such a density distributions. Therefore, the right-angle orientation of the two spin vectors at the bandhead corresponds to a minimal spatial overlap of the particle and hole wave-functions. One usually says that this is because the particle-hole interaction is repulsive, and tends to minimize the overlap. This reasoning may be misleading, because it overlooks the core, whereas the $P_2(\cos\theta)$ interaction describes precisely the influence of the core polarization, and does not have the sense of a direct force.

With the $P_2(\cos\theta)$ interaction at hand, one can address the fundamental question of how the energy changes with spin in a shears band. Although the quadrupole correlations smooth out the $E(I)$ dependence, the shears mechanism of generating the angular momentum is substantially different from that occurring for the collective rotation, and it is not obvious whether it should also lead to the rotational-like behavior of $E\sim I(I+1)$. In their model study, Macchiavelli and Clark assumed that the $P_2(\cos\theta)$ force is the only source of the energy change along the band and that the whole angular momentum comes from the shears closing, that is, $I=j$. Under these assumptions, the energy, $E$, of a level belonging to a shears band has the form of $P_2(\cos\theta)$,
\begin{equation}
\label{E3c2t12_eqn}
E\sim\frac{3\cos^2\theta-1}{2}~,
\end{equation}
whereas from the shears geometry shown in Fig.~\ref{shears_fig} one easily obtains
\begin{equation}
\label{ct_eqn}
\cos\theta=\frac{I^2-(j^p)^2-(j^h)^2}{2j^pj^h}~.
\end{equation}
Substitution of (\ref{ct_eqn}) to (\ref{E3c2t12_eqn}) yields a fourth-order $E(I)$ dependence, and the fourth-order term is not only a correction, because the quadratic term has a negative sign. For given values of $j^p$ and $j^h$, the shears angle, $\theta$, can be extracted from the experimental spins according to (\ref{ct_eqn}) and the obtained dependence of $E(\theta)$ can be compared with the form (\ref{E3c2t12_eqn}). Such an analysis was done for $^{198,199}$Pb \cite{Mac98b} and $^{142}$Gd \cite{Rza01a}, giving a satisfactory agreement with (\ref{E3c2t12_eqn}). The fourth-order dependence translates into non-constant moments of inertia. This aspect was exploited in \cite{Mac98c}, where a qualitative agreement with the moments of inertia for a band in $^{197}$Pb was found.

The important role of the quadrupole polarization and the quasi-classical behavior of the angular momenta of the high-$j$ nucleons make the magnetic rotation suitable for the mean-field description within the cranking model. In this approach, the alignment properties of the valence particles and holes are governed by the deformation of the mean potential. It turns out that the high-$j$ particles and holes align their mean angular momentum vectors, $\vec{j}^p$ and $\vec{j}^h$, on the short and on the long axis of the deformed nucleus, respectively, giving the perpendicular coupling at the bandhead. With increasing rotational frequency, the two blades gradually align towards the axis of rotation, in agreement with the shears picture. Since the valence nucleons provide components of the angular momentum on two different axes, the considered system is an example of nuclear rotation about a direction tilted with respect to the principal axes of the nucleus. The variant of the cranking model that allows for such a rotation is called Tilted-Axis Cranking (TAC), and was originally developed by Frauendorf \cite{Fra93a}. In fact, the very idea of the shears mechanism was first inspired by TAC considerations, and a great majority of experimental data on the magnetic bands has been successfully reproduced within this theory. From TAC calculations, $\beta$ deformations as small as $\sim0.05$ and rarely approaching $\sim0.19$ are assigned to the shears bands. It seems, however, that smaller deformations are needed to reproduce the properties of shears bands than those that come from the energy minimization. Recently, Frauendorf suggested \cite{Fra03a} that the direct interaction between the valence nucleons, absent in the mean field, may be of some importance, or that isovector deformations may play some role. Self-consistent methods are necessary to address the latter point. The Skyrme-Hartree-Fock (Skyrme-HF) TAC approach developed in this work is one of the first theoretical tools appropriate for a study of such effects.

In general, in generating the angular momentum there is a competition between closing of the shears and the collective rotation. Two elements are decisive here: deformation and pairing correlations. If deformation is small, there is almost no collective motion and the whole inertia comes from the aligning blades. As deformation grows, obviously the core rotation becomes more important. But it also turns out that the angular momenta of the valence nucleons get more and more rigidly fixed with respect to the nuclear shape, retain their perpendicular coupling, and thus do not produce any increase in spin. This pushes the balance towards the collective rotation and leads to a change in the total moment of inertia. That change is not necessarily drastic, because the $\cJ^{(2)}$ moments of the shears bands are comparable to those of the well-deformed rotational structures. But quantities like $B(M1)/B(E2)$ and $\cJ^{(2)}/B(E2)$ are radically influenced, mainly due to the increase in $B(E2)$. For a given deformation, inclusion of pairing in the calculations will generally reduce the collective inertia. It is argued in Section~\ref{pairin_sec} that by introducing additional mixing of the s.p.\ states, pairing can also soften the deformation alignment of the valence particles and holes and thus make the shears close faster. These arguments indicate that the pair correlations favor the shears mechanism over the core rotation as far as the generation of the angular momentum is concerned. Macchiavelli and Clark studied \cite{Mac99a} the shears-rotor competition within a classical model of two blades interacting via the $P_2(\cos\theta)$ force, coupled to and a spherical rotor. By minimizing the energy at a given spin they found the balance between the shears and rotor contributions. By scaling the model parameters to the lead region they arrived at a universal estimate that the shears mechanism is expected to dominate for $\epsilon<\sim0.12$. For the known magnetic bands, it is estimated that the core rotation contributes $\sim15\%$ of the total spin.


For more information on the magnetic rotation and the shears mechanism, see the review articles by Clark and Macchiavelli \cite{Cla00a} and by Frauendorf \cite{Fra01a}. A compilation by Amita {\it et al.} \cite{Ami01a} lists detailed experimental and theoretical data on all the known magnetic dipole bands (May 2001) together with complete bibliography.

\newpage

\section{Chiral rotation}
\label{chiral_sec}

The history of chirality began in 1848, when Louis Pasteur noticed that molecules of some substances exist in two species, which turn the light polarization plane either left or right. Half a century later, lord Kelvin called these forms {\it chiral}, from the greek word $\chi\varepsilon\iota\rho$, meaning {\it hand}, in analogy to the left and right hand. {\it I call any geometrical figure, or group of points, chiral, and say it has chirality, if its image in a plane mirror, ideally realized, cannot be brought to coincide with itself.} - Lord Kelvin, Baltimore Lectures on Molecular Dynamics and the Wave Theory of Light, 1904. Thus, a system is chiral if it is not symmetric with respect to mirror-reflection, $\hat{S}$, in any plane. However, the chirality of nuclear rotation does not concern the position vector, but rather the axial pseudo-vector of angular momentum. Therefore, it is represented by the operator $\hat{R}^T$, product of the time reversal and rotation through $180^\circ$. It is easily verified that $\hat{R}^T$ acts on the angular momentum like $\hat{S}$ acts on the position vector.

\begin{figure}[t]
\begin{center}
\includegraphics{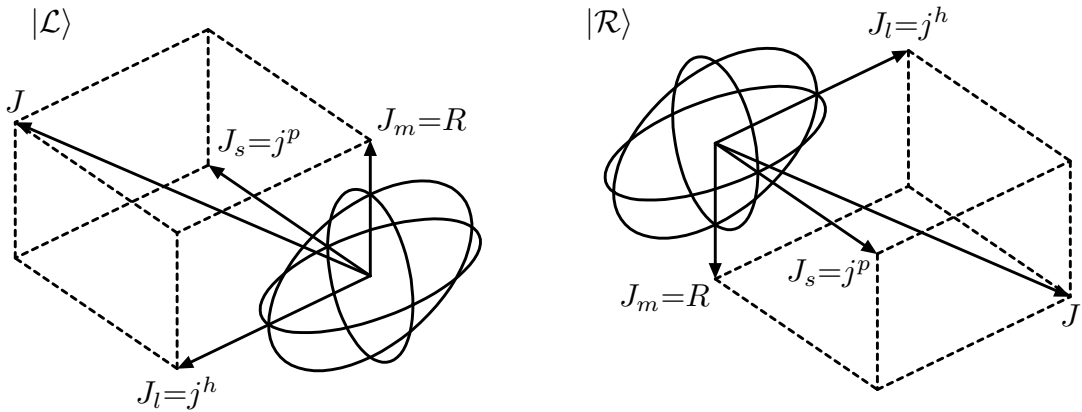}
\end{center}
\caption[Chirality of nuclear rotation.]{Triaxially deformed nucleus with the angular momentum, $\vec{J}$, having non-zero components, $\vec{J}_s$, $\vec{J}_m$, $\vec{J}_l$, on the short ($s$), medium ($m$), and long ($l$) axis. The left panel, $|\cL\rangle$, shows the left-handed orientation of those three vectors, and the right panel, $|\cR\rangle$, shows the right-handed orientation. The components $\vec{J}_s$, $\vec{J}_m$, $\vec{J}_l$ originate from the spin of the valence particle, $j^p$, from the collective angular momentum, $R$, and from the spin of the valence hole, $j^h$, respectively.}
\label{chiral_fig}
\end{figure}

The $\hat{R}^T$ symmetry was introduced into consideration in nuclear structure physics in 1997, by Frauendorf and Meng \cite{Fra97a}, who were inspired by a coincidence of two circumstances. One was that in 1987 Frisk and Bengtsson demonstrated \cite{Fri87a} that in triaxial nuclei mean-field cranking solutions can exist with the angular momentum vector, $\vec{J}$, having non-zero components, $\vec{J}_s$, $\vec{J}_m$, $\vec{J}_l$, on the short ($s$), medium ($m$), and long ($l$) axes, as illustrated in Fig.~\ref{chiral_fig}. Frauendorf and Meng pointed out that such an orientation of the angular momentum violates the $\hat{R}^T$ symmetry. Indeed, if vectors $\vec{J}_s$, $\vec{J}_m$, $\vec{J}_l$ form, say, a left-handed set\footnote{Of course, which handedness is which is a matter of convention. Here, the right-handed-screw orientation of space and the $s$-$m$-$l$ order of the axes is adopted.} (Fig.~\ref{chiral_fig}a), then their images in the mirror reflection, $\hat{R}^T$, will form a right-handed set (Fig.~\ref{chiral_fig}b). Like the left and right hand, the two systems cannot be superimposed by any rotation, and are thus chiral. Notice that the discussed most general orientation of $\vec{J}$ with respect to the triaxial shape also violates the signature symmetry, $\hat{R}$.

The second premise of Frauendorf and Meng was the observation, by Petrache {\it et al.} in 1996 \cite{Pet96a}, of an almost degenerate doublet of positive-parity $\Delta I=1$ bands in $^{134}$Pr. Appearance of such pairs can be explained in terms of the breaking of the chiral symmetry in the following way. Spectroscopic measurements provide us with eigenvalues of observables (spin, energy) that commute with $\hat{R}^T$. Therefore, quantum states observed in the laboratory do not violate $\hat{R}^T$. At a given spin, the two states of the doublet, $|+\rangle$ and $|-\rangle$, can be interpreted as
\begin{equation}
|+\rangle=\frac{1}{\sqrt{2}}(|\cL\rangle+|\cR\rangle)~, \qquad
|-\rangle=\frac{1}{\sqrt{2}}(|\cL\rangle-|\cR\rangle)~,
\end{equation}
where $|\cL\rangle$ and $|\cR\rangle$ are the left-handed and right-handed states represented in Fig.~\ref{chiral_fig}. Since states $|\cL\rangle$ and $|\cR\rangle$ are transformed onto one another by $\hat{R}^T$,
\begin{equation}
\hat{R}^T|\cL\rangle=|\cR\rangle~, \qquad \hat{R}^T|\cR\rangle=|\cL\rangle~,
\end{equation}
it is easily verified that $|+\rangle$ and $|-\rangle$ are, indeed, eigenstates of that operator,
\begin{equation}
\hat{R}^T|{\cal +}\rangle=|{\cal +}\rangle~, \qquad \hat{R}^T|{\cal -}\rangle=-|{\cal -}\rangle~.
\end{equation}
They are also orthogonal\footnote{More precisely, they can be always made orthogonal by an appropriate choice of phases of the states $|\cL\rangle$ and $|\cR\rangle$.}. The nuclear many-body Hamiltonian, $\hat{H}$, is invariant under $\hat{R}^T$, and therefore the states $|\cL\rangle$ and $|\cR\rangle$ have equal mean energies. At the same time, the matrix element of $\hat{H}$ between those states is, in general, non-zero, i.e.,
\begin{equation}
\langle\cL|\hat{H}|\cL\rangle=\langle\cR|\hat{H}|\cR\rangle=E~, \qquad |\langle\cL|\hat{H}|\cR\rangle|=\Delta\neq0~.
\end{equation}
As a consequence, the energies of the laboratory states, $|+\rangle$ and $|-\rangle$, read
\begin{equation}
\langle+|\hat{H}|+\rangle=E+\Delta~, \qquad \langle-|\hat{H}|-\rangle=E-\Delta~.
\end{equation}
Their energy splitting is, of course, due to the interaction between the left- and right-handed states. One can see that doublet bands like those observed in $^{134}$Pr may be attributed to the breaking of the chiral symmetry.

One can thus suppose that the appearance of the doublet bands is connected to the existence of triaxially deformed states with the angular momentum vector lying outside any principal plane. Frauendorf and Meng pointed out that in odd-odd nuclei around $^{134}$Pr such states can arise in the following way. In this region, a configuration is easily available, in which one proton particle occupies the lowest substate of the $h_{11/2}$ orbital, and simultaneously one neutron hole is left in the highest $h_{11/2}$ substate. The former drives the nucleus towards elongated shapes, while the latter towards disc-like forms. The interplay of these opposite tendencies may yield stable triaxiality. In the triaxially deformed potential, the considered particle and hole align their angular momenta, $j^p$ and $j^h$, along the short and long axis of the nucleus, respectively, providing the $\vec{J}_s$ and $\vec{J}_l$ components of $\vec{J}$. As expected from the hydrodynamical irrotational-flow model \cite{Boh75a}, the moment of inertia with respect to the medium axis is the largest, which energetically favors the collective rotation about this direction. Thus, the component $\vec{J}_m$ comes from the collective angular momentum, $R$. This is illustrated in Fig.~\ref{chiral_fig}.

Since the first observation in $^{134}$Pr, about 12 similar structures have been found in odd-odd nuclei in the $A\sim130$ region. They are all attributed to the discussed configuration of $\pi h_{11/2}^1~\nu h_{11/2}^{-1}$. Recently, first observations of possibly chiral bands in the even-odd $^{135}$Nd \cite{Zhu03a} and in the even-even $^{136}$Nd \cite{Mer02a} were reported in the same region. There, the proposed active configurations are $\pi h_{11/2}^2~\nu h_{11/2}^{-1}$ and $\pi h_{11/2}^2~\nu h_{11/2}^{-1}$, respectively. Peng, Meng, and Zhang predicted theoretically \cite{Pen03a} that another island of chirality, associated mainly with the configuration $\pi g_{9/2}^1~\nu g_{9/2}^{-1}$, may exist around $A\sim100$. Indeed, experimental work is underway for $^{102-106}$Rh \cite{Sta03a} and gives promising results. It seems that a pair of bands found in $^{104}$Rh \cite{Koi03b,Vam04a} provides the best known example of a chiral doublet, because the chiral partners are situated closest to each other. In this case, the $\pi g_{9/2}^{-1}~\nu h_{11/2}$ configuration is proposed. Latest lifetime measurements in $^{132}$La, by Grodner {\it et al.} \cite{Gro04a}, provided the first direct information about the absolute $B(E2)$ and $B(M1)$ values in the chiral bands; see Section~\ref{prechi_sec}. A complete (January 2004) review of experimental and theoretical investigations of the chiral bands in particular nuclides is given in Tabs.~\ref{chifir_tab} and \ref{chisec_tab}. See also Figs.~\ref{n75sch_fig} and \ref{lasche_fig} for sample level schemes and Figs.~\ref{csener_fig}--\ref{ppener_fig} for spin-energy plots.

Certainly, an important question about the observed doublet bands is whether the chiral rotation is the only possible explanation of their appearance. A more standard picture would be that they are in fact four $\Delta I=2$ bands corresponding to the two signature states of the two odd nucleons, one proton and one neutron. The energy separation between the bands would then be just the signature splitting. However, the calculated signature splitting turned out to be a few times larger than the observed distance \cite{Sta01b}. Also the possibility that the yrare partner band is a gamma-vibrational excitation built on the yrast band has been ruled out for the same reason. However, recently Pasternak \cite{Pas04a} managed to reproduce all the three bands observed in $^{132}$La (see Section~\ref{prechi_sec}) in a very simple classical model that invokes only planar rotation. Yet, some of the model parameters were fitted to each band separately.

Two theoretical tools have been used for the description of the chiral rotation: various versions of the Particle-Rotor Model (PRM), originally developed by Bohr and Mottelson \cite{Boh75a} and the mean-field Tilted Axis Cranking (TAC) model, which is discussed in Section~\ref{tac_sec} and used in this work. In the PRM approach, the nucleus is represented by a rotating deformed structureless core and a few valence particles interacting with it, and possibly among themselves. The valence nucleons are usually treated as pure particles or holes, while calculations with pairing appeared only recently \cite{Koi03a}. The core is usually taken as a triaxial rigid quantum rotor (Davydov-Fillipov model \cite{Dav58a}), characterized by three moments of inertia, $\cJ_s$, $\cJ_m$, $\cJ_l$, and described by the Hamiltonian
\begin{equation}
\label{prm_eqn}
\hat{H}_{rot}=\frac{\hat{R}_s^2}{2\cJ_s}+\frac{\hat{R}_m^2}{2\cJ_m}+\frac{\hat{R}_l^2}{2\cJ_l}~,
\end{equation}
where $\hat{R}_s$, $\hat{R}_m$, $\hat{R}_l$ are the intrinsic-frame components of the core angular momentum. In all available PRM studies, the moments of inertia, $\cJ_k$, are calculated from the irrotational-flow formula \cite{Boh75a}, 
\begin{equation}
\label{irrflo_eqn}
\cJ_k=4B\beta^2\sin^2\left(\gamma+\frac{2k\pi}{3}\right)~, \qquad k=1,2,3~,
\end{equation}
where $k=1,2,3$ corresponds to the short, medium, and long axis\footnote{The irrotational-flow formula, as it is given in Eq.~(\ref{irrflo_eqn}), and with such an assignment of $k$ to the three intrinsic axes, is valid for the $\gamma$ range from $0$ to $60^\circ$, which is used in the present work.}, respectively, and $B$ is a mass parameter. The important point is that PRM gives wavefunctions of good angular momentum, and thus describes the system in the laboratory frame. Therefore, it directly yields the splitting between the bands and the transition probabilities. Due to the presence of the rigid core, PRM does not take into account polarization of the nucleus by valence particles, nor change of shape induced by rotation. These effects are properly included in the mean-field TAC model, especially in its self-consistent version. The mean-field wavefunctions, in turn, explicitly break the chiral symmetry, and correspond to the $|\cL\rangle$ and $|\cR\rangle$ states in the intrinsic frame. This makes the physical interpretation more straightforward, but neither the splitting nor the transition probabilities can be directly calculated. The PRM and TAC methods are to an extent complementary. A more detailed comparison of the two models in their assumptions and results can be found in \cite{Fra96b,Fra97a}. It is noteworthy that the salient features of the chiral rotation can be elucidated within a very simple classical model of two gyroscopes, representing the valence particle and hole, coupled to a triaxial rigid body standing for the core. This model is discussed in \cite{Olb02b,Dim02a,Olb04a}, and in Section~\ref{clasic_sec}.

\begin{figure}[t]
\begin{center}
\includegraphics{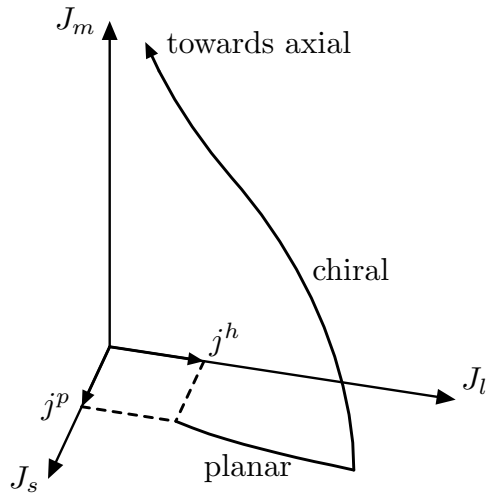}
\end{center}
\caption[Trajectory of the spin vector along a chiral band.]{Schematic drawing of the trajectory followed by the total angular momentum vector, $\vec{J}$, along planar and chiral bands. Indices $s$, $m$, $l$ refer to the short, medium and long intrinsic axis, respectively. Arrows marked as $j^p$ and $j^h$ denote the initial angular momenta of the valence proton particle and neutron hole. Regions of planar, chiral and axial rotation are marked.}
\label{evolut_fig}
\end{figure}

\begin{figure}[t]
\begin{center}
\includegraphics{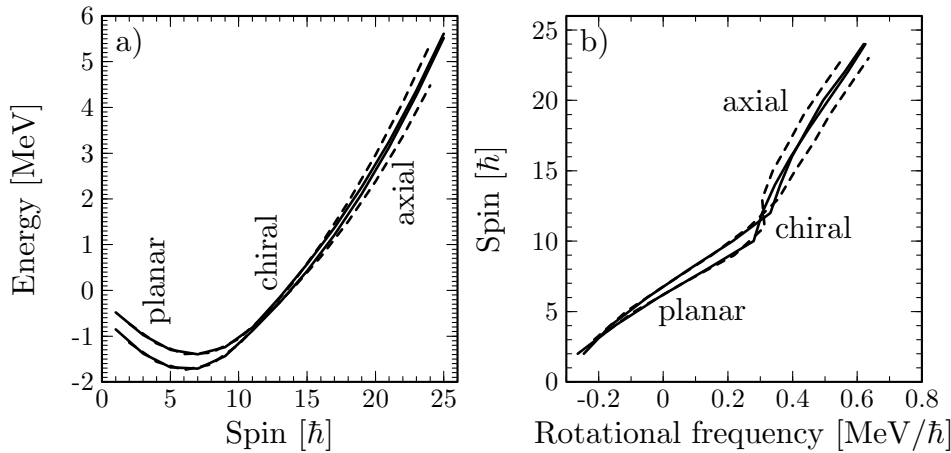}
\end{center}
\caption[A generic chiral band from PRM.]{A chiral doublet band for the $\pi g_{9/2}^1~\nu g_{9/2}^{-1}$ configuration in the $A\sim100$ region, calculated within the PRM \cite{Pen03a}. a) energy in function of spin, b) spin versus rotational frequency calculated as $\omega(I)=[E(I+1)-E(I-1)]/2$. Solid and dashed lines connect the points corresponding to odd and even spins, respectively. The regions of planar, chiral and axial rotation are marked.}
\label{prm_fig}
\end{figure}

The following general structure of a chiral band could be established from the theoretical studies. Figure~\ref{evolut_fig} shows a schematic trajectory of the mean angular momentum vector, $\vec{J}$, in the intrinsic frame, and Fig~\ref{prm_fig}b gives the $I(\omega)$ dependence calculated within the PRM \cite{Pen03a} for a generic $\pi g_{9/2}^1~\nu g_{9/2}^{-1}$ chiral band in the $A\sim100$ region. As already discussed, the valence particle and hole tend to align their spins, $\vec{j}^p$ and $\vec{j}^h$, along the short and long axes of the triaxial shape, respectively. At low spins, the total $\vec{J}$ comes mostly from those valence nucleons, and remains in the plane spanned by the short and long axes (marked as 'planar' in Fig~\ref{evolut_fig}). This is called {\it planar rotation}, in which the chiral symmetry is not broken. In this part of the band, the dependence of spin on rotational frequency is approximately linear (marked as 'planar' in Fig~\ref{prm_fig}b), with the second moment of inertia close to the moments, $\cJ_s$ and $\cJ_l$, for the short and long axes (in most studies, $\gamma\approx30^\circ$ is assumed, for which $\cJ_s\approx\cJ_l$). At a certain value of the rotational frequency, the spin vector moves out of the $s$-$l$ plane into one of the octants of the intrinsic frame, and the rotation becomes chiral (marked as 'chiral' in Fig~\ref{evolut_fig}). In the current study, an analytical estimate for that critical frequency has been found in terms of the classical model; see \cite{Olb02b,Olb04a} and Section~\ref{clasic_sec}. In the chiral regime, the spin-frequency plot is also a straight line (marked as 'chiral' in Fig~\ref{prm_fig}b), but $\cJ^{(2)}$ is now approximately equal to the moment of inertia, $\cJ_m$, associated with the medium axis. This results in a kink in Fig.~\ref{prm_fig}b around spin $11\,\hbar$. As the angular momentum vector traverses the selected octant further along the band, it aligns more and more with the medium axis (marked as 'towards axial' in Fig~\ref{evolut_fig}). Simultaneously, contributions from the valence particle and hole become negligibly small as compared to the total spin, and one effectively arrives at the axial rotation about the medium axis. In this regime, not only the chiral symmetry, but also the signature is unbroken. In view of such a scenario, it is stressed in the literature that the chiral rotation is a transient phenomenon, existing only in a limited range of angular momentum.

Figure~\ref{evolut_fig} shows the mean intrinsic-frame trajectory of the angular momentum vector along a chiral band. Classically, each point of that trajectory corresponds to a uniform rotation of the nucleus about a fixed direction, without any wobbling. Within the TAC model, the trajectory follows the minimum of the energy surface with increasing rotational frequency. The two chiral partners can be viewed as two lowest quantum states in the potential represented by that surface. In the single planar minimum existing at low spins, the two states are interpreted as zero- and one-phonon excitations separated by a finite energy interval. They correspond to oscillations of the spin vector about the planar equilibrium, which has a classical analogy in non-uniform rotations \cite{Dim02a}. In the chiral regime, there is the left and right minimum, of equal depths. If the barrier between them is high enough, the two lowest eigenstates are nearly degenerate, and one can talk about strong chiral symmetry breaking. For low barriers, the spin vector can tunnel between the minima, and the splitting between the band states is non-zero\footnote{However, in order to actually calculate the chiral splitting in such an approach, one would need the mass parameter in addition to the potential surface, and such calculations have not been done.}. This kind of weak symmetry breaking has been dubbed chiral vibrations \cite{Sta01b}. As the axial rotation is approached, the weakened breaking of both the chiral symmetry and the signature manifests itself, in particular, by the onset of the signature splitting. That splitting is absent for the planar and chiral rotation, because the signature is strongly broken in those modes. The above features appear clearly in the structure of the chiral band, as calculated in the PRM model. This can be seen from Fig.~\ref{prm_fig}, showing the plots of energy versus spin and spin versus rotational frequency.

The evolution from planar to chiral to axial finds a manifestation in the electromagnetic properties of the chiral bands, too. Those were examined in \cite{Fra97a,Pen03a,Pen03b}, mainly in the frame of the PRM model. For planar rotation, the transitions within each partner band are strong, the transitions from the yrare to the yrast band are weak, and there are practically no decays in the opposite direction. In the chiral regime, all reduced probabilities assume moderate values, and the intraband and interband transition strengths are comparable. As the planar rotation is approached, the $B(E2)$ within the two partners become equal and slowly increase, while the interband $E2$ decays disappear. The $B(M1)$ show the characteristic odd-even staggering, absent at lower spins. Reference \cite{Koi03a} gives a discussion of the staggering in $B(M1)/B(E2)$ and $B(M1)_{in}/B(M1)_{out}$ (interband/intraband) in connection with the signature splitting and inversion.

The appearance of the chiral bands is considered a strong evidence for the existence of triaxial deformations; certainly, the chiral geometry cannot occur in an axially symmetric nucleus. It is clear from the irrotational-flow formula (\ref{irrflo_eqn}) that $\gamma=30^\circ$ gives the largest moment of inertia with respect to the medium axis ($\cJ_m=4\cJ_s=4\cJ_l$) and thus favors the aplanar orientation of the angular momentum. The PRM study by Starosta {\it et al.} \cite{Sta01a,Sta02a} showed that the value of $\gamma$ practically does not influence the alignments of the valence particle and hole on the short and long axis. However, those alignments are only well defined for sufficiently large $\beta$ deformation. References \cite{Sta01a,Sta02a} investigated the influence of $\gamma$ and $\beta$ on the mean value of the so-called orientation operator, $\hat{\sigma}=(\vec{j}^p\times\vec{j}^h)\hat{\vec{R}}$, which measures the aplanarity of $\vec{j}^p$, $\vec{j}^h$, and $\vec{R}$, i.e., the degree of chirality, in a sense. The average value of $\hat{\sigma}$ increases with $\beta$ and has a maximum for $\gamma=30^\circ$. This is reflected in the splitting between the chiral partners, which decreases with $\beta$, and is minimal for $\gamma=30^\circ$. However, it is known that the flow of matter in nuclei is not purely irrotational, i.e., the curl of the velocity field is not exactly zero. Thus, the moments of inertia may deviate from the irrotational-flow values (\ref{irrflo_eqn}) assumed in the PRM studies, especially if pairing is weak, and the dependence of the discussed observables on $\gamma$ may not be so straightforward.

As far as the mean-field methods are concerned, certainly taking into account the interaction between the left- and right-handed minima would be desirable, preferably in conjunction with the projection onto good chirality before variation. This would be the first tool capable of calculating the chiral splitting and the transition probabilities in a fully microscopic way. In the PRM domain, efforts are being undertaken to include pairing in the calculations \cite{Koi03a}, to use $\gamma$-soft cores, and to take into account the direct residual interaction between the valence nucleons \cite{Rai03a,Rai03b}. Such an interaction is supposed to be responsible for the stabilization of the chiral geometry in nuclei at the limits of the islands of chirality. Experimentally, lifetime measurements would certainly be of great use.

\begin{table}[t]
\caption[Experimental and theoretical investigations of candidate chiral bands.]{Experimental and theoretical investigations of candidate chiral rotational bands; this Table is continued in Table~\ref{chisec_tab}. Acronyms in column Theory stand for calculations in the following models. TAC: phenomenological Tilted-Axis Cranking model of Frauendorf {\it et al.}; PRM: Particle-Rotor Model used by Peng {\it et al.} \cite{Pen03a,Pen03b} or its different version used by Hartley {\it et al.} \cite{Har01a}; CPH: Core-Particle-Hole model, a version of PRM developed by Starosta {\it et al.}; CQP: Core-Quasi-Particle model, a version of CPH that includes pairing; SHF: Skyrme-Hartree-Fock results of the present work; see Tab~\ref{crifre_tab} for details. The symbol $\nu$ in the column Pairing means that pairing was included only for neutrons.}
\label{chifir_tab}
\begin{center}
\begin{tabular}{r|ccccccc|c}
\hline\hline
           & Exp. & The. & Conf.                            & Pair.   & Def. & Tri. & $\omega_{crit}$ & Ref. \\
\hline
$A\sim100$ &     & PRM & $\pi g_{9/2}^1~\nu g_{9/2}^{-1}$   & no      & $\beta\sim0.12$  & $30^\circ$ & $\approx0.3$ & \cite{Pen03a} \\
\hline
$^{100}$Rh &     &     & $\pi g_{9/2}^{-1}~\nu h_{11/2}^1$  &         &                  &            & & \cite{Vam04a} \\
$^{102}$Rh &     &     & $\pi g_{9/2}^{-1}~\nu h_{11/2}^1$  &         &                  &            & & \cite{Vam04a} \\
$^{103}$Rh &     &     &                                    &         &                  &            & & \cite{Sta03a} \\
$^{104}$Rh &     &     &                                    &         &                  &            & & \cite{Koi03b} \\
           &     & PRM & $\pi g_{9/2}^{-1}~\nu h_{11/2}^1$  & no      & $\beta=0.12$     & $30^\circ$ & & \cite{Pen03a} \\
           &     & PRM & $\pi g_{9/2}^{-1}~\nu h_{11/2}^1$  & no      & $\beta=0.25$     & $25^\circ$ & $\approx0.3$ & \cite{Pen03b} \\
           & EXP &     & $\pi g_{9/2}^{-1}~\nu h_{11/2}^1$  &         &                  &            & & \cite{Vam04a} \\
$^{105}$Rh &     &     &                                    &         &                  &            & & \cite{Sta03a} \\
$^{106}$Rh &     & PRM & $\pi g_{9/2}^{-1}~\nu h_{11/2}^1$  & no      & $\beta=0.25$     & $25^\circ$ & $\approx0.3$ & \cite{Pen03b} \\
           &     &     & $\pi g_{9/2}^{-1}~\nu h_{11/2}^1$  &         &                  &            & & \cite{Vam04a} \\
$^{108}$Rh &     & PRM & $\pi g_{9/2}^{-1}~\nu h_{11/2}^1$  & no      & $\beta=0.25$     & $25^\circ$ & $\approx0.3$ & \cite{Pen03b} \\
\hline
$^{110}$Ag &     & PRM & $\pi g_{9/2}^{-1}~\nu h_{11/2}^1$  & no      & $\beta=0.25$     & $25^\circ$ & $\approx0.3$ & \cite{Pen03b} \\
\hline
$^{118}$I  & EXP &     & $\pi g_{9/2}^{-1}~\nu h_{11/2}$    &         &                  &            & & \cite{Sta01a} \\
\hline
$^{124}$Cs &     &     &                                    &         &                  &            & & \cite{Koi03a} \\
$^{126}$Cs & EXP &     & $\pi h_{11/2}^1~\nu h_{11/2}^{-1}$ &         &                  &            & & \cite{Li02a} \\
           &     &     &                                    &         &                  &            & & \cite{Koi03a} \\
$^{128}$Cs & EXP &     & $\pi h_{11/2}^1~\nu h_{11/2}^{-1}$ &         &                  &            & & \cite{Koi01a} \\
           & EXP & CQP & $\pi h_{11/2}^1~\nu h_{11/2}^{-1}$ & yes     & $\beta=0.19$     & $27^\circ$ & & \cite{Koi03a} \\
$^{130}$Cs & EXP &     & $\pi h_{11/2}^1~\nu h_{11/2}^{-1}$ &         &                  &            & & \cite{Sta01a} \\
           & EXP & TAC & $\pi h_{11/2}^1~\nu h_{11/2}^{-1}$ & yes     & $\epsilon=0.16$  & $39^\circ$ & & \cite{Sta01b} \\
           & EXP &     & $\pi h_{11/2}^1~\nu h_{11/2}^{-1}$ &         &                  &            & & \cite{Koi03a} \\
           &     & PRM & $\pi h_{11/2}^1~\nu h_{11/2}^{-1}$ & no      & $\epsilon=0.16$  & $39^\circ$ & & \cite{Pen03a} \\
           &     & SHF & $\pi h_{11/2}^1~\nu h_{11/2}^{-1}$ & no      & $\beta=0.24$     & $48^\circ$ & $\approx0.4$ & present \\
$^{132}$Cs & EXP &     & $\pi h_{11/2}^1~\nu h_{11/2}^{-1}$ &         &                  &            & & \cite{Koi03a} \\
           & EXP & TAC & $\pi h_{11/2}^1~\nu h_{11/2}^{-1}$ & no      & $\epsilon=0.16$  & $36^\circ$ & $\approx0.18$ & \cite{Rai03a} \\
           & EXP & CPH & $\pi h_{11/2}^1~\nu h_{11/2}^{-1}$ &         &                  &            & & \cite{Rai03b} \\
\end{tabular}
\end{center}
\end{table}

\begin{table}[t]
\caption[Continuation of Table~\ref{chifir_tab}.]{Continuation of Table~\ref{chifir_tab}.}
\label{chisec_tab}
\begin{center}
\begin{tabular}{r|ccccccc|c}
\hline\hline
           & Exp. & The. & Conf.                            & Pair.   & Def. & Tri. & $\omega_{crit}$ & Ref. \\
\hline
$^{130}$La & EXP &     & $\pi h_{11/2}^1~\nu h_{11/2}^{-1}$ &         &                  &            & & \cite{Koi01a} \\
$^{132}$La & EXP & CPH & $\pi h_{11/2}^1~\nu h_{11/2}^{-1}$ & no      &                  &            & & \cite{Sta01a} \\
           & EXP & TAC & $\pi h_{11/2}^1~\nu h_{11/2}^{-1}$ & yes     & $\epsilon=0.175$ & $32^\circ$ & & \cite{Sta01b} \\
           & EXP & CPH & $\pi h_{11/2}^1~\nu h_{11/2}^{-1}$ & no      & $\beta=0.23$     & $21^\circ$ & & \cite{Sta02a} \\
           & EXP &     & $\pi h_{11/2}^1~\nu h_{11/2}^{-1}$ & \multicolumn{4}{c|}{Lifetimes measured}   & \cite{Gro04a} \\
           &     & PRM & $\pi h_{11/2}^1~\nu h_{11/2}^{-1}$ & no      & $\epsilon=0.175$ & $32^\circ$ & & \cite{Pen03a} \\
           &     & SHF & $\pi h_{11/2}^1~\nu h_{11/2}^{-1}$ & no      & $\beta=0.26$     & $46^\circ$ & $\approx0.5$ & present \\
$^{134}$La & EXP & PAC & $\pi h_{11/2}^1~\nu h_{11/2}^{-1}$ &         & $\epsilon=0.14$  & $30^\circ$ & & \cite{Bar01a} \\
\hline
$^{132}$Pr & EXP &     & $\pi h_{11/2}^1~\nu h_{11/2}^{-1}$ &         &                  &            & & \cite{Koi01a} \\
$^{134}$Pr &     & TAC & $\pi h_{11/2}^1~\nu h_{11/2}^{-1}$ & yes     & $\epsilon=0.175$ & $27^\circ$ & 0.3 & \cite{Dim00a} \\
           & EXP &     & $\pi h_{11/2}^1~\nu h_{11/2}^{-1}$ &         &                  &            & & \cite{Sta01a} \\
           & EXP & TAC & $\pi h_{11/2}^1~\nu h_{11/2}^{-1}$ & yes     & $\epsilon=0.175$ & $27^\circ$ & & \cite{Sta01b} \\
           & EXP & CPH & $\pi h_{11/2}^1~\nu h_{11/2}^{-1}$ & no      & $\beta=0.25$     & $35^\circ$ & & \cite{Sta02a} \\
           &     & PRM & $\pi h_{11/2}^1~\nu h_{11/2}^{-1}$ & no      & $\epsilon=0.175$ & $27^\circ$ & & \cite{Pen03a} \\
           &     & SHF & $\pi h_{11/2}^1~\nu h_{11/2}^{-1}$ & no      & $\beta=0.26$     & $58^\circ$ & & present \\
           &     & SHF & $\pi h_{11/2}^1~\nu h_{11/2}^{-1}$ & no      & $\beta=0.23$     & $22^\circ$ & $\approx0.9$ & present \\
$^{136}$Pm & EXP &     & $\pi h_{11/2}^1~\nu h_{11/2}^{-1}$ &         &                  &            & & \cite{Bea01a} \\
           & EXP &     & $\pi h_{11/2}^1~\nu h_{11/2}^{-1}$ &         &                  &            & & \cite{Sta01a} \\
           & EXP & TAC & $\pi h_{11/2}^1~\nu h_{11/2}^{-1}$ & yes     & $\epsilon=0.195$ & $27^\circ$ & & \cite{Sta01b} \\
           & EXP & TAC & $\pi h_{11/2}^1~\nu h_{11/2}^{-1}$ &         & $\epsilon=0.194$ & $25^\circ$ & $<0.2$ & \cite{Hec01a} \\
           & EXP & PRM & $\pi h_{11/2}^1~\nu h_{11/2}^{-1}$ & no      & $\epsilon=0.194$ & $25^\circ$ & & \cite{Har01a} \\
           &     & PRM & $\pi h_{11/2}^1~\nu h_{11/2}^{-1}$ & no      & $\epsilon=0.195$ & $27^\circ$ & & \cite{Pen03a} \\
           &     & SHF & $\pi h_{11/2}^1~\nu h_{11/2}^{-1}$ & no      & $\beta=0.25$     & $53^\circ$ & & present \\
           &     & SHF & $\pi h_{11/2}^1~\nu h_{11/2}^{-1}$ & no      & $\beta=0.22$     & $19^\circ$ & $\approx0.6$ & present \\
\hline
$^{138}$Eu & EXP &     & $\pi h_{11/2}^1~\nu h_{11/2}^{-1}$ &         &                  &            & & \cite{Bea01a} \\
           &     & TAC & $\pi h_{11/2}^1~\nu h_{11/2}^{-1}$ & yes     & $\epsilon=0.225$ & $20^\circ$ & & \cite{Sta01b} \\
           & EXP & TAC & $\pi h_{11/2}^1~\nu h_{11/2}^{-1}$ &         & $\epsilon=0.202$ & $24^\circ$ & $<0.2$ & \cite{Hec01a} \\
\hline
$^{135}$Nd & EXP & TAC & $\pi h_{11/2}^2~\nu h_{11/2}^{-1}$ & $\nu$   & $\epsilon=0.20$  & $30^\circ$ & $\approx0.45$ & \cite{Zhu03a} \\
$^{136}$Nd & EXP &     & $\pi h_{11/2}^1~\nu h_{11/2}^{-2}$ &         &                  &            & & \cite{Mer02a} \\
\hline
$^{188}$Ir &     & TAC &                                    & yes     & $\epsilon=0.21$  & 40$^\circ$ & $<0.2$ & \cite{Dim00a} \\
\end{tabular}
\end{center}
\end{table}

\chapter{Theoretical tools}
\label{thetol_cha}

In this Chapter, some general information about the spontaneous symmetry breaking in the mean field is recalled, mainly in the context of rotational bands. While the Tilted-Axis Cranking (TAC) method has already been widely discussed and used in its version based on phenomenological potentials, this work introduces one of its first applications to the self-consistent Hartree-Fock (HF) calculations. Therefore, the TAC basics are recapitulated with particular emphasis on the differences between the phenomenological and self-consistent implementations. It is also described how the values of the time-odd coupling constants of the HF energy functional are chosen in the present calculations to investigate the role of the time-odd nucleonic densities and fields in the shears and chiral solutions. After a concise review of other computer codes capable of performing symmetry-unrestricted mean-field calculations, the program \pr{HFODD}, used in this work, is presented.

\section{Spontaneous symmetry breaking}
\label{spobre_sec}

This Section summarizes the basics of the HF method, and explains how the spontaneous symmetry breaking intervenes in the mean-field approach.

The variational HF method consists, in its standard version, in minimizing the expectation value of the many-body Hamiltonian,
\begin{equation}
\hat{H}=\hat{T}+\hat{V}~,
\end{equation}
in the trial class of Slater determinants. Here, $\hat{T}$ is the kinetic energy, and $\hat{V}$ is the effective two-body interaction. A Slater determinant, $|\Psi\rangle$, is uniquely characterized by its hermitian, projective density matrix,
\begin{equation}
\rho_{\alpha\beta}=\langle\Psi|a_\beta^+a_\alpha|\Psi\rangle~.
\end{equation}
For a given density, one introduces the so-called s.p.\ Hamiltonian, $\hat{h}[\rho]$, which is the functional derivative of the energy, $E=\langle\Psi|\hat{H}|\Psi\rangle$, with respect to the density,
\begin{equation}
h[\rho]_{\beta\alpha}=\frac{\partial E}{\partial\rho_{\alpha\beta}}~.
\end{equation}
By evaluating the derivative one obtains
\begin{equation}
\hat{h}[\rho]=\hat{T}+\hat{\Gamma}[\rho]~,
\end{equation}
where $\hat{\Gamma}[\rho]$ is the mean potential generated by averaging the two-body interaction with the density distribution,
\begin{equation}
\Gamma[\rho]_{\mu\nu}=\sum_{\alpha\beta}V_{\mu\beta\nu\alpha}\rho_{\alpha\beta}~.
\end{equation}
The HF equations state that a local energy minimum is attained for such a density matrix, which commutes with the s.p.\ Hamiltonian it induces,
\begin{equation}
\label{hfeqns_eqn}
\left[\rho,\hat{h}[\rho]\right]=0~.
\end{equation}
The last equation tells us, in particular, that the density matrix and the s.p.\ Hamiltonian have common eigenstates. Therefore, the Slater determinant corresponding to the HF solution represents a set of particles moving in the mean potential $\hat{\Gamma}[\rho]$. Although it is known that a Slater determinant provides a poor approximation to the nuclear stat, the obtained approximation to the s.p.\ density, $\rho$, is much better. Therefore, like in the Density Functional Theory, of which the HF method is a particular case, a reliable physical result is contained in the density.

The nuclear many-body Hamiltonian, $\hat{H}$, is invariant under a number of symmetry operations, $\hat{S}$,
\begin{equation}
[\hat{H},\hat{S}]=0~.
\end{equation}
Normally, these {\it conserved} symmetries include the translation, rotation, plane reflection, time reversal, and particle-number symmetries. This list can be supplemented with the approximate isospin symmetry, which is broken mostly by the Coulomb interaction, and by the Galilean invariance, which is only a symmetry of the interaction $\hat{V}$. In different physical conditions, e.g., when the system is subjected to an external electromagnetic field, some of those symmetries are {\it unconserved}. The nuclear states belong to irreducible representations of the group formed by the conserved symmetries of the Hamiltonian $\hat{H}$. In general, those representations may be multi-dimensional, like for the angular-momentum eigenstates with $L>0$. Such states are said to be {\it covariant} with the group operations. States that are bases of one-dimensional representations, like the angular-momentum eigenstates with $L=0$, are called {\it invariant}\footnote{Also for one-dimensional representations different than the totally symmetric one.}. In the present work, only conservation of single dichotomic symmetries, like plane reflections, is analyzed. A dichotomic symmetry is a two-element group containing the given symmetry operation and the neutral element. Such a group has only two irreducible representations, both one-dimensional, the symmetric and the anti-symmetric one.

In the variational HF approach, one does not calculate exact eigenstates of $\hat{H}$, but looks for their approximation among possibly simple states, the Slater determinants. If, for instance, there are strong quadrupole correlations in the exact solution, the only way the simple determinant can account for them may be by a static quadrupole deformation, which violates the rotational invariance. Thus, in general, the mean-field solution does not possess the symmetries of the many-body Hamiltonian, which means that the density matrix, $\rho$, corresponding to the minimum energy, does not commute with the symmetry operator, $\hat{S}$,
\begin{equation}
[\rho,\hat{S}]\neq0~;
\end{equation}
then one says that the symmetry $\hat{S}$ is spontaneously {\it broken} in the HF solution. But it may also remain {\it unbroken}, which depends on the constituents of the system and on the interaction $\hat{V}$. It is clear that symmetries broken in the mean field are reminiscent of the presence of certain correlations in the exact solution. Although the quantum mechanics formally requires a proper symmetry behavior, one may note that the mean-field description eventually becomes exact in the limit of large systems, like macroscopic crystals or even molecules. Thus, it is also sufficient for the description of certain properties of nuclei. If information of a more quantal character is required, like the excitation spectrum, one may need to restore the broken symmetries, e.g., by projecting the mean-field solutions onto good angular momentum. See \cite{Rin00a} for a more thorough discussion.

An important result about the symmetry breaking in the HF approach is provided by the theorem of self-consistent symmetries \cite{Rin00a}, which requires that if $\hat{S}$ is a symmetry of the many-body Hamiltonian, $\hat{H}$, then
\begin{equation}
\label{selsym_eqn}
\hat{S}^+\hat{h}[\rho]\hat{S}=\hat{h}[\hat{S}^+\rho\hat{S}]~.
\end{equation}
In particular, if $\rho$ is invariant under $\hat{S}$ then so is $\hat{h}$. On the other hand, if $\hat{h}$ commutes with $\hat{S}$ then its s.p.\ eigenstates can be chosen as eigenstates of $\hat{S}$, which means that also $\rho$ is invariant under $\hat{S}$. Consequently, the s.p.\ density and the s.p.\ Hamiltonian have the same symmetries in a self-consistent solution. In the present study, the condition of the theorem are always satisfied, because the Skyrme interaction is rotationally invariant, and only subgroups of the rotation group are considered here. 

The exact solutions are often referred to as {\it laboratory} states, which are, in general, covariant with the symmetry group of the many-body Hamiltonian. One says that spontaneous symmetry breaking occurs in the {\it intrinsic frame}, which actually alludes to the mean-field solutions. In Section~\ref{chiral_sec}, the states $|+\rangle$ and $|-\rangle$ of a chiral doublet are the laboratory states, and the left- and right-handed states $|\cL\rangle$ and $|\cR\rangle$ are mean-field solutions, while the passage from the latter to the former is the projection onto good chirality.

\section{Symmetries and rotational bands}

Relation between the symmetries of the nucleus and the structure of rotational bands is presented in this Section. The symmetry group $D_{2h}^T$, which plays a crutial role in the present analysis, is defined.

For all quantum systems whose rotational excitations are observed, like molecules or nuclei, one can find universal connections between the symmetries in the intrinsic frame and the sequence of spin and parity of levels within a rotational band. This can be done by checking what values of angular momentum and parity can be projected out of a mean-field solution with given symmetries unbroken. Frauendorf \cite{Fra01a,Fra01b} summarized the results of such an analysis in a table which is reproduced in Table~\ref{symmet_tab}. It can be seen that three symmetry operations are relevant for the classification of the bands, namely, the space inversion or parity, $\hat{P}$ (reflection in a point), the signature, $\hat{R}$ (rotation through $180^\circ$), and the chiral symmetry, $\hat{R}^T$ (product of the signature and the time reversal). Not all of the 15 mathematically conceivable kinds of bands listed in the Table have been experimentally found, and the interesting question whether they all appear in nuclei remains open.

\begin{table}[t]
\caption[Types of rotational bands and symmetries of the nucleus.]{Symmetries of the nucleus in the intrinsic frame and sequences of spins and parities in the corresponding rotational bands. The parity ($\hat{P}$), signature ($\hat{R}_z$), and chiral symmetry ($\hat{R}^T_y$) operators are defined in the text. Symbols U and B stand for unbroken and broken, respectively; an operator as entry means that this operator gives the same result as the one given in the header of the column. $I^\pm$ means that there are two degenerate states of opposite parity. The $2$ means that there are two degenerate states with the same $I^\pm$. In rows I-V, the parity of the band can be positive or negative, although the latter possibility is not explicitly indicated. From \cite{Fra01a,Fra01b}.}
\label{symmet_tab}
\begin{center}
\begin{tabular}{c|ccc|c}
\hline\hline
& $\hat{P}$          & $\hat{R}_z$ & $\hat{R}^T_y$ & Level sequence \\
\hline
I    & U             & U           & U  	   & $I^+,(I+2)^+,(I+4)^+$ \\
II   & U             & B           & U  	   & $I^+,(I+1)^+,(I+2)^+$ \\
III  & U             & B           & B  	   & $2I^+,2(I+1)^+,2(I+2)^+$ \\
IV   & U             & U           & B  	   & $2I^+,2(I+2)^+,2(I+4)^+$ \\
V    & U             & B           & $\hat{R}_z$   & $I^+,(I+1)^+,(I+2)^+$ \\
VI   & B             & U           & U  	   & $I^\pm,(I+2)^\pm,(I+4)^\pm$ \\
VII  & B             & B           & U  	   & $I^\pm,(I+1)^\pm,(I+2)^\pm$ \\
VIII & B             & U           & B  	   & $2I^\pm,2(I+2)^\pm,2(I+4)^\pm$ \\
IX   & B             & B           & $\hat{R}_z$   & $I^\pm,(I+1)^\pm,(I+2)^\pm$ \\
X    & $\hat{R}_z$   & B           & U  	   & $I^+,(I+1)^-,(I+2)^+$ \\
XI   & $\hat{R}_z$   & B           & B  	   & $2I^+,2(I+1)^-,2(I+2)^+$ \\
XII  & $\hat{R}^T_y$ & U           & B  	   & $I^\pm,(I+2)^\pm,(I+4)^\pm$ \\
XIII & $\hat{R}^T_y$ & B           & B  	   & $I^\pm,(I+1)^\pm,(I+2)^\pm$ \\
XIV  & $\hat{R}_z$   & B           & $\hat{R}_z$   & $I^+,(I+1)^-,(I+2)^+$ \\
XV   & B             & B           & B  	   & $2I^\pm,2(I+1)^\pm,2(I+2)^\pm$ \\
\end{tabular}
\end{center}
\end{table}

Together with the important time-reversal operation, the symmetries appearing in Table~\ref{symmet_tab} generate a group called $D_{2h}^T$ \cite{Dob00b}, which is of crucial importance in the present analysis. The group comprises\footnote{The identity element, $1$, the change of sign, $-1$, and the products of $-1$ with other elements of the group are not mentioned.}:
\begin{itemize}
\item Three signature operations, $\hat{R}_i=\exp(-i\pi\hat{J}_i)$, $i=x,y,z$, which are rotations through $180^\circ$ about the three cartesian axes. Here, $\hat{J}_i$ denotes components of the angular momentum operator.
\item Space invertion, $\hat{P}$, sometimes called parity, changing $\vec{r}$ into $-\vec{r}$.
\item Three simplex operations, $\hat{S}_i=\hat{P}\exp(-i\pi\hat{J}_i)$, which are reflections in the planes perpendicular to the $i=x,y,z$ axes.
\item Antilinear time-reversal operator, $\hat{T}=\exp(-i\pi\hat{s}_y)\hat{K}$, where $\hat{s}_y$ is the $y$ component of the total intrinsic spin operator, and $\hat{K}$ is the complex conjugation in the space representation.
\item $T$-signatures, $\hat{R}^T_i=\hat{T}\hat{R}_i$, T-parity, $\hat{P}^T=\hat{T}\hat{P}$, and $T$-simplexes, $\hat{S}^T_i=\hat{T}\hat{S}_i$, where $i=x,y,z$.
\end{itemize}
The $T$-signatures are the chiral operators discussed in Section~\ref{chiral_sec}.

When enumerating the $D_{2h}^T$ symmetries it is necessary to introduce names of the three cartesian axes. But obviously, the question in the mean-field calculations is not, e.g., whether the signature with respect to some particular a priori given axis is broken or not. One says that the signature is broken uniquely if there is no axis $i$ such that $\hat{R}_i$ is the symmetry of the density. Otherwise, the signature is not broken. The same pertains to the simplex, $T$-signature, and $T$-simplex symmetries. In the following, the terms broken and unbroken are always understood in this sense, when referring to the signature, simplex, $T$-signature, and $T$-simplex symmetries, unless a particular axis is specified.

\section{Tilted-Axis Cranking}
\label{tac_sec}

To obtain HF solutions describing states with higher angular momentum, particularly the levels of a rotational band, one usually imposes a linear constraint on the total angular momentum, and minimizes the mean value of the quantity called Routhian,
\begin{equation}
\label{HHwJ_eqn}
\hat{H}'=\hat{H}-\vec{\omega}\hat{\vec{J}}~,
\end{equation}
where the Lagrange multiplier $\vec{\omega}$ is called {\it rotational frequency}, and $\hat{\vec{J}}$ is the angular-momentum operator. This approach is called {\it cranking} approximation. Since $\hat{H}$ is rotationally invariant, the solutions obtained for the same length, but different directions of $\vec{\omega}$ differ only by their orientation in space. Therefore, only the length, $\omega$, has a physical meaning. The Kerman-Onishi theorem \cite{Ker81a} states that, at the minimum, the total angular momentum vector, $\vec{J}=\langle\hat{\vec{J}}\rangle$, is parallel to $\vec{\omega}$. Indeed, this configuration obviously minimizes the mean value of the cranking term, $-\vec{\omega}\hat{\vec{J}}$, while $\hat{H}$ is rotationally invariant. It comes from the variation of the many-body Routhian, $\hat{H}'$, that in the cranking model the s.p.\ Hamiltonian, $\hat{h}$, is replaced by the s.p.\ Routhian,
\begin{equation}
\hat{h}'=\hat{h}-\vec{\omega}\hat{\vec{J}}~,
\end{equation}
both in the HF equations (\ref{hfeqns_eqn}) and in the theorem of self-consistent symmetries (\ref{selsym_eqn}).

If a given symmetry is not broken in a mean-field solution, expectation values of observables represented by operators odd under this symmetry vanish. Here, we are interested in properties of the $D_{2h}^T$ symmetries with respect to the mean angular momentum vector and electric quadrupole moment of the mass distribution, because they will play a fundamental role in the search for broken signature and chiral symmetries. These properties can be summarized as (see \cite{Dob00b} for a detailed derivation):
\begin{itemize}
\item If the time reversal, $\hat{T}$, is not broken, all components of the angular momentum vanish, while the quadrupole moments are not affected.
\item Unbroken parity, $\hat{P}$, does not constrain neither the angular momentum nor the quadrupole moments.
\item If $\hat{R}_i$ or $\hat{S}_i$ is unbroken, the angular momentum can have a non-zero component only on the $i$-th axis, which in turn is a principal axis of the quadrupole tensor.
\item If $\hat{R}^T_i$ or $\hat{S}^T_i$ is unbroken, the axis $i$ is a principal axis of the quadrupole tensor, while the angular-momentum vector is confined to the plane perpendicular to that axis.
\end{itemize}
These conclusions are valid, of course, for any arbitrarily oriented axis $i$, not only the $x$, $y$, $z$ axes of some particular frame. It follows from the third bullet that if the angular momentum has non-zero components on at least two principal axes of the quadrupole tensor, then there does not exist any axis $i$ such that $\hat{R}_i$ or $\hat{S}_i$ be an unbroken symmetry. The last point tells us that if the angular momentum has non-zero components on all the three principal axes of the quadrupole tensor, then there is no such axis $i$ that $\hat{R}^T_i$ or $\hat{S}^T_i$ be unbroken.

It is clear from the above properties of the $D_{2h}^T$ symmetries that whenever there exist an axis $i$ such that either $\hat{R}_i$, $\hat{S}_i$, $\hat{R}^T_i$, or $\hat{S}^T_i$ is an unbroken symmetry, then $i$ is a principal axis of the quadrupole tensor. So, only the principal axes and planes of this tensor are candidates to being symmetry axes and symmetry planes. It is, therefore, useful to consider the {\it intrinsic frame} of reference, defined by the principal axes of the quadrupole moment. To denote the axes of the intrinsic frame, the symbols $s$, $m$, $l$, are used, in view of the short, medium and long axes of a triaxially deformed shape. Let us also introduce the {\it program frame}, which is defined by the axes used for solving the mean-field equations, e.g., in a computer code\footnote{It would be unfortunate to call it laboratory frame, because in the common interpretation of the cranking model one would say that the program frame rotates in the laboratory system. The program and intrinsic frames do not move with respect to each other, they differ only by the orientation of their axes.}. They are denoted as $x$, $y$, $z$. In general, the axes of the two frames do not coincide.

If one wants to check by using the mean-field methods whether a given symmetry is spontaneously broken or not in the considered system, one looks for the minimum of the Routhian among trial solutions both breaking and not breaking the symmetry in question. Then one checks, to which of those two categories the minimum belongs. These are the {\it symmetry-unrestricted} calculations. One can also {\it impose a given symmetry}, which amounts to searching for the Routhian minimum only among solutions that do not break that symmetry. When performing calculations with imposed $D_{2h}^T$ symmetries one can, without loss of generality, impose the symmetries associated with the axes of the program frame, e.g., $\hat{R}_x$, which is technically much simpler. In fact, imposing either $\hat{R}_i$, $\hat{S}_i$, $\hat{R}^T_i$ or $\hat{S}^T_i$, where $i=x,y,z$, causes that one of the intrinsic axes coincides with the program $i$-th axis. In the following, the term {\it imposed} always has this meaning.

If some symmetries are imposed in the calculations, notably the cranking term, $-\vec{\omega}\hat{\vec{J}}$, must conform to them. This means that all the above restrictions concerning the mean angular momentum, $\vec{J}$, equally touch the cranking frequency, $\vec{\omega}$. For example, solutions invariant under the time reversal may only result if no cranking is applied. These remarks allow us to introduce the following widely used classification of calculations with imposed symmetries. Whenever a signature $\hat{R}_i$ or simplex $\hat{S}_i$ symmetry is imposed, one can talk about {\it one-dimensional cranking}, because the applied rotational frequency, $\vec{\omega}$, together with all resulting (single-particle (s.p.) and total) angular momenta vectors, point along the unique, $i$-th axis. A name {\it Principal-Axis Cranking} (PAC) is also adopted, because the $i$-th axis is then a principal axis of the mass distribution (its quadrupole moment). In PAC calculations, the Kerman-Onishi condition ($\vec{J}\parallel\vec{\omega}$) is satisfied identically. In the {\it Tilted-Axis Cranking} (TAC), all signature and simplex symmetries are broken, and therefore the frequency, $\vec{\omega}$, and the mean angular momentum, $\vec{J}$, vectors can become tilted with respect to the principal axes. If either a $T$-signature or $T$-simplex is still imposed, these vectors remain in a single plain, and hence the name of {\it planar} or {\it two-dimensional cranking} (2D) is used. If not, $\vec{\omega}$ and the resulting angular momentum, $\vec{J}$, can take any direction, and one talks about {\it three-dimensional cranking} (3D). In both the 2D and 3D cases, still the parity can be either broken or not.

To solve the HF equations, one often uses the iterative method \cite{Rin00a}. It follows from the theorem of self-consistent symmetries (see Section~\ref{spobre_sec}) that if a symmetry is not broken in a certain iteration, it will remain so in all subsequent ones, and consequently in the final result. Therefore, to impose a symmetry it is enough to start the iterations from an $\hat{h}$ even under this symmetry and also to apply the cranking frequency $\vec{\omega}$ in a proper direction, as discussed above\footnote{It is still possible that an initially unbroken symmetry breaks due to numerical noise. To avoid such a noise, in practical calculations the terms of $\hat{h}'$ odd under the considered symmetry are simply disregarded.}. To render the calculations symmetry-unrestricted, either the initial $\hat{h}$ or the cranking term must violate the considered symmetry. If the symmetry is not dynamically broken in the physical solution, it will recover in the course of the iterations. In the present calculations, a $D_{2h}^T$-symmetric Nilsson \cite{Nil55a} potential is always used as the first approximation to $\hat{h}$, unless the iterations are restarted from some previously obtained solution. Therefore, the $D_{2h}^T$ symmetries can be broken only by the cranking term. If one wants to perform symmetry-unrestricted calculations for a non-rotating state, one may nevertheless apply a small cranking frequency in the first iteration, only to break the $D_{2h}^T$ symmetries, and then continue with no cranking.

To relate the discussion to the magnetic and chiral rotation, after having performed a symmetry-unrestricted calculation one should verify whether the signature or chiral symmetry is broken by considering the mean angular momentum vector in the intrinsic frame. If it has non-zero components on more than one intrinsic axis, then the signature symmetry is broken. If all three components are non-zero, also the chiral symmetry is broken. However, to make one sure that these symmetries are unbroken is not so straightforward, because there may a priori exist other factors violating them, different that the angular momentum coupling, e.g., exotic deformations. A detection of unbroken symmetries based on the values of the variational parameters (e.g., elements of the density matrix) would be doable, but difficult. Another way is to perform a separate calculation imposing the symmetry in question and then to check, whether the results are equivalent, i.e., whether they differ at most by orientation in space. This can be verified by looking at the total energy, contributions thereto from the different terms of the Skyrme functional, density integrals, various observables, and s.p.\ energies.

A phenomenological approximation to the HF approach consists in replacing $\hat{\Gamma}$ with a model potential, like the Nilsson \cite{Nil55a} or Woods-Saxon \cite{Woo54a} one, which depends on several parameters, mainly multipole deformations, $\alpha_{\lambda\mu}$, of the nuclear surface. Then, the expectation value of the Routhian $\hat{h}'$ (corrected for the liquid drop energy via the Strutinsky renormalization \cite{Rin00a}) is minimized with respect to parameters of the potential.

It is well understood that only the relative orientation of the angular momentum vector and the intrinsic axes carries a physical information. Their position with respect to the program frame is meaningless and can be arbitrary. One can take advantage thereof in the phenomenological method, where the orientation of the mean potential is under control via the variational parameters, like the multipole deformations $\alpha_{\lambda\mu}$ of the nuclear surface, and can be fixed so that the $\alpha_{2\mu}$ tensor be diagonal in the program frame simply by setting $\alpha_{21}=\im\alpha_{22}=0$. With this condition satisfied, the intrinsic frame coincides with the program frame, facilitating many computational details. The minimization of the expectation value of the Routhian $\hat{h}'$ at a given magnitude of $\omega$ is then performed by varying the direction of $\vec{\omega}$ and all $\alpha_{\lambda\mu}$ except $\alpha_{21}$ and $\im\alpha_{22}$. The position of $\vec{\omega}$ with respect to the coinciding intrinsic and program frames is often described in the spherical coordinates, $\vartheta$, $\varphi$, called {\it tilt angles}. It is in fact this procedure that is usually understood under the term TAC and that has been described many times in the literature, see e.g. the seminal papers by Frauendorf {\it et al.} \cite{Fra91a,Fra92a,Fra93a,Fra97c,Fra00a} and \cite{Dim00b}.

In the HF method, however, the Euler angles defining the orientation of the intrinsic axes are not free variational parameters, but complicated functions of the densities, that change from one iteration to another according to the HF recipe. Thus, the orientation cannot be easily controlled. The only feasible way to keep the nucleus in the axes of the program frame is by imposing dynamical constraints on the off-diagonal components of the quadrupole tensor and requiring that they vanish. Then, one can vary the cranking frequency vector like in the phenomenological approach. It is actually the only way to proceed if the energy dependence on the tilt angles, $\vartheta$, $\varphi$, is sought. It has been checked, however, that multipole constraints strong enough to confine the nucleus may easily lead to divergenciesm, at least in the present calculations. If only the minimum is of interest, a more natural and faster way is to fix the rotational frequency vector with respect to the program frame and let the mean potential reorient and conform to it self-consistently in the course of the iterations. It is but this procedure that corresponds to the minimization of the given Routhian (\ref{HHwJ_eqn}). The direction of the rotational frequency vector in the program frame no more plays any role, while the solution now does become tilted with respect to that frame. This method has been adopted in the present work.

Some quantities, like mean angular momenta and multipole moments, carry a clear information only when expressed in the intrinsic frame of the nucleus. But they are first calculated in the program frame. Since the two no more coincide, it is necessary to find the principal axes of the nuclear shape by diagonalization of the electric quadrupole tensor and to transform the considered quantities into the corresponding frame by use of the Wigner matrices.

To examine deformed nuclei one usually choses deformed bases. However, deformation of the basis improves the description only if it suites the shape of the nucleus; otherwise it may spoil the result. For instance, deformation of the Harmonic-Oscillator (HO) basis amounts to stretching the cartesian coordinates associated with the axes of the program frame, and to taking different numbers of quanta in each direction. This is not appropriate for representing the TAC solutions, whose principal axes are arbitrarily tilted with respect to the program frame. A deformed basis would drive such a solution like a force to a certain orientation different from the self-consistent one, determined by the direction of the cranking vector. In particular, the Kerman-Onishi condition would be violated. Therefore, in self-consistent TAC calculations, the s.p.\ space spanned by the basis should be rotationally invariant. It means that the basis should have the three characteristic frequencies equal and comprise only entire HO shells. To obtain a reasonable description for deformed nuclei, the only way is to use large bases.

It follows from the theorem of self-consistent symmetries that the s.p.\ Routhian and the density matrix are invariant under the same symmetry operations; see Section~\ref{spobre_sec}. However, one should be aware of one exception to that rule, that concerns the time-reversal symmetry in case the time-odd terms of the mean field are artificially set to zero in the calculations. This simplification is always the case in the phenomenological mean-field approach, where time-odd terms simply do not appear in the model potential. With the time-odd fields cut off (and with no cranking), the Routhian is explicitly time-even, while the density will still acquire time-odd components whenever there is an unpaired particle; see Appendix~\ref{eveden_app}. The spectrum of the time-even Routhian exhibits the two-fold Kramers degeneracy, and the state of the unpaired particle can be any unitary combination of the states in the Kramers pair; the physicist must take the decision. The main point of this discussion is that this freedom of choice influences the time-odd densities, while the time-even ones remain undisturbed. This lemma is proven in Appendix~\ref{eveden_app}. Together with the time-odd densities, also the time-odd observables, notably the mean angular momentum, depend on that choice. The time-even quantities, like the quadrupole moments, are uniquely determined, because they depend only on time-even densities. These conclusions are valid for both the phenomenological and the self-consistent implementation of the mean-field method. However, in the self-consistent approach, in each iteration there is a feedback from the density to the mean field. One may doubt, therefore, whether the choice of state for the unpaired particle alters the generated mean field. This is not the case, because the time-even fields originate only from the time-even density, which are not disturbed, while the time-odd fields are suppressed.

\section{The HF time-odd coupling constants}
\label{timodd_sec}

One important point in the investigations of the magnetic and chiral rotation is to examine the role of various time-odd observables, like current and magnetic-moment distributions. Their impact on the results is governed by the time-odd coupling constants of the HF energy functional. After introducing necessary definitions of the local densities, energy functional, mean fields, and local gauge invariance, this Section explains how the values of the time-odd coupling constants are set in the present calculations.

In the HF approximation, the total energy is, in general, a functional of the non-local one-body density,
\begin{equation}
\rho_\alpha(\vec{r}\sigma,\vec{r}'\sigma')=\langle\Psi|a_{\vec{r}'\sigma'\alpha}^+a_{\vec{r}\sigma\alpha}|\Psi\rangle~,
\end{equation}
where $|\Psi\rangle$ is the many-body wavefunction, $\vec{r}$ and $\sigma$ are the position and spin variables, and $\alpha=n,p$ for neutrons and protons, respectively. The density $\rho_\alpha(\vec{r}\sigma,\vec{r}'\sigma')$ can be written as a sum of the scalar, $\rho_\alpha(\vec{r},\vec{r}')$, and vector, $\vec{s}_\alpha(\vec{r},\vec{r}')$, terms,
\begin{equation}
\rho_\alpha(\vec{r},\vec{r}')=\sum_\sigma\rho_\alpha(\vec{r}\sigma,\vec{r}'\sigma')~, \qquad
\vec{s}_\alpha(\vec{r},\vec{r}')=\sum_{\sigma\sigma'}\rho_\alpha(\vec{r}\sigma,\vec{r}'\sigma')\vec{\sigma}_{\sigma'\sigma},
\end{equation}
as
\begin{equation}
\rho_\alpha(\vec{r}\sigma,\vec{r}'\sigma')=\textstyle{\frac{1}{2}}\rho_\alpha(\vec{r},\vec{r}')\delta_{\sigma\sigma'}+\textstyle{\frac{1}{2}}\vec{s}_\alpha(\vec{r},\vec{r}')\cdot\vec{\sigma}_{\sigma\sigma'}~.
\end{equation}
The HF equations are usually solved within the Local-Density Approximation (LDA), in which the total energy depends only on local ($\vec{r}=\vec{r}'$) densities, and on their derivatives at $\vec{r}=\vec{r}'$ up to the second order. One has to consider the following real nucleonic densities \cite{Eng75a}:
\begin{equation}
\label{locden_eqn}
\begin{array}{lrcl}
\mbox{Particle:} & \rho_\alpha(\vec{r}) & = & \rho_\alpha(\vec{r},\vec{r})~, \\
\mbox{Spin:} & \vec{s}_\alpha(\vec{r})& = & \vec{s}_\alpha(\vec{r},\vec{r})~, \\
\mbox{Kinetic:} & \tau_\alpha(\vec{r}) & = & \left[\vec{\nabla}\cdot\vec{\nabla}'\rho_\alpha(\vec{r},\vec{r}')\right]_{\vec{r}=\vec{r}'}~, \\
\mbox{Vector kinetic:} & \vec{T}_\alpha(\vec{r}) & = & \left[\vec{\nabla}\cdot\vec{\nabla}'\vec{s}_\alpha(\vec{r},\vec{r}')\right]_{\vec{r}=\vec{r}'}~, \\
\mbox{Momentum:} & \vec{j}_\alpha(\vec{r}) & = & \frac{1}{2i}\left[(\vec{\nabla}-\vec{\nabla}')\rho_\alpha(\vec{r},\vec{r}')\right]_{\vec{r}=\vec{r}'}~, \\
\mbox{Spin-current:} & J_{\mu\nu,\alpha}(\vec{r}) & = & \frac{1}{2i}\left[(\nabla_\mu-\nabla_\mu')s_{\nu,\alpha}(\vec{r},\vec{r}')\right]_{\vec{r}=\vec{r}'}~,
\end{array}
\end{equation}
of which $\rho_\alpha$, $\tau_\alpha$, and $\ten{J}_\alpha=J_{\mu\nu,\alpha}$ are time-even, and $\vec{s}_\alpha$, $\vec{T}_\alpha$, and $\vec{j}_\alpha$ are time-odd. For each of the local densities, one defines the isoscalar and isovector density, e.g.,
\begin{equation}
\rho_0=\rho_n+\rho_p \qquad \rho_1=\rho_n-\rho_p
\end{equation}
The isoscalar and isovector terms are denoted by subscript $t=0,1$, respectively.

The interaction part of the HF LDA energy functional can be expressed as a space integral,
\begin{equation}
\cE=\sum_t\int d^3\vec{r}\left(\cH_t^\mathrm{even}+\cH_t^\mathrm{odd}\right)~,
\end{equation}
of energy densities, $\cH_t^\mathrm{even}$ and $\cH_t^\mathrm{odd}$, which depend on the time-even and time-odd nucleonic densities, respectively. They are defined as
\begin{equation}
\begin{array}{rcl}
\cH_t^\mathrm{even}(\vec{r}) & = & C_t^\rho\rho_t^2+C_t^{\Delta\rho}\rho_t\Delta\rho_t+C_t^\tau\rho_t\tau_t+C_t^J\ten{J}_t^2+C_t^{\nabla J}\rho_t\vec{\nabla}\cdot\vec{J}_t~, \\
\cH_t^\mathrm{odd}(\vec{r}) & = & C_t^s\vec{s}_t^2+C_t^{\Delta s}\vec{s}_t\cdot\Delta\vec{s}_t+C_t^T\vec{s}_t\cdot\vec{T}_t+C_t^j\vec{j}_t^2+C_t^{\nabla j}\vec{s}_t\cdot(\vec{\nabla}\times\vec{j}_t)~,
\end{array}
\end{equation}
where the square of the tensor density is defined as $\ten{J}_t^2=\sum_{\mu\nu}J_{\mu\nu,t}^2$, and its vector part, $\vec{J}_t$, is defined as $J_{\lambda,t}=\sum_{\mu\nu}\epsilon_{\lambda\mu\nu}J_{\mu\nu,t}$. The coupling constants $C_t$ appearing in $\cH_t^\mathrm{even}$ and $\cH_t^\mathrm{odd}$ are conventionally called time-even and time-odd, respectively. They are functions of the parameters of the Skyrme force \cite{Dob95a}, and may depend on density if the force is density-dependent. Note, however, that in the spirit of the Density-Functional Theory the energy functional does not need to come from any interaction, and the coupling constants $C_t$ play the role of fundamental parameters.

Upon variation of the interaction energy, $\cE$, with respect to local densities, one obtains the neutron and proton Skyrme-HF s.p.\ mean potentials (see Section~\ref{spobre_sec}),
\begin{equation}
\Gamma_n=\Gamma_0^\mathrm{even}+\Gamma_0^\mathrm{odd}+\Gamma_1^\mathrm{even}+\Gamma_1^\mathrm{odd} \qquad
\Gamma_p=\Gamma_0^\mathrm{even}+\Gamma_0^\mathrm{odd}-\Gamma_1^\mathrm{even}-\Gamma_1^\mathrm{odd}~,
\end{equation}
where the time-even and time-odd fields, $\Gamma_t^\mathrm{even}$ and $\Gamma_t^\mathrm{odd}$, read
\begin{equation}
\label{fields_eqn}
\begin{array}{rcl}
\Gamma_t^\mathrm{even} & = & -\vec{\nabla}\cdot\left[M_t\vec{\nabla}\right]+U_t+\frac{1}{2i}\left(\ten{\nabla\sigma}\cdot\ten{B}_t+\ten{B}_t\cdot\ten{\nabla\sigma}\right)~, \\
\Gamma_t^\mathrm{odd} & = & -\vec{\nabla}\cdot\left[\left(\vec{\sigma}\cdot\vec{C}_t\right)\vec{\nabla}\right]+\vec{\sigma}\cdot\vec{\Sigma}_t+\frac{1}{2i}\left(\vec{\nabla}\cdot\vec{I}_t+\vec{I}_t\cdot\vec{\nabla}\right)~.
\end{array}
\end{equation}
They depend on the nucleonic densities through the potential functions,
\begin{equation}
\label{potfun_eqn}
\begin{array}{rcl}
U_t(\vec{r}) & = & 2C_t^\rho\rho_t+2C_t^{\Delta\rho}\Delta\rho_t+C_t^\tau\tau_t+C_t^{\nabla J}\vec{\nabla}\cdot\vec{J}_t+U_t'~, \\
\vec{\Sigma}_t(\vec{r}) & = & 2C_t^s\vec{s}_t+2C_t^{\Delta s}\Delta\vec{s}_t+C_t^T\vec{T}_t+C_t^{\nabla j}\vec{\nabla}\times\vec{j}_t~, \\
M_t(\vec{r}) & = & C_t^\tau\rho_t~, \\
\vec{C}_t(\vec{r}) & = & C_t^T\vec{s}_t~, \\
\ten{B}_t(\vec{r}) & = & 2C_t^J\ten{J}_t-C_t^{\nabla J}\ten{\nabla}\rho_t~, \\
\vec{I}_t(\vec{r}) & = & 2C_t^j\vec{j}_t+C_t^{\nabla j}\vec{\nabla}\times\vec{s}_t~.
\end{array}
\end{equation}
The tensor gradient operators in (\ref{fields_eqn}) and (\ref{potfun_eqn}) are defined as $(\ten{\nabla\sigma})_{\mu\nu}=\nabla_\mu\sigma_\nu$ and $\nabla_{\mu\nu}=\sum_\lambda\epsilon_{\mu\nu\lambda}\nabla_\lambda$. The term $U_t'$ represents the rearrangement terms resulting from the density dependence of the coupling constants, $C_t$. In standard parametrizations of the Skyrme interaction, which are used in the present work, only $C_t^\rho$ and $C_t^s$ depend on the isoscalar particle density, $\rho_0$, and then
\begin{equation}
U_0'(\vec{r})=\sum_{t'}\left(\frac{\partial C_{t'}^\rho}{\partial\rho_0}\rho_{t'}^2+\frac{\partial C_{t'}^s}{\partial\rho_0}\vec{s}_{t'}^2\right) \qquad
U_1'(\vec{r})=0~.
\end{equation}

If the energy functional is derived from the Skyrme interaction, there are unique relations between the time-even and time-odd coupling constants, $C_t$. In particular,
\begin{equation}
\label{eveodd_eqn}
C_t^j=-C_t^\tau~, \qquad C_t^J=-C_t^T~, \qquad C_t^{\nabla j}=+C_t^{\nabla J}~.
\end{equation}
However, it turns out \cite{Dob95a} that the three relations (\ref{eveodd_eqn}) have a deeper sense, namely, they are necessary and sufficient conditions for the invariance of the energy functional under a local gauge transformation of the $A$-body wave-function, $|\Psi\rangle$,
\begin{equation}
\label{locgau_eqn}
|\Psi'\rangle=\exp\left(i\sum_{j=1}^A\phi(\vec{r}_j)\right)|\Psi\rangle~,
\end{equation}
where $\phi(\vec{r})$ is an arbitrary real function of position. In particular, for $\phi=\vec{p}\cdot\vec{r}/\hbar$, the gauge transformation (\ref{locgau_eqn}) corresponds to a Galilean boost with velocity $\vec{v}=\vec{p}/m$, where $m$ is the particle mass.

In the present calculations, the SLy4 \cite{Cha97a} and SkM* \cite{Bar82a} parametrizations of the Skyrme force were used. Like in the original fits, the time-even coupling constant $C_t^J$ and the time-odd constant $C_t^T$ were always set to zero. This does not hamper the local gauge invariance. Apart from $C_t^J$, all other time-even coupling constants were taken as they come from the parameters of the Skyrme force. To examine the role of the time-odd densities and fields, attempts were undertaken to obtain each solution with three sets of values of the time-odd coupling constants. In each case, values of selected constants were taken as they come from the Skyrme force, while those of the remaining ones were set to zero. The constants selected in each set are listed in Table~\ref{timodd_tab}. The density-dependent and independent components of $C_t^s$ were suppressed or not simultaneously. These three types of calculations are denoted throughout the text as $N$, $G$, and $T$. In the first case, $N$, the mean field contained only time-even components, like the phenomenological mean potentials, and the local gauge invariance was violated. In the set $G$, only those time-odd coupling constants were not set to zero, that must be related to the time-even coupling constants according to (\ref{eveodd_eqn}) for the energy functional to be invariant under the local gauge transformations (\ref{locgau_eqn}). Finally, in the set $T$ all the time-odd coupling constants were taken as they come from the Skyrme force, with the exception of $C_t^T$. In the following Chapters, setting or not selected time-odd coupling constants to zero is referred to as excluding or including the corresponding time-odd terms of the mean field, as given by Eqs. (\ref{fields_eqn}, \ref{potfun_eqn}). 

\begin{table}
\caption[HF time-odd coupling constants.]{Three sets of values of the time-odd coupling constants used in the present calculations. For each set, values of the listed constants were taken as they come from the Skyrme force, while the remaining constants were set to zero.}
\label{timodd_tab}
\begin{center}
\begin{tabular}{c|r}
\hline\hline
Set symbol & Coupling constants \\
\hline
$N$ & \\
$G$ & $C_t^j$, $C_t^{\nabla j}$ \\
$T$ & $C_t^s$, $C_t^{\Delta s}$, $C_t^j$, $C_t^{\nabla j}$
\end{tabular}
\end{center}
\end{table}

The results obtained for each of the sets, $N$, $G$, $T$, are compared for particular bands in Chapters~\ref{magnet_cha} and \ref{chiral_cha}. A general conclusion is that the inclusion of the time-odd fields does not change the solutions qualitatively. Switching them on and off influences mostly the moments of inertia. Additionally, convergence problems arise at low cranking frequencies for the $G$, and particularly the $T$ set, and no solutions could be obtained in several cases.

\section{Symmetry-unrestricted mean-field codes}
\label{codes_sec}

This Section gives a review of computer codes that can perform symmetry-unrestricted mean-field calculations, particularly within the TAC model, and that existed prior to the present study.

Imposed symmetries in mean-field calculations allow to reduce the sizes of matrices or of the space lattice, and thus to gain on the computer memory and CPU time. Before the interest in magnetic bands emerged at the beginning of the 1990's, the signature or simplex symmetries were inherently imposed in the majority of mean-field codes. One of the first numerical TAC calculations were carried out in 1987 by Frisk and Bengtsson \cite{Fri87a} by using the Nilsson potential \cite{Nil55a} with pairing. The phenomenological TAC code originally written by Frauendorf \cite{Fra93a,Dim00b} is now widely used for description of experimental data on the magnetic and chiral bands. It allows for including monopole pairing correlations with quasiparticle excitations. Recently, Dudek and Schunck \cite{Dud04a} have written a relativistic Woods-Saxon TAC program, and first results are about to appear soon. One of the first symmetry-unrestricted calculations within realistic self-consistent methods were performed in 1991 by Umar {\it et al.} \cite{Uma91a}. Without cranking and pairing, they solved the Skyrme-HF equations in the space representation, by using the Spline Collocation Method. About the year 2000, Yamagami and Matsuyanagi developed a symmetry-unrestricted Skyrme-HF program working on a Cartesian mesh with the imaginary-time method \cite{Yam00a,Yam01a}. They include either only cranking or only pairing with the density-dependent delta force and without quasiparticle excitations. Calculations of Madokoro {\it et al.} for a magnetic band in $^{\rm 84}$Rb \cite{Mad00a} were done in 2000 by use of a Relativistic Mean Field TAC code imposing only the parity, working in the harmonic-oscillator basis and not including pairing. For the purpose of the present work, the author developed a Skyrme-HF TAC program, \pr{HFODD} version (v2.05c), that works in the harmonic-oscillator basis and allows for breaking of all the $D_{2h}^T$ symmetries. Currently, the pair correlations cannot be taken into account in the TAC mode. This program is described in the next Section.

\section{The program \pr{HFODD} (v2.05c)}
\label{hfodd_sec}

In order to perform self-consistent calculations for the magnetic and chiral rotation, the author of this work developed a Skyrme-HF computer code, \pr{HFODD} version (v2.05c). In the previous versions, (v1.60r) and (v1.75r) \cite{Dob00a}, the simplex symmetry was imposed, thus not allowing for TAC calculations. Removing this restriction was the principal improvement introduced during the present work, and led to the version (v2.05c), which was used for obtaining results presented here. A brushed-up version, (v2.08i), was published in \cite{Dob04a}. This Section gives an overview of the program's main features, with particular emphasis on those that have been introduced for the purpose of the present investigations.

The code \pr{HFODD} solves the nuclear HF equations for the Skyrme effective interaction by using the iterative method. The s.p.\ wavefunctions are expanded onto the deformed Cartesian Harmonic-Oscillator (HO) basis. From the s.p.\ wavefunctions, the HF local densities (\ref{locden_eqn}), and then the mean potentials (\ref{fields_eqn}) are calculated on a Cartesian mesh. By using the Gauss-Hermite quadrature, matrix elements of those potentials between the HO states are calculated. The matrix of the s.p.\ Hamiltonian, obtained in this way, is then diagonalized, to find the s.p.\ wavefunctions in the next iteration. The Nilsson potential \cite{Nil55a} is implemented as a starting point. All densities (\ref{locden_eqn}) appearing in the Skyrme functional are calculated, and all time-even and time-odd fields (\ref{fields_eqn}) are taken into account. Several parametrizations of the force are available and can be modified. The generalized spin-orbit terms \cite{Rei95a} can be included in the Skyrme functional. Appropriate time-odd coupling constants of the energy functional can be determined either from the Skyrme force or from the Landau parameters \cite{Ben02a}. For the center-of-mass motion, either the standard one-body correction before variation or a two-body correction after variation can be used. Apart from such basic quantities like s.p.\ and total energies, the code calculates the angular momentum vectors, electric, magnetic, and surface multipole moments, mean-square radii and Bohr deformation parameters. Linear and quadratic constraints on spin as well as quadratic constraints on the electric and surface moments are implemented. With the simplex imposed as an unbroken symmetry, the program can also work in the HFB mode, with either the fixed-gap or density-dependent delta pairing. The Fermi level is found automatically by using the equivalent-spectrum method \cite{Dob84a}. The quasi-particle blocking is not yet programmed.

The user can choose the pattern of imposed symmetries by independently imposing or not the $\hat{S}_y$, $\hat{R}_y$, $\hat{P}$, $\hat{T}$, $\hat{S}^T_x$, $\hat{S}^T_y$, and $\hat{S}^T_z$ symmetries. Keeping in mind that the unbroken symmetries must form a group, 34 combinations listed in Table~\ref{hfosym_tab} are permitted. Some of them differ only by an exchange of the Cartesian axes associated with the signature and simplex symmetries, and are thus physically equivalent. Altogether, 19 out of the 28 non-equivalent subgroups\footnote{Including the trivial subgroups.} of $D_{2h}^T$ can be selected as imposed symmetries. Imposing the missing 9 subgroups can easily be implemented provided there is a physics motivation to study such new cases. When either of the three symmetries, $\hat{S}_y$, $\hat{R}_y$, or $\hat{P}$ is imposed, the s.p.\ Routhian acquires a specific block-diagonal form, and diagonalization of smaller matrices results in a faster execution time. In version (v2.08i), this is implemented for $\hat{S}_y$ and/or $\hat{P}$ imposed, but in case $\hat{R}_y$ alone is imposed it is not implemented yet. The three $T$-simplexes and the time reversal cannot be used in a similar way, because they are represented by anti-linear operators. However, when the time reversal and simplex or signature are imposed, the Kramers degeneracy allows for diagonalization of matrices only in one simplex or signature, which reduces the numerical effort by half. The code stores the Skyrme local densities on a Cartesian mesh, and takes advantage of imposed symmetries, both represented by linear and anti-linear operators, to reduce the number of lattice points on which the densities are calculated.

\begin{table}
\caption[Symmetry patterns allowed in the code \pr{HFODD}.]{Complete list of imposed (I) or non-imposed (N) symmetries allowed in the code \pr{HFODD} version (v2.08i). The first column enumerates all the possible symmetry patterns, while the second column enumerates non-equivalent subgroups (see text).}
\label{hfosym_tab}
\begin{center}
\begin{tabular}{rr|ccccccc|l}
\hline\hline
\multicolumn{2}{c|}{Number} & \multicolumn{7}{c|}{Symmetries} & \multicolumn{1}{c}{Conserved group}  \\
& & $\hat{S}_y$ & $\hat{R}_y$ & $\hat{P}$ & $\hat{T}$ & $\hat{S}_y^T$ & $\hat{S}_x^T$ & $\hat{S}_z^T$ & \\
\hline
 1 &  1 & I & I & I & I & I & I & I & $\hat{R}_{xyz}~\hat{P}~\hat{S}_{xyz}~\hat{T}~\hat{R}^T_{xyz}~\hat{P}^T~\hat{S}^T_{xyz}$ \\
 2 &  2 & I & I & I & I & I & N & N & $\hat{R}_y~\hat{P}~\hat{S}_y~\hat{T}~\hat{R}^T_y~\hat{P}^T~\hat{S}^T_y$ \\
 3 &  3 & I & I & I & N & N & I & I & $\hat{R}_y~\hat{P}~\hat{S}_y~\hat{R}^T_{zx}~\hat{S}^T_{zx}$ \\
 4 &  4 & I & I & I & N & N & N & N & $\hat{R}_y~\hat{P}~\hat{S}_y$ \\
 5 &  5 & I & N & N & I & I & I & N & $\hat{R}_z~\hat{S}_{xy}~\hat{T}~\hat{R}^T_z~\hat{S}^T_{xy}$ \\
 6 &  5 & I & N & N & I & I & N & I & $\hat{R}_x~\hat{S}_{yz}~\hat{T}~\hat{R}^T_x~\hat{S}^T_{yz}$ \\
 7 &  6 & I & N & N & I & I & N & N & $\hat{S}_y~\hat{T}~\hat{S}^T_y$ \\
 8 &  7 & I & N & N & N & N & I & N & $\hat{S}_y~\hat{R}^T_z~\hat{S}^T_x$ \\
 9 &  7 & I & N & N & N & N & N & I & $\hat{S}_y~\hat{R}^T_x~\hat{S}^T_z$ \\
10 &  8 & I & N & N & N & N & N & N & $\hat{S}_y$ \\
11 &  5 & N & I & N & I & N & I & I & $\hat{R}_y~\hat{S}_{zx}~\hat{T}~\hat{R}^T_y~\hat{S}^T_{zx}$ \\
12 &  9 & N & I & N & I & N & N & N & $\hat{R}_y~\hat{T}~\hat{R}^T_y$ \\
13 & 10 & N & I & N & N & I & I & I & $\hat{R}_{xyz}~\hat{P}^T~\hat{S}^T_{xyz}$ \\
14 & 11 & N & I & N & N & I & N & N & $\hat{R}_y~\hat{P}^T~\hat{S}^T_y$ \\
15 & 12 & N & I & N & N & N & I & I & $\hat{R}_y~\hat{S}^T_{zx}$ \\
16 & 13 & N & I & N & N & N & N & N & $\hat{R}_y$ \\
17 &  2 & N & N & I & I & N & I & N & $\hat{R}_x~\hat{P}~\hat{S}_x~\hat{T}~\hat{R}^T_x~\hat{P}^T~\hat{S}^T_x$ \\
18 &  2 & N & N & I & I & N & N & I & $\hat{R}_z~\hat{P}~\hat{S}_z~\hat{T}~\hat{R}^T_z~\hat{P}^T~\hat{S}^T_z$ \\
19 & 14 & N & N & I & I & N & N & N & $\hat{P}~\hat{T}~\hat{P}^T$ \\
20 &  3 & N & N & I & N & I & I & N & $\hat{R}_z~\hat{P}~\hat{S}_z~\hat{R}^T_{xy}~\hat{S}^T_{xy}$ \\
21 &  3 & N & N & I & N & I & N & I & $\hat{R}_x~\hat{P}~\hat{S}_x~\hat{R}^T_{yz}~\hat{S}^T_{yz}$ \\
22 & 15 & N & N & I & N & I & N & N & $\hat{P}~\hat{R}^T_y~\hat{S}^T_y$ \\
23 & 15 & N & N & I & N & N & I & N & $\hat{P}~\hat{R}^T_x~\hat{S}^T_x$ \\
24 & 15 & N & N & I & N & N & N & I & $\hat{P}~\hat{R}^T_z~\hat{S}^T_z$ \\
25 & 16 & N & N & I & N & N & N & N & $\hat{P}$ \\
26 &  6 & N & N & N & I & N & I & N & $\hat{S}_x~\hat{T}~\hat{S}^T_x$ \\
27 &  6 & N & N & N & I & N & N & I & $\hat{S}_z~\hat{T}~\hat{S}^T_z$ \\
28 & 17 & N & N & N & I & N & N & N & $\hat{T}$ \\
29 & 12 & N & N & N & N & I & I & N & $\hat{R}_z~\hat{S}^T_{xy}$ \\
30 & 12 & N & N & N & N & I & N & I & $\hat{R}_x~\hat{S}^T_{yz}$ \\
31 & 18 & N & N & N & N & I & N & N & $\hat{S}^T_y$ \\
32 & 18 & N & N & N & N & N & I & N & $\hat{S}^T_x$ \\
33 & 18 & N & N & N & N & N & N & I & $\hat{S}^T_z$ \\
34 & 19 & N & N & N & N & N & N & N &
\end{tabular}
\end{center}
\end{table}

Arbitrary s.p.\ configurations can be required in the code \pr{HFODD}, and the way of selecting them is related to the pattern of imposed symmetries. If $\hat{P}$ and $\hat{R}_y$ are imposed, then the s.p.\ states are divided into four groups characterized by the parity and signature eigenvalues. Those groups are referred to as {\it parity-signature} blocks. The code does not put the particles just on the energetically lowest s.p.\ states, but the user must specify the numbers of particles in each block separately. Then, in each block, the lowest levels are occupied, which gives the {\it reference} configuration, from which the particle-hole excitation are counted afterwards. The same is implemented for imposing either of $\hat{P}$, $\hat{R}_y$, or $\hat{S}_y$ alone, only that in these cases there are two blocks instead of four. They are called {\it parity}, {\it signature}, and {\it simplex} blocks, respectively. If no symmetries are imposed, obviously there is only one block. The three $T$-simplexes and the time reversal do not play any role in this mechanism, because anti-linear operators do not provide quantum numbers. In the following, the parity-signature reference configuration is specified as ($n_{++}$, $n_{+-}$, $n_{-+}$, $n_{--}$), where, e.g., $n_{+-}$ is the number of occupied states of parity $+1$ and signature $-i$, etc. Similarly, the parity, signature, and simplex reference configurations are given as ($n_{+}$, $n_{-}$). States in parity-signature blocks are denoted as, e.g., 18--\,+, which refers to the eighteenth state of negative parity and positive signature, counting from one in energetical order. Sometimes, a pair of states, like 18--\,+ and 18--\,--, is denoted as 18--\,$\pm$. The notation for the parity, signature, and simplex blocks is similar, e.g., 28\,--. Particle-hole excitations are defined by selecting from which state in which symmetry block a particle is removed and onto which state in which block it is put. This is denoted as, e.g., (18--\,+\,$\rightarrow$\,19--\,+) for the parity-signature case and, e.g., (28\,--\,$\rightarrow$\,29\,--) for the parity, signature, or simplex imposed. The use of symmetry blocks makes it easier to follow a given configuration diabatically, because crossings of levels belonging to different blocks do not cause any confusion as to which of them should be occupied. If two levels from the same block are about to cross near the Fermi surface, the user may require that a state with a lower/higher value of a certain s.p.\ observable be occupied. In the present context, such diabatic blocking concerns primarily the projection of the s.p.\ angular momentum on the cranking frequency vector.

The code allows for TAC calculations, and there are no restrictions on the direction of the angular frequency vector, except for those following from the imposed symmetries. According to the Kerman-Onishi theorem \cite{Ker81a}, in TAC calculations all self-consistent solutions have their spin and cranking frequency vectors, $\vec{J}$ and $\vec{\omega}$, parallel. In practice, $\vec{J}$ approaches $\vec{\omega}$ very slowly in the course of the iterations, and up to several thousands of iterations are necessary to obtain a converged result. To avoid this inconvenience, a quadratic constraint on the vector product, $\vec{\omega}\times\vec{J}$, has been introduced, constraining that quantity to zero. However, this does not speed up the convergence significantly, and may easily lead to divergencies. In version (v2.08i), a different prescription was implemented, which is much more stable and substantially reduces the required number of iterations. Namely, in each iteration, the rotational frequency vector is explicitly reset to be parallel to the current total angular momentum vector, while its length, $\omega$, is kept unchanged. Since $\vec{\omega}$ is not constant in this method, it does not correspond to minimization of any a priori given Routhian. However, it is the Routhian with the last direction of $\vec{\omega}$ that will have been minimized once the converged solution is found. This method was invented after the completion of the present calculations, and was not yet implemented in version (v2.05c). This is the only difference between the two versions that matters for the present study.

Some quantities, like the multipole moments, only carry a clear information if calculated with respect to the center of mass. As far as the symmetry patterns used in the code (see Table~\ref{hfosym_tab}) are concerned, if neither the parity nor $T$-parity is imposed, the center-of-mass of the solution may move away from the origin of the program frame. The code then finds its position and recalculates the quantities in question in the center-of-mass frame. It is also useful to compute the spin vectors and all types of multipole moments in the intrinsic frame of the nucleus, defined by the principal axes of the electric quadrupole tensor. Unless any two of the simplex symmetries, $\hat{S}_x$, $\hat{S}_y$, $\hat{S}_z$, $\hat{S}^T_x$, $\hat{S}^T_y$, $\hat{S}^T_z$, are imposed, the principal axes may become tilted with respect to the Cartesian axes of the program frame; see Section~\ref{tac_sec}. The code then finds the directions of the principal axes by diagonalization of the quadrupole tensor, and transforms all necessary quantities to the intrinsic system of reference. If the parity is not imposed, the solution may acquire non-zero linear momentum. The code computes its value and corrects the mean angular momenta for that translational motion.

\chapter{Properties of the $h_{11/2}$ valence nucleons}
\label{valenc_cha}

Behavior of the shears and chiral bands is, to a big extent, determined by the rotational properties of the valence high-$j$ particles and holes. In the cases studied here, we are primarily interested in properties of the $h_{11/2}$ orbital. In this Chapter, some general properties of the $h_{11/2}$ states are analyzed, first within self-consistent one-dimensional cranking method, and then analytically. It is demonstrated that those properties are affected by the triaxial deformation and that the concerned valence nucleons are expected to align their angular momenta rather stiffly along the principal axes of the triaxial shape. It is also argued that pairing correlations may soften those alignments.

\section{Symmetries of the non-rotating solutions}
\label{symmet_sec}

In the following two Chapters, Hartree-Fock (HF) shears solutions in $^{142}$Gd and chiral solutions in $^{130}$Cs, $^{132}$La, $^{134}$Pr, $^{136}$Pm are presented. Calculations were carried out with two Skyrme-force parametrizations, SLy4 \cite{Cha97a} and SkM* \cite{Bar82a}, with and without the HF time-odd fields included; see Section~\ref{timodd_sec}. Since there is no physical indication for the breaking of parity in either of the shears or chiral bands, this symmetry was imposed in the calculations. For all considered nuclides, symmetry-unrestricted calculations (modulo the imposed parity) were performed to check, whether some other symmetries are spontaneously broken in the non-rotating states. Whenever the time-odd fields were absent (either switched off or self-consistently zero in the ground state of the even-even $^{142}$Gd), it was found that the mean field possessed the $D_{2h}^T$ symmetry. This was inferred from comparison of the symmetry-unrestrained results with those obtained with the $D_{2h}^T$ symmetry imposed\footnote{In \pr{HFODD}, for configurations with unpaired nucleons, the $D_{2h}^T$ symmetry can be imposed on the mean field by imposing $\hat{P}$, $\hat{S}_y$, $\hat{S}^T_x$, $\hat{S}^T_z$, and switching the time-odd fields off.}, as described in Section~\ref{tac_sec}.

\section{Single-particle PAC Routhians for triaxial nuclei}
\label{pacrth_sec}

All the HF solutions considered in the present work are triaxial, and this property determines many of their features. It is instructive to examine their rotations about the three principal axes first. This amounts to applying Principal-Axis Cranking (PAC), with both parity and signature imposed. For non-rotating states with the time-odd terms absent, the mean field is anyway symmetric with respect to the whole $D_{2h}^T$ group, and imposing either of the three signatures does not change it. Although the associated Slater determinants can be different for different signatures imposed, due to different signature states chosen for unpaired particles, this does not affect the mean field; see Section~\ref{tac_sec} for details. Thus, it is justified to say that three different rotations of the same triaxial object are considered. With the time-odd fields present, the symmetry-unrestrained solutions violate all the three signatures. Therefore, imposing any of them necessarily leads to a substantially different solution, and different for each of the signatures imposed. However, the energies, deformations, and s.p.\ energies are very similar, and the effects from the time-odd fields are not so strong. In the code \pr{HFODD}, it is technically most convenient to perform PAC about the $y$ axis\footnote{Although PAC about the $x$ and $z$ axes is also implemented, conservation of the associated simplex or signature symmetries does not yield block-diagonal matrices in the basis used \cite{Dob04a}.}. To examine rotation about each given intrinsic axis, the non-rotating solutions were first turned in space, by use of appropriate quadrupole constraints, so that the required principal axis coincide with the program $y$ axis.

Sample self-consistent s.p.\ PAC Routhians for $^{130}$Cs, $^{132}$La, $^{134}$Pr, $^{136}$Pm, and $^{142}$Gd are collected in Appendix~\ref{routhi_app}. The s.p.\ levels are numbered so that, e.g., 14--+ denotes the fourteenth level of negative parity and positive signature; see Section~\ref{hfodd_sec}. In non-rotating solutions, the s.p.\ levels originating from the 
spherical $h_{11/2}$ orbital have numbers 14--$\pm$ to 19--$\pm$ (each level comprising two states of opposite signatures), for both protons and neutrons. This was checked in calculations with constraints on the quadrupole moment by the observation that the s.p.\ energies of the levels 14--$\pm$ to 19--$\pm$ converge to the same value with decreasing $\beta$, and no crossings with other negative-parity levels occur\footnote{With the exception of the triaxial minima in $^{134}$Pr and $^{136}$Pm; see Section~\ref{chimin_sec}.} in function of $\beta$. In the spherical case, the $h_{11/2}$ s.p.\ levels unambiuously have numbers 14--$\pm$ to 19--$\pm$ in all realistic potentials, because $h_{11/2}$ is an intruder orbital, well separated from all other negative-parity states. 

For cranking about the short axis, the lowest sublevels of the $h_{11/2}$ orbital (14--$\pm$) split strongly in function of $\omega$, while the highest (19--$\pm$) do not. For the long axis, it is precisely the opposite, while for the medium axis only the intermediate levels split. Routhians for cranking about the medium axis are given only in Figs.~\ref{la_pro_fig} and \ref{la_neu_fig}, for the sample case of $^{132}$La. In fact, the intermediate $h_{11/2}$ levels split for cranking about each of the three axes, but only weakly. Slopes of the PAC Routhians, $e'$, translate into the s.p.\ alignments, $j_i$, on the $i$-th axis according to the well-known formula
\begin{equation}
\label{jidedwi_eqn}
j_i=-\frac{\mathrm{d}e'}{\mathrm{d}\omega_i}~,
\end{equation}
which holds exactly for fixed mean potentials and approximately if the potential changes with rotational frequency due to the HF self-consistency or TRS minimization. Therefore, the lowest $h_{11/2}$ sublevels align their angular-momenta vectors on the short axis, and the highest on the long one. The corresponding alignments are approximately equal to $11/2\,\hbar$. The intermediate states have non-zero alignments on all the three axes. Although these results are based on PAC calculations, they are confirmed in the full Tilted-Axis Cranking (TAC) results of Chapters~\ref{magnet_cha} and \ref{chiral_cha}. For the $D_{2h}^T$-symmetric mean fields, the alignments obtained from one-dimensional cranking acquire a very specific meaning, which is discussed in the next Section.

\section{$D_2^T$-symmetric mean field}
\label{cond2t_sec}

In all the present HF solutions without the time-odd fields, as well as in virtually all phenomenological studies of the magnetic and chiral rotation, the mean-field Hamiltonian, $\hat{h}$, is symmetric under the $D_{2h}^T$ group. In such a case, important properties of the s.p.\ states can be inferred from pure symmetry considerations. Eigenstates of a $D_{2h}^T$-symmetric potential exhibit the two-fold Kramers degeneracy, and this Section is focused on properties of a single Kramers pair. In fact, to obtain the results of this Section, it is enough to assume that $\hat{h}$ is symmetric with respect to $D_2^T$, a subgroup of $D_{2h}^T$ which consists of the time reversal, the three signature, and the three $T$-signature operations.

A complete information about the matrix elements of the angular-momentum operator, $\hat{\vec{J}}$, between the states of a Kramers pair, ($|\mu\rangle$,~$|\bar\mu\rangle$), is contained in the real {\it alignment vector}, $\vec{J}^\mu$, and the complex {\it decoupling vector}, $\vec{D}^\mu$, of the state $|\mu\rangle$,
\begin{equation}
\vec{J}^\mu=\langle\mu|\hat{\vec{J}}|\mu\rangle~, \qquad
\vec{D}^\mu=\langle\mu|\hat{\vec{J}}|\bar\mu\rangle~.
\end{equation}
The alignment and decoupling vectors, $\vec{J}^{\bar\mu}$, $\vec{D}^{\bar\mu}$, of the state $|\bar\mu\rangle$ are related to $\vec{J}^\mu$ and $\vec{D}^\mu$, by Eq.~(\ref{tirejd_eqn}); see Appendix~\ref{polvec_app}. For any Hamiltonian, $\hat{h}$, invariant under the $D_2^T$ group, one can choose the states of the considered Kramers pair as eigenstates of either of the three signatures, $\hat{R}_i$, where $i=x,y,z$. This results in three formally different pairs, ($|\mu_i\rangle$,~$|\bar\mu_i\rangle$), that correspond to just three different bases in the same two-dimensional eigenspace of $\hat{h}$. It is shown in Appendix~\ref{polvec_app}, that there are only three independent real parameters that determine all components of the alignment and decoupling vectors associated with the three examined pairs. These are the "diagonal" components $J^{\mu_x}_x$, $J^{\mu_y}_y$, $J^{\mu_z}_z$, where the lower indices refer to the Cartesian components of $\vec{J}$. One has
\begin{equation}
\label{JxMx00_eqn}
\vec{J}^{\mu_x}=(J^{\mu_x}_x,0,0)~, \qquad \vec{D}^{\mu_x}=(0,J^{\mu_y}_y,-iJ^{\mu_z}_z)~,
\end{equation}
\begin{equation}
\label{Jy0My0_eqn}
\vec{J}^{\mu_y}=(0,J^{\mu_y}_y,0)~, \qquad \vec{D}^{\mu_y}=(-iJ^{\mu_x}_x,0,J^{\mu_z}_z)~,
\end{equation}
\begin{equation}
\label{Jz00Mz_eqn}
\vec{J}^{\mu_z}=(0,0,J^{\mu_z}_z)~, \qquad \vec{D}^{\mu_z}=(J^{\mu_x}_x,-iJ^{\mu_y}_y,0)~.
\end{equation}
Invariance of $\hat{h}$ under $D_2^T$ does not restrict the values of $J^{\mu_x}_x$, $J^{\mu_y}_y$, $J^{\mu_z}_z$. In the often considered case of axial symmetry, say with respect to the $z$ axis,
\begin{equation}
\label{axial_eqn}
(J^{\mu_x}_x,J^{\mu_y}_y,J^{\mu_z}_z)=\left\{
\begin{array}{lll}
(J^{\mu_\perp}_\perp,J^{\mu_\perp}_\perp,J^{\mu_\parallel}_\parallel) & \mathrm{for} & J^{\mu_\parallel}_\parallel=1/2 \\
(0,0,J^{\mu_\parallel}_\parallel)                                     & \mathrm{for} & J^{\mu_\parallel}_\parallel=3/2,~5/2,~... \\
\end{array}\right.
\end{equation}
The component $J^{\mu_\parallel}_\parallel$ is quantized due to the axial symmetry, and the still arbitrary component $J^{\mu_\perp}_\perp$ is usually called {\it decoupling parameter}; see Appendix~\ref{polvec_app} for details.

Suppose that the states in the considered Kramers pair have no coupling to other states through the angular-momentum operator. Under this assumption, the 3D cranking for this pair becomes a two-dimensional diagonalization problem, that can be solved analytically. Of course, it concerns the non-selfconsistent cranking, where the Hamiltonian $\hat{h}$ in the Routhian $\hat{h}'$ is fixed. For a degenerate Kramers pair, $\hat{h}$ reduces to its eigenvalue, $e$. Taking, e.g., the states $|\mu_z\rangle$, $|\bar\mu_z\rangle$ as a basis one has, according to (\ref{Jz00Mz_eqn}),
\begin{equation}
\hat{h}'=\hat{h}-\vec{\omega}\hat{\vec{J}}=\left[
\begin{array}{cc}
e & 0 \\
0 & e
\end{array}
\right]-\left[
\begin{array}{cc}
\omega_zJ^{\mu_z}_z                      &  \omega_xJ^{\mu_x}_x-i\omega_yJ^{\mu_y}_y \\
\omega_xJ^{\mu_x}_x+i\omega_yJ^{\mu_y}_y & -\omega_zJ^{\mu_z}_z
\end{array}
\right]~.
\end{equation}
For the eigenstates of $\hat{h}'$, one obtains the following components of the mean angular momentum vector, $\vec{J}$,
\begin{equation}
J_i=\pm\frac{\omega_i(J^{\mu_i}_i)^2}{\left(\omega_x^2(J^{\mu_x}_x)^2+\omega_y^2(J^{\mu_y}_y)^2+\omega_z^2(J^{\mu_z}_z)^2\right)^{1/2}}~.
\end{equation}
They differ by sign for the two eigenstates. This dependence has a singularity at $\omega=0$, and is not linear in the general case. Consider the following two extreme situations. If $J^{\mu_x}_x=J^{\mu_y}_y=J^{\mu_z}_z=g$, then $J_i=g\omega_i/\omega$, and $\vec{J}$ always orients itself along $\vec{\omega}$, already for infinitesimal $\omega$. This can be called {\it soft alignment}. On the other hand, when only one of the parameters $J^{\mu_x}_x$, $J^{\mu_y}_y$, $J^{\mu_z}_z$ is non-zero, say $J^{\mu_j}_j$, then $J_i=J^{\mu_j}_j\delta_{ij}$, and $\vec{J}$ is independent of $\vec{\omega}$ (unless $\omega=0$). We call this {\it stiff alignment} on the $j$-th axis.

In axial nuclei, precisely one two-fold degenerate substate of each deformation-split $j$-shell has $J^{\mu_\parallel}_\parallel=1/2$ and $J^{\mu_\perp}_\perp\neq0$, which represents the soft alignment. According to Eq.~(\ref{axial_eqn}), all other necessarily have a vanishing decoupling parameter, and are thus rigidly aligned with the symmetry axis. For prolate shapes, the lowest-energy substate has $J^{\mu_\parallel}_\parallel=1/2$, and is soft, while for oblate shapes it is the highest substate. In triaxial nuclei, values of the parametrs $J^{\mu_i}_i$, where $i=s,m,l$ corresponds to the short, medium, and long principal axes, are equal to the alignments obtained from the one-dimensional cranking about the three axes. Indeed, for cranking about the axis $i$, the s.p.\ states are eigenstates of $\hat{R}_i$. Taking into account the PAC results of the previous Section, one can see that for the lowest $h_{11/2}$ substates of a triaxial nucleus only $J^{\mu_s}_s$ is non-zero, while for the highest substates only $J^{\mu_l}_l$ does not vanish. These alignments are thus stiff. The Routhians of these states are not much curved, which confirms that their angular-momentum coupling to other states is rather weak. Note that there are no states with stiff alignment on the medium axis. The response to rotation of the middle $h_{11/2}$ substates is soft, because all the three parameters, $J^{\mu_s}_s$, $J^{\mu_m}_m$, $J^{\mu_l}_l$, are non-zero, and complicated, because these states interact with one another, what can be seen from the strong bending of their PAC Routhians.

Although the presented results come from a non-selfconsistent cranking (with fixed $\hat{h}$), the full HF calculations in Chapters~\ref{magnet_cha} and \ref{chiral_cha} show that at least the deformation does not change much with rotational frequency, particularly for the chiral bands. Therefore, the non-selfconsistent cranking should be a good approximation here. In both the magnetic and chiral bands, the active high-$j$ particles occupy the lowest substates of a given orbital, while the valence holes occupy the highest substates. The principal conclusion of the last two Sections is, therefore, that those valence particles and holes align their individual angular momenta on the short and long axes of the triaxial shape, respectively, and that those alignments are stiff. These results, based mostly on symmetry considerations, agree well with the full HF TAC results of Chapters~\ref{magnet_cha} and \ref{chiral_cha}. According to the presented considerations, the triaxial deformation should affect the blades closing in the shears bands. Indeed, for prolate or oblate shapes, respectively the lowest particle or the highest hole would rather have a soft alignment.

\section{Influence of the pairing correlations}
\label{pairin_sec}

Besides deformation, pairing correlations may affect properties of the shears and chiral mean-field solutions. Since pairing is not included in the present HF calculations, this Section aims at least at a model study of its possible role. It is well known that pairing reduces the collective moments of inertia. Here, the interest is rather in the influence that pairing may exert on the alignment properties of the valence particles and holes.

\begin{figure}
\begin{center}
\includegraphics{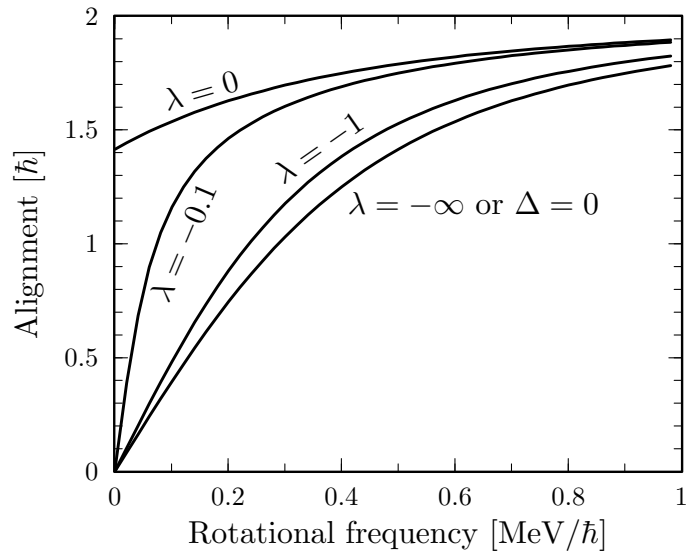}
\end{center}
\caption[Influence of pairing on the single-particle alignments.]{Alignment of the canonical particle with $v^2=1$ for sample values of the Fermi energy, $\lambda$, and for the case with no pairing ($\Delta=0$), obtained from one-dimensional cranking in a simple model (see text). Values of $\lambda$ and $\Delta$ are given in MeV.}
\label{pairing_fig}
\end{figure}

Non-selfconsistent one-dimensional cranking about the $y$ axis is considered for a system of two Kramers pairs, ($|\mu_y\rangle$, $|\bar\mu_y\rangle$) and ($|\nu_y\rangle$, $|\bar\nu_y\rangle$), whose states have good signature $\hat{R}_y$. Let $e_\mu<e_\nu$ be the energies of those two pairs, taken at $\omega_y=0$, and $g_y$ be the matrix element of the $y$-th component, $\hat{J}_y$, of the angular-momentum operator between $|\mu_y\rangle$ and $|\nu_y\rangle$. It is assumed for simplicity that the considered states have zero alignment on the $y$ axis at $\omega_y=0$. In the basis of the states $|\mu_y\rangle$, $|\bar\mu_y\rangle$, $|\nu_y\rangle$, $|\bar\nu_y\rangle$, the no-pairing Routhian for this model takes the form
\begin{equation}
\label{he1e1e2e2_eqn}
h'=\left[
\begin{array}{cccc}
e_1          & 0           & -\omega_yg_y & 0 \\
0            & e_1         & 0            & \omega_yg_y \\
-\omega_yg_y & 0           & e_2          & 0 \\
0            & \omega_yg_y & 0            & e_2
\end{array}
\right]~.
\end{equation}
The interest here is in the alignment, $j_y$, of a single particle that at $\omega_y=0$ occupies the state $|\mu_y\rangle$. By diagonalization of (\ref{he1e1e2e2_eqn}) and, e.g., by using formula (\ref{jidedwi_eqn}), one obtains
\begin{equation}
j_y=\frac{2\omega_yg_y^2}{\sqrt{(e_\nu-e_\mu)^2+4\omega_y^2g_y^2}}~.
\end{equation}
This dependence is traced in Fig.~\ref{pairing_fig} for $e_\mu=-1\,\mathrm{MeV}$, $e_\nu=+1\,\mathrm{MeV}$ and $g_y=2~\hbar$.

Inclusion of pairing in the simplest fixed-delta form leads to the quasi-particle Routhian
\begin{equation}
\cH'=\left[
\begin{array}{cccc|cccc}
e_1          & 0           & -\omega_yg_y & 0           & 0           & -\Delta      & 0           & 0 \\
0            & e_1         & 0            & \omega_yg_y & \Delta      & 0            & 0           & 0 \\
-\omega_yg_y & 0           & e_2          & 0           & 0           & 0            & 0           & -\Delta \\
0            & \omega_yg_y & 0            & e_2         & 0           & 0            & \Delta      & 0 \\
\hline
0            & \Delta      & 0            & 0           & -e_1        & 0            & \omega_yg_y & 0 \\
-\Delta      & 0           & 0            & 0           & 0           & -e_1         & 0           & -\omega_yg_y \\
0            & 0           & 0            & \Delta      & \omega_yg_y & 0            & -e_2        & 0 \\
0            & 0           & -\Delta      & 0           & 0           & -\omega_yg_y & 0           & -e_2
\end{array}
\right]~.
\end{equation}
For a qualitative study, a gap parameter of $\Delta=1\,\mathrm{MeV}$ is taken. In order to talk about an unpaired particle in the Hartree-Fock-Bogolyubov formalism, one has to consider a one-quasiparticle excitation and identify the particle in question with the canonical state with occupation $v^2=1$. Precisely one such state must appear in the canonical spectrum for one-quasiparticle excitation. For $\omega_y=0$, that state necessarily coincides with some eigenstate of the particle-hole Routhian (\ref{he1e1e2e2_eqn}), but it is not so for $\omega_y\neq0$. In other words, in rotating solutions pairing introduces extra admixtures to the blocked particle, that can change its alignment properties. To describe a particle initially put on $|\mu_y\rangle$, in the following such a quasi-particle is blocked that for $\omega_y\rightarrow0$ the fully occupied canonical state continuously pass onto $|\mu_y\rangle$. The model was solved numerically. Figure~\ref{pairing_fig} shows the alignment, $j_y$, of the considered canonical particle in function of the rotational frequency, $\omega_y$, for a few sample values of the Fermi energy, $\lambda$. If $\lambda\ll e_\mu$, the result is obviously identical to that without pairing. As the Fermi energy increases towards $\lambda=0$, the process of alignment gets faster. It is illustrated for $\lambda=-1\,\mathrm{MeV}$ and $\lambda=-0.1\,\mathrm{MeV}$. If $\lambda=0$, already for infinitesimal rotational frequencies a non-zero value of $j_y$ is achieved. For this value of $\lambda$, the Fermi level lies exactly at half way between $e_\mu$ and $e_\nu$. For $\lambda>0$ the fully occupied canonical particle initially anti-aligns its angular momentum, and only for higher rotational frequencies the positive $j_y$ is regained. This is not marked in Fig.~\ref{pairing_fig} because such a situation does not seem to be the case, neither for the magnetic nor for the chiral rotation.

Although no quantitative conclusions can be drawn from this simple model, it becomes clear that inclusion of pairing correlations may render the valence particles and holes more susceptible to rotational alignment. Together with the reduction of the total moments of inertia, it may affect the shears-rotor competition in the magnetic bands and the value of the chiral critical frequency, as well as the structure of chiral bands; see Chapter~\ref{magchi_cha}.

\chapter{Hartree-Fock shears results in $^{142}$Gd}
\label{magnet_cha}

In the present work, the $\pi h_{11/2}^2~\nu h_{11/2}^{-2}$ single-particle (s.p.) configuration in $^{142}$Gd is taken as an example for the study of the shears mechanism. It is particularly simple to handle, because it involves only the well-separated $h_{11/2}$ intruder orbital, and can be considered a model shears configuration. Additionally, a magnetic band supposed to be built on that structure has been observed experimentally in $^{142}$Gd. The presented results show the first fully self-consistent Hartree-Fock (HF) Tilted-Axis Cranking (TAC) solutions, and confirm the important role of the shears mechanism. However, the agreement with experiment is not yet satisfactory. Possible reasons are discussed, but further research is needed to clarify this point. A preliminary report on these results was given in \cite{Olb02a}.

\section{Previous studies in $^{142}$Gd}

\begin{figure}
\begin{center}
\includegraphics{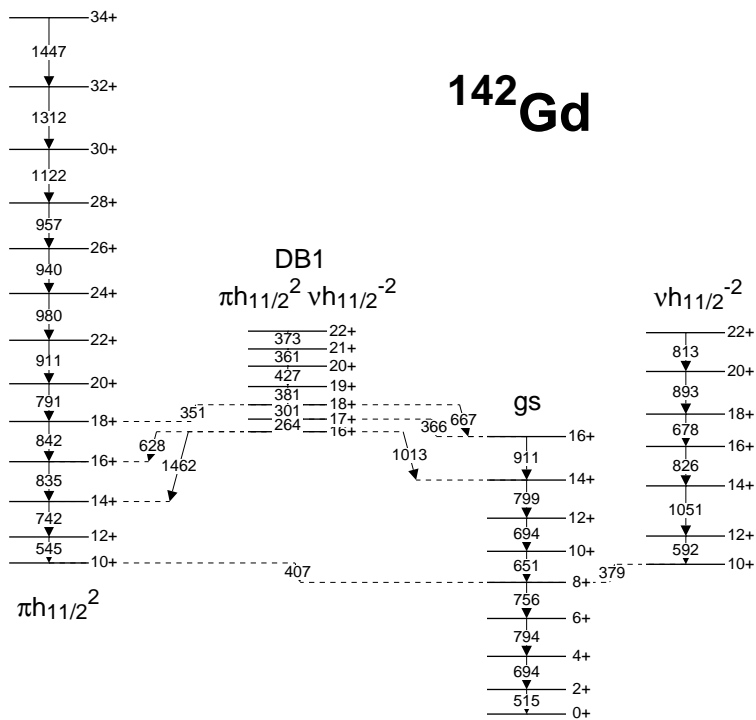}
\end{center}
\caption[Partial level scheme of $^{142}$Gd.]{Partial level scheme of $^{142}$Gd. The magnetic dipole band DB1 \cite{Lie02a} is shown together with the ground-state sequence and the two $\Delta I=2$ bands \cite{Sta88a} built on the $\pi h_{11/2}^2$ and $\nu h_{11/2}^{-2}$ configurations.}
\label{gd142scheme_fig}
\end{figure}

Magnetic bands in $^{142}$Gd were first investigated by Sugawara {\it et al.} \cite{Sug97a}, but the present analysis is based on the most recent EUROBALL III measurements by Lieder {\it et al.} \cite{Lie02a}. In that experiment, four magnetic bands, denoted DB1, DB2, DB3, DB4, were observed, and configurations of $\pi h_{11/2}^2~\nu h_{11/2}^{-2}$, $\pi h_{11/2}^1g_{7/2}^{-1}~\nu h_{11/2}^{-2}$, $\pi h_{11/2}^2~\nu h_{11/2}^{-4}$ and $\pi h_{11/2}^1g_{7/2}^{-1}~\nu h_{11/2}^{-4}$ were assigned to them, respectively. The present work focuses on the DB1 structure. Figure~\ref{gd142scheme_fig} shows a partial level scheme of $^{142}$Gd, including the DB1 and the ground-state band. Two $\Delta I=2$ bands built on the $\pi h_{11/2}^2$ and $\nu h_{11/2}^{-2}$ configurations \cite{Sta88a} are also displayed, because if the $\pi h_{11/2}^2$ and $\nu h_{11/2}^{-2}$ excitations are combined together, they yield the $\pi h_{11/2}^2~\nu h_{11/2}^{-2}$ configuration of the band DB1. The $\pi h_{11/2}^2$ and $\nu h_{11/2}^{-2}$ bandheads are $10^+$ isomeric states, which points to the parallel coupling of the angular momenta of the two $h_{11/2}$ proton particles and neutron holes. Such a coupling is characteristic for the shears mechanism; see Section~\ref{magnet_sec}.

\begin{figure}
\begin{center}
\includegraphics{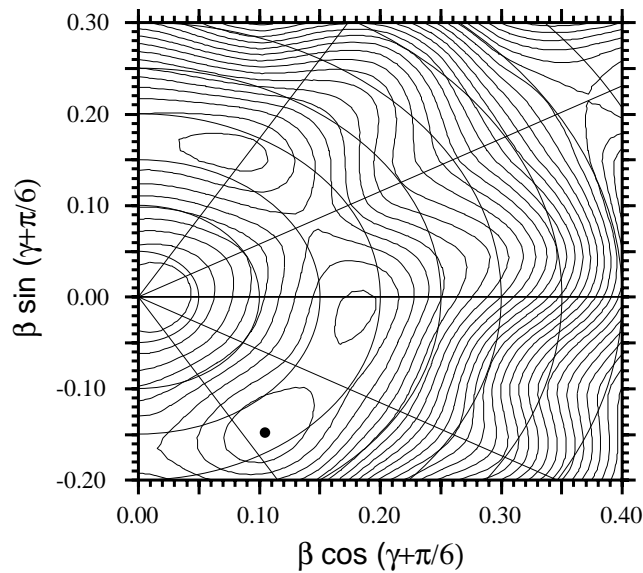}
\end{center}
\caption[TRS for the ground state in $^{142}$Gd.]{Energy surface for the non-rotating ground state in $^{142}$Gd, obtained from the TRS calculations taking into account the pairing correlations.}
\label{gd142trs_fig}
\end{figure}

It was discussed in Section~\ref{magnet_sec} that deformation has a strong influence on the shears mechanism. There is no direct experimental data on the deformation of $^{142}$Gd, but several theoretical predictions for the ground state are available. The Extended Thomas-Fermi method with the Strutinsky Integral \cite{Abo92a} gives $\beta_2=-0.21$, the Folded Yukawa potential combined with the Droplet Model \cite{Mol95a} yields $\beta_2=-0.156$ and the Relativistic Mean Field \cite{Lal96a} gives $\beta_2=-0.158$. These theoretical calculations were limited to axially-symmetric shapes. In the present work, Total Routhian Surface (TRS) calculations allowing for triaxial deformations were performed with the pairing correlations included; see Appendix~\ref{detals_app} for details. The obtained energy surface is shown in Fig.~\ref{gd142trs_fig}. The minimum corresponds to a triaxial shape with $\gamma\approx30^\circ$. At $\gamma=0^\circ$ and $\gamma=60^\circ$ there are saddle-points, and the one for $\gamma=60^\circ$ has lower energy, in agreement with the previous predictions of oblate shapes. The triaxial minimum is rather shallow in the $\gamma$ direction, so one can talk about $\gamma$-softness. Both the minima and the saddle-points have the absolute deformation of $\beta\approx0.181$, placing the TRS prediction between those of \cite{Mol95a,Lal96a} and that of \cite{Abo92a}. For the magnetic bands in $^{142}$Gd, phenomenological TAC calculations allowing for triaxial shapes were performed in the experimental paper \cite{Lie02a}, by using the code of Frauendorf; see Section~\ref{codes_sec}. Energy minimization gave  $\beta\approx0.15$ and $\gamma\approx60^\circ$ for all the four bands, but $\beta\approx0.10$ and $\gamma=60^\circ$ was taken for the cranking description of DB1 and DB3. A satisfactory agreement with experiment has been obtained for both the energies and the $B(M1)/B(E2)$ ratios.

\section{HF solutions in $^{142}$Gd}
\label{shemin_sec}

This Section describes the HF ground-state and shears solutions in $^{142}$Gd. Since this is the first application of the Skyrme-HF TAC method developed in this work, below it is explained step-by-step how those solutions were found. Refer to Section~\ref{hfodd_sec} for the treatment of symmetries and s.p.\ configurations in the code \pr{HFODD}.

The HF calculations in $^{142}$Gd were performed with two Skyrme pa\-ram\-e\-trizations, SLy4 \cite{Cha97a} and SkM* \cite{Bar82a}; see Appendix~\ref{detals_app} for all details. To investigate the role of the HF time-odd densities and fields, three different sets of time-odd fields, $N$ (no time-odd fields), $G$ (time-odd fields responsible for a gauge invariance of the force), and $T$ (all Skyrme time-odd fields), were included; see Section~\ref{timodd_sec}. The study of the shears mechanism was preceded by a search for the ground-state minimum. Only the parity was imposed as an unbroken symmetry. Several parity reference configurations (see Section~\ref{hfodd_sec}) were considered that seemed reasonable from the standard Nilsson diagrams, with no particle-hole excitations required. Cranking constraint was applied in the first iteration to make the calculations symmetry-unrestricted, and then released, as described in Section~\ref{tac_sec}. The lowest minimum was found for the parity configuration listed in the first row, right column of Table~\ref{shecon_tab}. With the SLy4 force, that solution corresponds to a HF vacuum, and for SkM* there is one empty positive-parity state under the last occupied negative-parity state. It is known that the order of s.p.\ levels may vary from force to force. Deformation of $\beta=0.19$, $\gamma=40^\circ$ for SLy4 and $\beta=0.18$, $\gamma=39^\circ$ for SkM* was obtained. These values of $\beta$ are very close to the TRS result for the ground state in $^{142}$Gd.

\begin{table}[h]
\caption[Single-particle configurations for the HF solutions in $^{142}$Gd.]{Parity-signature and parity single-particle configurations (see Section~\ref{hfodd_sec}) for protons ($\pi$) and neutrons ($\nu$) used to obtain the solutions in $^{142}$Gd at zero cranking frequency.}
\label{shecon_tab}
\begin{center}
\begin{tabular}{ll|c|cc|cc}
\hline\hline
           &      &       & \multicolumn{2}{c|}{Parity-Signature}               & \multicolumn{2}{c}{Parity} \\
           &      &       & Ref.          & Exc.	                        & Ref.    & Exc. \\
\hline
$^{142}$Gd & SLy4 & $\pi$ & (18,18,14,14) &  	                                & (36,28) & \\
Ground     & SkM* & $\nu$ & (21,21,18,18) &  	                                & (42,36) & \\
\hline
$^{142}$Gd & SLy4 & $\pi$ & (18,18,14,14) & (14--\,--\,$\rightarrow$\,15--\,--) & (36,28) & (28\,--\,$\rightarrow$\,29\,--) \\
Shears     & SkM* & $\nu$ & (21,21,18,18) & (18--\,+\,$\rightarrow$\,19--\,+)   & (42,36) & (36\,--\,$\rightarrow$\,37\,--) \\
\end{tabular}
\end{center}
\end{table}

It should be expected that the ground state of the even-even $^{142}$Gd with quadrupole deformation is symmetric with respect to the whole $D_{2h}^T$ group, not only the parity. If that was the case, each Kramers pair of a given parity would comprise two states of opposite signatures, and the considered parity configuration would translate into a parity-signature configuration given in the first row, left column of Table~\ref{shecon_tab}. Indeed, $D_{2h}^T$-imposing calculations with such a configuration were performed, and an identical result was obtained; see Section~\ref{tac_sec}.

The $D_{2h}^T$-imposing solution was then used to examine one-dimensional rotations about the three principal axes; see Section~\ref{pacrth_sec}. Figure~\ref{gd_a0a_fig} shows the s.p.\ Routhians from the Principal-Axis Cranking (PAC) calculations with the SLy4 force and no time-odd fields. As it was discussed in Section~\ref{pacrth_sec}, the s.p.\ levels 14--$\pm$ to 19--$\pm$ (see Section~\ref{hfodd_sec}) can be recognized as belonging to the $h_{11/2}$ orbital for both kinds of nucleons. One can see, therefore, that in the considered HF solution there are two $h_{11/2}$ proton particles occupying the lowest two-fold degenerate substate of the orbital and two $h_{11/2}$ neutron holes on its highest substate.

The considered shears band was calculated with only the parity imposed. In the language of the parity configurations, the $h_{11/2}$ orbital comprises states from the one described by 27\,-- to that described by 38\,--, both for neutrons and protons. With no time-odd fields, the $\pi h_{11/2}^2~\nu h_{11/2}^{-2}$ shears configuration was obtained from the ground state by exciting one proton from the 28\,-- onto the 29\,-- state, and simultaneously promoting one neutron hole\footnote{Since the code \pr{HFODD} works in the language of particles rather than holes, this was actually done by removing one neutron particle from 36\,--, and putting it onto 37\,--.} from 37\,-- onto 36\,--, see the second row of Table~\ref{shecon_tab}. This led to a converged solution for zero cranking frequency. Still with the time-odd fields switched off, a similar test as for the ground state was performed, showing that the mean field in the $\pi h_{11/2}^2~\nu h_{11/2}^{-2}$ solution breaks none of the $D_{2h}^T$ symmetries with the parity-signature configuration as specified in the second row, left column of Table~\ref{shecon_tab}.

To obtain the two-dimensional TAC solutions, a cranking frequency of $0.05\,\mathrm{MeV}/\hbar$ with equal components on the short and long axes of the nucleus was applied, starting the iterations from the previously converged non-rotating state. For the sake of simplicity, the mean field obtained with the $D_{2h}^T$ symmetry imposed was used as a starting point, which guaranteed that the principal axes initially coincided with the program axes; see Section~\ref{tac_sec}. As expected, a perpendicular coupling of the valence-nucleon angular momenta was obtained. The band was calculated with the cranking-frequency step of $0.05\,\mathrm{MeV}/\hbar$, starting the iterations for each subsequent value from the converged solution for the previous one.

The two occupied proton $h_{11/2}$ states are marked with filled circles in Fig.~\ref{gd_00p_fig}, showing the s.p.\ Routhians for the shears band with $N$ time-odd fields. These are just the lowest two $h_{11/2}$ states with positive alignments on the cranking vector, as reflected by their negative slopes. At low frequencies, the two states have numbers 27\,-- and 29\,--, as discussed previously. At $\omega\approx0.15\,\mathrm{MeV}/\hbar$, there is a crossing of the down-sloping 29\,-- Routhian with the up-sloping 28\,-- Routhian. For the configuration to be followed diabatically, the proton particle has to be moved from 29\,-- onto 28\,--, and one deals from now on with a vacuum state in the negative-parity block. The analogous holds for the two neutron holes.

Obtaining the HF TAC solutions in the presence of the time-odd terms of the Skyrme mean field is more cumbersome than without those terms. In general, the inclusion of the time-odd fields splits the Kramers pairs even in non-rotating states, which may change the order of the s.p.\ levels. In particular, some empty states from above the Fermi surface may come below the highest occupied states, which formally corresponds to particle-hole excitations. Since one cannot foresee which such excitations are a priori needed, it is difficult to find the s.p.\ configuration corresponding to the required physical state at zero cranking frequency. It is easier to proceed in case the signature or simplex symmetry is imposed, because such symmetry provides an additional quantum number, which allows to distinguish between s.p.\ states. Moreover, practice shows that calculations with the time-odd fields are prone to divergencies at low cranking frequencies.

In order to circumvent the above problems, calculations with the time-odd fields were not started from the zero cranking frequency. Instead, iterations for the $G$ and $T$ fields were restarted at each cranking frequency from the corresponding converged solution with the $N$ fields, with the s.p.\ configuration unchanged. Because of shifts in the s.p.\ energies introduced by the time-odd fields, this did not work in cases when levels cross either due to those fields, or in function of the rotational frequency. Yet, converged results were obtained for higher frequencies, where the proton and neutron configurations are already the s.p.\ vacua in negative parity. These fragments of bands were then extended towards lower frequencies step by step, by changing the configuration by hand whenever a crossing was about to occur, or by using the diabatic blocking on the total alignment on the cranking vector; see Section~\ref{hfodd_sec}. In spite of that, convergence could not be achieved in some cases. See Fig.~\ref{gddefo_fig} for the frequency ranges, in which the $G$ and $T$ solutions did converge.

\section{The shears mechanism in $^{142}$Gd}

The key point of the present study is to examine the shears mechanism in $^{142}$Gd. It is particularly interesting to see how the aligning angular momenta of the valence nucleons and the collective rotation compete in generating the total spin. These issues are addressed in this Section.

Below, spins of the two valence neutron $h_{11/2}$ holes are defined as the spins of the two empty $h_{11/2}$ states, taken with the minus sign. To ensure that the total neutron spin is equal to the sum of the two holes' and the core's contributions, the core is defined as all the occupied neutron states plus the two empty $h_{11/2}$ states. The division of the proton s.p.\ states into the two valence $h_{11/2}$ particles and the remaining core is obvious.

\begin{figure}
\begin{center}
\includegraphics{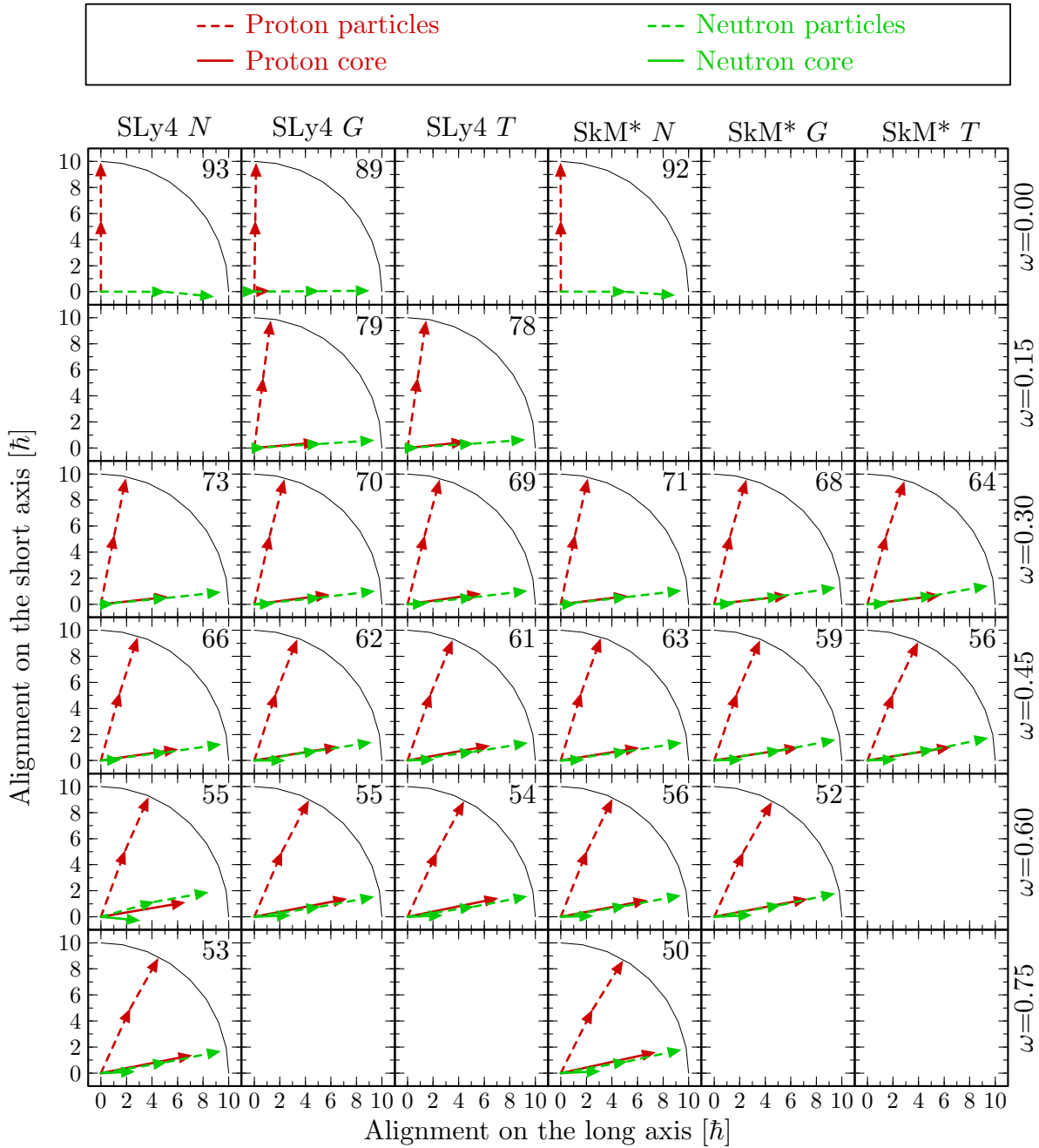}
\end{center}
\caption[HF TAC results for the shears closing in $^{142}$Gd.]{Angular momentum vectors of the valence nucleons and of the proton and neutron cores along the shears $\pi h_{11/2}^2~\nu h_{11/2}^{-2}$ band in $^{142}$Gd. HF results with the SLy4 and SkM* forces are shown for the $N$, $G$, $T$ time-odd fields included. For the proton particles, the arrow anchored at the origin stands for the energetically lower of the two concerned s.p.\ states, while the inverse holds for the neutron holes. Arc of radius $10\hbar$ is drawn in each panel, and the angle between the proton and neutron blade is given. Rotational frequencies are given in $\mathrm{MeV}/\hbar$.}
\label{shemec_fig}
\end{figure}

The angular momentum vectors of the valence nucleons and of the proton and neutron cores are shown in Fig.~\ref{shemec_fig} for several values of the rotational frequency. The angular momenta of the two $h_{11/2}$ proton particles are nearly parallel (stretched coupling), and their vector sum (particle blade) has length of approximately 10$\hbar$. The same holds for the neutron holes. The stretched coupling and the length of the two blades is retained up to high cranking frequencies. In the non-rotating state (bandhead), the particle and the hole blade aligns with the short and the long intrinsic axis, respectively. In the absence of collective rotation, that perpendicular orientation yields the total bandhead spin of about 14$\hbar$. With increasing rotational frequency, the two blades align towards each other (shears closing). These results are perfectly consistent with the shears scenario proposed in the literature and discussed in Section~\ref{magnet_sec}. At $\omega=0.75\,\mathrm{MeV}/\hbar$, the angle between the blades is equal to $53^\circ$ for SLy4 and $50^\circ$ for SkM*. At that point, the angular momentum gain due to the shears closing (with respect to the bandhead spin) reaches about $5\hbar$. Note that the maximum gain, corresponding to the complete closure, would be not more than $6\hbar$ for two blades of $10\hbar$ each, initially forming a right angle.

In competition with the shears mechanism, also the core generates its angular momentum. Both the proton and neutron cores tend to align their spins with the long axis, because the moment of inertia with respect to that axis is larger than with respect to the short axis. This is expected from the irrotational-flow formula (\ref{irrflo_eqn}) for $\gamma>30^\circ$, and can be also checked by considering one-dimensional cranking about the two axes, like in Section~\ref{pac_sec}. Interestingly, the contribution from the neutron core is a few times smaller than that from the proton core. At $\omega=0.75\,\mathrm{MeV}/\hbar$, the spin generated by the two cores equals $\approx10\hbar$, two times more than the contribution from the shears mechanism. This is in accord with the estimate of Macchiavelli and Clark \cite{Mac99a} (see Section~\ref{magnet_sec}) that deformations smaller than $\epsilon\approx0.12$ are needed for the shears mechanism to dominate. It can be seen from Fig.~\ref{shemec_fig} that the behavior of the shears blades and of the two cores is very similar for the SLy4 and SkM* Skyrme parametrizations, and independent of the time-odd fields included.

\begin{figure}
\begin{center}
\includegraphics{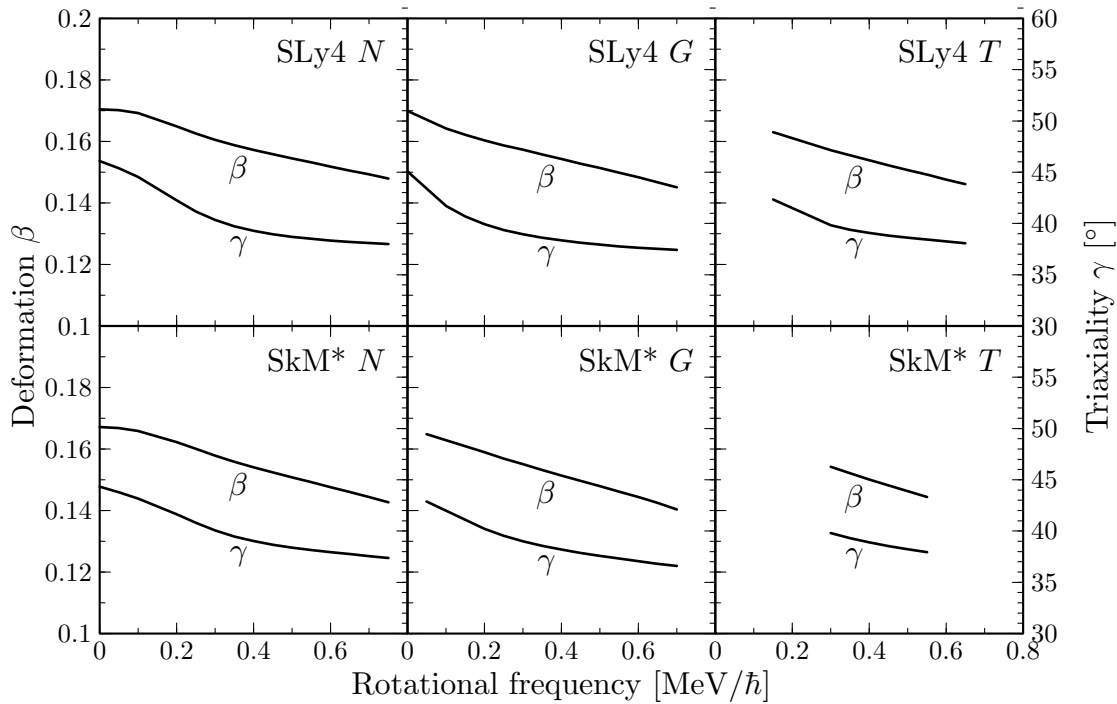}
\end{center}
\caption[Deformation along the HF shears band in $^{142}$Gd.]{The $\beta$ and $\gamma$ deformations along the shears $\pi h_{11/2}^2~\nu h_{11/2}^{-2}$ band in $^{142}$Gd. HF results with the SLy4 and SkM* forces are shown for the $N$, $G$, $T$ time-odd fields included.}
\label{gddefo_fig}
\end{figure}

At $\omega=0$, the shears solution has the deformation of $\beta\approx0.17$, $\gamma\approx45^\circ$, with negligible differences between the forces used. Thus, the $\pi h_{11/2}^2~\nu h_{11/2}^{-2}$ excitation slightly reduces the  value of $\beta$ as compared to the ground state. The deformation further diminishes with increasing rotational frequency, which is illustrated in Fig.~\ref{gddefo_fig}, and falls below $0.15$ with $\omega$ approaching $0.75\,\mathrm{MeV}/\hbar$. Such a decrease of deformation is characteristic for the shears bands; see Section~\ref{magnet_sec}. In the present results, however, the solution simultaneously becomes more and more triaxial, as it can be seen from the evolution of $\gamma$ shown in the same Figure. The closing blades composed of the two pairs of high-$j$ nucleons have quite a strong polarization effect on the core.

\section{Comparison with experiment}

\begin{figure}[h]
\begin{center}
\includegraphics{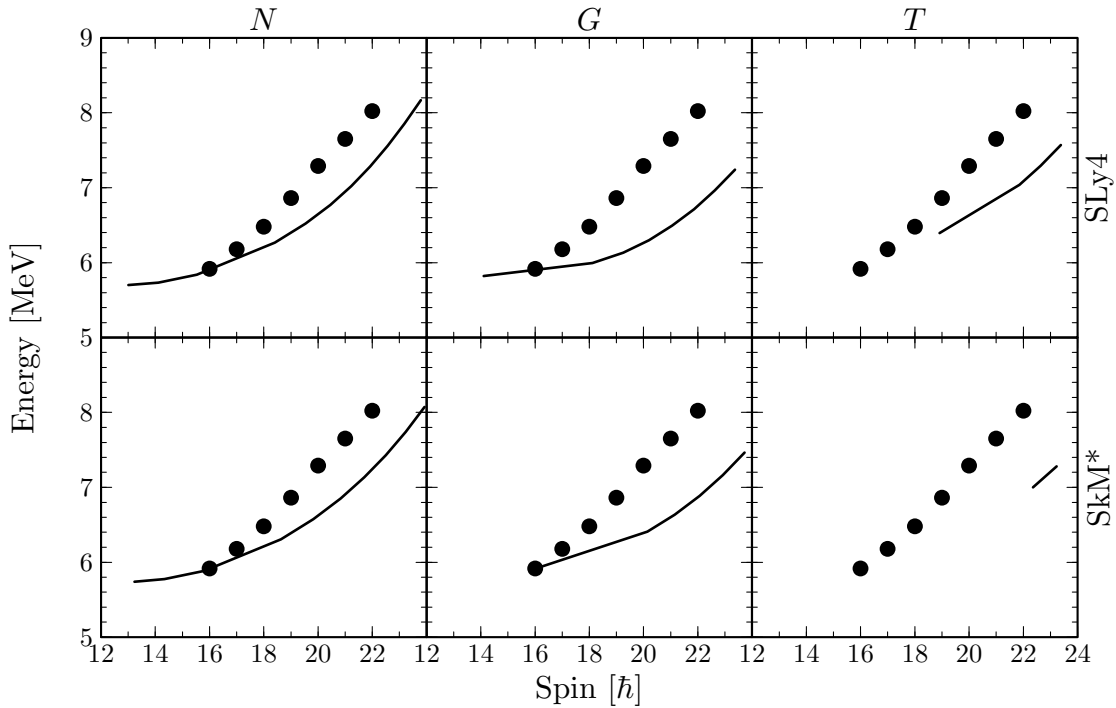}
\end{center}
\caption[HF and experimental energies for the $\pi h_{11/2}^2~\nu h_{11/2}^{-2}$ shears band in $^{142}$Gd.]{Calculated energies for the $\pi h_{11/2}^2~\nu h_{11/2}^{-2}$ shears band in $^{142}$Gd (lines), compared with the experimental data for the DB1 band (points). Hartree-Fock results with the SLy4 and SkM* forces are shown for the $N$, $G$, $T$ time-odd fields included. The energy scale refers to the experimental energy measured with respect to the ground state. For the $N$ and $G$ fields, the calculated curves are offset in energy so that they coincide with the experimental bandhead at spin $16\hbar$. For the $T$ fields, the offset is arbitrary.}
\label{gdener_fig}
\end{figure}

In Fig.~\ref{gdener_fig}, the energies calculated for the $\pi h_{11/2}^2~\nu h_{11/2}^{-2}$ band are compared with the experimental data for the band DB1. Clearly, in the calculations the spin is generated at a too low energy expense in the vicinity of the experimental bandhead. Since the shears closing is rather slow in the HF solutions, the discrepancy should rather be attributed to an overestimated collective inertia. The effect is most pronounced for the $G$ set of the time-odd fields, which systematically yields the largest values of the moments of inertia; also for the $N=75$ isotones studied in Section~\ref{crifre_sec}. Apart from the time-odd fields, the moment of inertia depends strongly on the deformation and on the intensity of the pairing correlations; see Section~\ref{pairin_sec}. Since the $\pi h_{11/2}^2~\nu h_{11/2}^{-2}$ shears configuration is a two-quasiparticle state in each kind of nucleons, the pairing effects should be significantly weakened, and one would expect the pure HF approximation to provide a correct description. From the theoretical point of view, there is no apparent reason to surmise that the calculated deformation is wrong, all the more since the predictions for the ground state are consistent with the TRS results. It must be conceded, however, that the quoted estimates of \cite{Mol95a,Lal96a} give a considerably lower $\beta$ for the ground state.

Whether the combined effects of the lack of pairing in the present calculations and of a possibly overestimated deformation may account for the observed deviation from the experimental data should be a subject of further research. Certainly, there are still several unclear points about the theoretical description of the magnetic bands in $^{142}$Gd, particularly concerning the deformation. In \cite{Lie02a}, a considerably smaller $\epsilon$ had to be used to reproduce the DB1 and DB3 bands than it came from the energy minimization. From the level scheme in Fig.~\ref{gd142scheme_fig} it is clear that the bands built on the $\pi h_{11/2}^2$ and $\nu h_{11/2}^{-2}$ configurations are rather well-deformed structures, similar to the ground-state band. Why would the combination of these two excitations diminish the deformation so much? It should be noted here that recently Pasternak {\it et al.} \cite{Pas04a} obtained a surprisingly good description of the magnetic bands in $^{142}$Gd within a very simple classical model. It would be instructive to see whether such solutions have their self-consistent mean-field counterparts.

\chapter{Hartree-Fock chiral results}
\label{chiral_cha}

Investigation of the chiral rotation in this thesis was directly stimulated by the first experimental observations of candidate chiral bands, in $^{130}$Cs, $^{132}$La, $^{134}$Pr, and $^{136}$Pm \cite{Sta01b}. These nuclei were naturally chosen for the analysis in this work. An exploratory study performed within the Principal-Axis Cranking (PAC) revealed that a very appealing and accurate model for the chiral rotation can be given in the classical framework of two gyroscopes coupled to a triaxial rotor. That model is presented in Section~\ref{clasic_sec} and leads to one of the principal conclusions of this dissertation, that chiral rotation can only exist above some critical value of the rotational frequency or spin. This result is confirmed by the full Skyrme-Hartree-Fock (Skyrme-HF) solutions that are presented later in this Chapter, and constitute the first fully self-consistent proof of the chiral symmetry breaking in rotating nuclei. However, the chiral bands could be obtained only in $^{132}$La from among the studied isotones. A brief report on the results obtained in $^{132}$La is given in \cite{Olb04a}, which is a revised version of the e-print published earlier in \cite{Olb02b}.

\section{Previous studies in $^{130}$Cs, $^{132}$La, $^{134}$Pr, and $^{136}$Pm}
\label{prechi_sec}

\begin{figure}
\begin{center}
\includegraphics{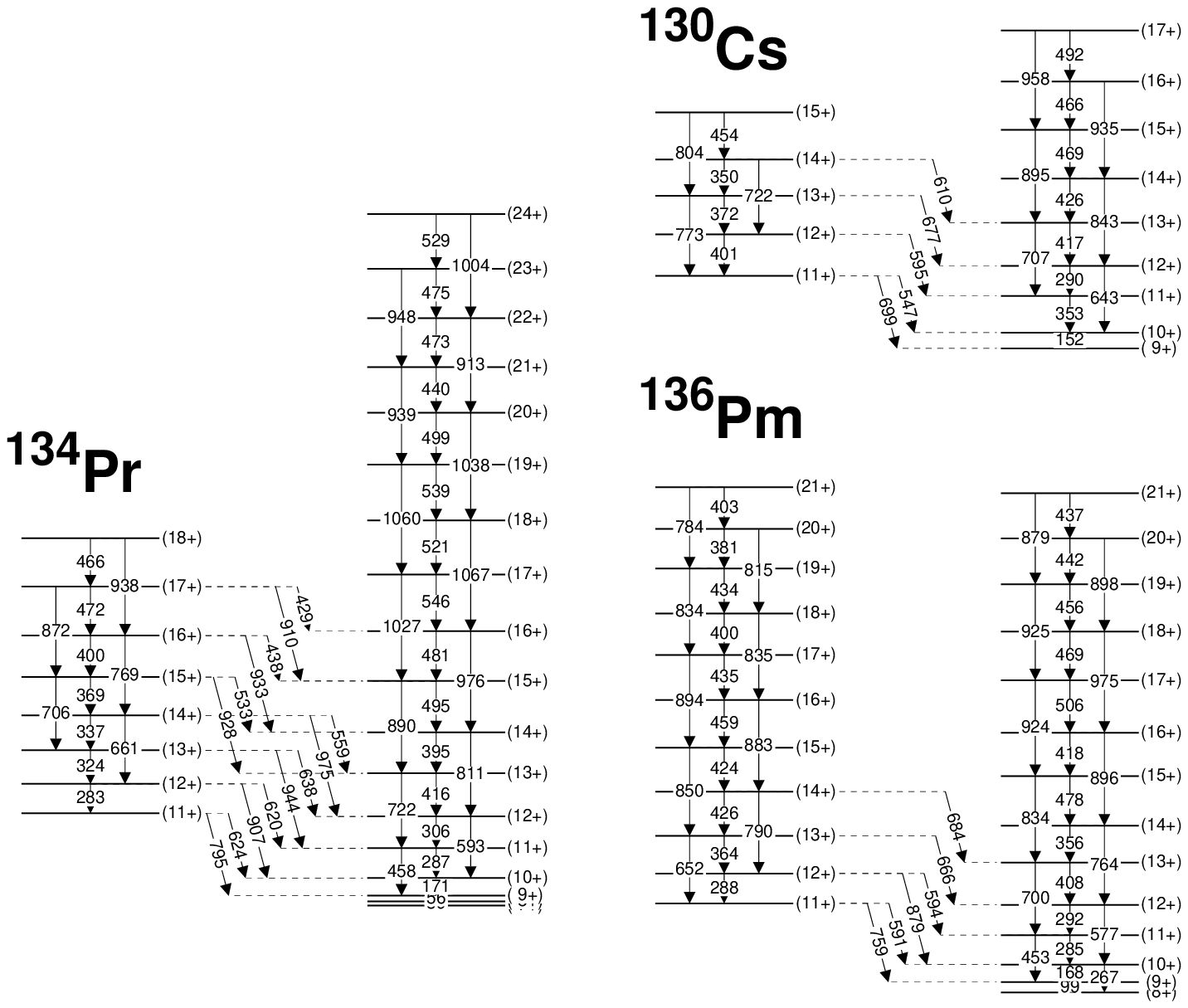}
\end{center}
\caption[Level schemes of the chiral doublets in $^{130}$Cs, $^{134}$Pr, $^{136}$Pm.]{Partial level schemes of $^{130}$Cs \cite{Koi03a}, $^{134}$Pr \cite{Sta01b,Rob03a} and $^{136}$Pm \cite{Har01a}, showing the yrast positive-parity bands (to the right) and supposed chiral partners (to the left).}
\label{n75sch_fig}
\end{figure}

\begin{figure}
\begin{center}
\includegraphics{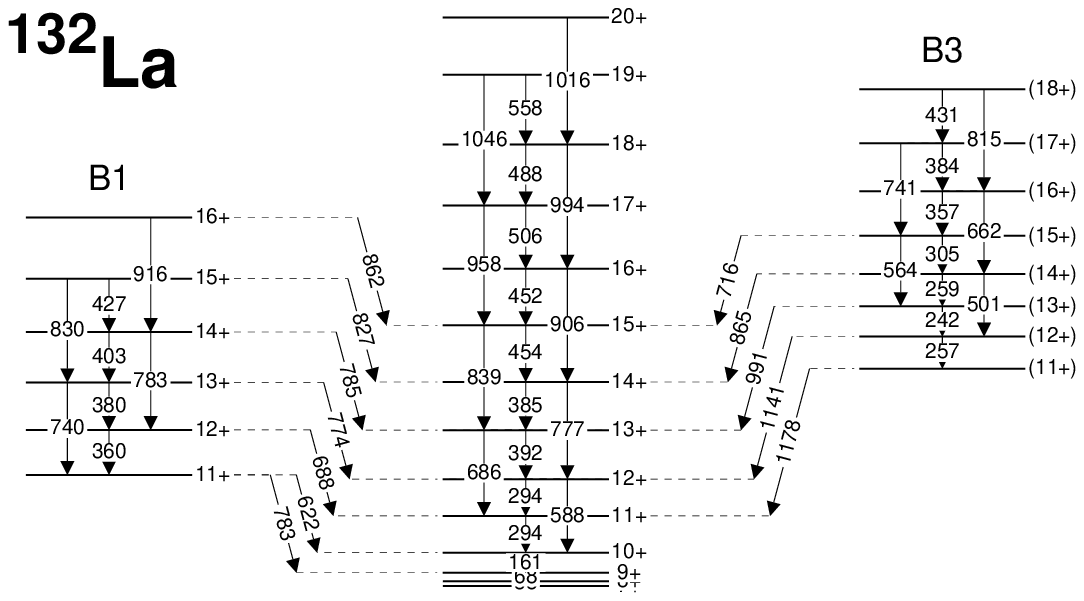}
\end{center}
\caption[Partial level scheme of $^{132}$La.]{Partial level scheme of $^{132}$La, showing the yrast band (middle) and its two candidate chiral partners. The band B1 is the previously known partner \cite{Sta02a}, while B3 is the recently discovered third band \cite{Gro04a}.}
\label{lasche_fig}
\end{figure}

As it was discussed in Section~\ref{chiral_sec}, the $\pi h_{11/2}^1~\nu h_{11/2}^{-1}$ configuration is in the literature commonly adopted for the chiral doublet bands in the $N=75$ isotones. Level schemes of those bands are displayed in Figs.~\ref{n75sch_fig} and \ref{lasche_fig}. Experimental data were taken from the most recent measurements, namely from \cite{Koi03a} for $^{130}$Cs, from \cite{Sta02a,Tim03a,Gro04a} for $^{132}$La, from \cite{Sta01b,Rob03a} for $^{134}$Pr, and from \cite{Har01a} in the case of $^{136}$Pm.
The names of bands in $^{132}$La, B1 and B3, are reproduced from \cite{Gro04a}. Spin assignments are still tentative, apart from the two lowest bands in $^{132}$La. Note that the bands referred to as yrast in the present work are only yrast in the positive parity. The low-spin regions of those bands are difficult to explore using $\gamma$-spectroscopy, because transition energies become very low. However, recent highbrow experiments in $^{132}$La \cite{Tim03a} and $^{134}$Pr \cite{Rob03a} suggested that the bandhead spins are $7^+$, which is consistent with the orthogonal angular momentum coupling of the $h_{11/2}$ proton particle and neutron hole. See Tables~\ref{chifir_tab} and \ref{chisec_tab} for a review of experimental and theoretical investigations of the chiral rotation in the $N=75$ isotones.

For all of nuclei $^{130}$Cs, $^{132}$La, $^{134}$Pr, and $^{136}$Pm phenomenological Tilted-Axis Cranking (TAC) calculations were performed in \cite{Sta01b}, and in \cite{Dim00a} also for $^{134}$Pr and in \cite{Hec01a} for $^{136}$Pm. All these calculations were performed by using the TAC code of Frauendorf. The $\epsilon$ deformations of $0.16$, $0.175$, $0.175$, $0.195$ and triaxialities $\gamma$ of $39^\circ$, $32^\circ$, $27^\circ$, $\approx26^\circ$ were found, respectively for $^{130}$Cs, $^{132}$La, $^{134}$Pr, $^{136}$Pm. Chiral solutions were obtained in a limited range of angular frequency. The lower limits correspond to the critical frequency, $\omega_{crit}$, discussed in this Chapter. Only \cite{Dim00a} and \cite{Hec01a} quote their values, which are $\omega_{crit}=0.3\,\mathrm{MeV}/\hbar$ for $^{134}$Pr and $\omega_{crit}<0.2\,\mathrm{MeV}/\hbar$ for $^{136}$Pm. Comparison with experimental energies is given only for $^{134}$Pr \cite{Sta01b,Dim00a}, where the mean trend is reproduced.

For all the considered $N=75$ isotones, calculations within the Particle-Rotor Model (PRM) were performed as well \cite{Pen03a}. A different version of the model was used in \cite{Har01a} for $^{136}$Pm. The same deformations were used as in the TAC calculations described in the previous Paragraph. A satisfactory agreement with experimental data was obtained. Probably the most sophisticated implementation of the PRM was employed in $^{132}$La \cite{Sta01a,Sta02a} and $^{134}$Pr \cite{Sta02a}. Deformations of $\beta=0.23$, $\gamma=21^\circ$ and $\beta=0.25$, $\gamma=35^\circ$ were adopted, respectively, and good agreement was obtained both for energies and $B(M1)/B(E2)$. Note that the values of $\beta$ used in the latter calculations are significantly larger than in the former ones.

Recently, a lifetime measurement in $^{132}$La \cite{Gro04a} provided the first data on the absolute values of $B(M1)$ and $B(E2)$ in the proposed chiral bands. That experiment revealed that the band B1, so far taken for the chiral partner, has the intraband $B(E2)$ values an order of magnitude lower than the yrast band and than those predicted by the PRM. Moreover, a new band, called B3, was discovered, that cannot be ruled out as the chiral partner, either; see Fig.~\ref{laener_fig}. Its electromagnetic properties are more similar to those of the yrast band. Energetically, it is more distant form the yrast band than B1, but the spin assignments are tentative, and there are some indications \cite{Gro04b} that they should be shifted up by one unit. In such a case, the band B3 would become almost exactly degenerate with the yrast sequence at spin $18~\hbar$. Certainly, further experimental research is needed to clarify this point.

\section{HF minima in the $N=75$ isotones}
\label{chimin_sec}

The present HF study of $^{130}$Cs, $^{132}$La, $^{134}$Pr, and $^{136}$Pm is limited to the chiral $\pi h_{11/2}^1~\nu h_{11/2}^{-1}$ configuration. This Section describes the structure of the minima that were found for that configuration before application of cranking.

Calculations were performed with two Skyrme parametrizations, SLy4 \cite{Cha97a} and SkM* \cite{Bar82a}; see Appendix~\ref{detals_app} for all details. To investigate the role of the HF time-odd densities and fields, three different sets of time-odd fields, $N$ (no time-odd fields), $G$ (time-odd fields responsible for a gauge invariance of the force), and $T$ (all Skyrme time-odd fields), were included; see Section~\ref{timodd_sec}. The $\pi h_{11/2}^1~\nu h_{11/2}^{-1}$ configuration is energetically most favored in the mean field for the single $h_{11/2}$ proton particle occupying the lowest substate of the orbital, and the neutron hole occupying the highest substate. Only such a case was considered, and the HF solutions were sought much like those in $^{142}$Gd, see Section~\ref{shemin_sec}, with the parity imposed. It was inferred from comparison of the symmetry-unrestricted and $D_{2h}^T$-imposing calculations that the $D_{2h}^T$ symmetry was not spontaneously broken in all solutions where no time-odd fields were included. Table~\ref{chicon_tab} gives the parity and parity-signature configurations (see Section~\ref{hfodd_sec}) that were found for the studied states.

In $^{130}$Cs and $^{132}$La, deformations of $\beta=0.24-0.26$ and $\gamma\approx45^{\circ}-49^{\circ}$ were obtained, depending on the force and time-odd fields. In the $N=75$ isotones, 25 neutrons are left above the $N=50$ spherical shell. Within the considered configuration, 11 of them are put on the $h_{11/2}$ orbital, and the remaining 14 occupy the positive-parity single-particle (s.p.) states above $N=50$. Presumably, they entirely fill the $g_{7/2}$ and $d_{5/2}$ orbitals, and thus exert no shape-driving force, although the positive-parity orbitals are obviously mixed by deformation.

In the $Z=55$ nucleus $^{130}$Cs, one of the 5 protons outside the $Z=50$ shell occupies the $h_{11/2}$ orbital, and the other 4 the $g_{7/2}$ orbital. Addition of two protons ($^{132}$La) also results in their location on that orbital. Although there obviously exist some mixing between the $g_{7/2}$ and $d_{5/2}$ proton orbitals, they seem to be quite well separated and can be easily recognized among the s.p.\ Routhians, see Figs.~\ref{cs_a0a_fig} and \ref{la_pro_fig}, \ref{la_neu_fig}, thanks to their alignment properties on the short and long axes. The $g_{7/2}$ orbital is composed of the levels 13+$\pm$, 14+$\pm$, 15+$\pm$, and 16+$\pm$ (see Section~\ref{hfodd_sec}), and $d_{5/2}$ comprises levels 17+$\pm$, 18+$\pm$, and 19+$\pm$.

In both $^{134}$Pr and $^{136}$Pm, two minima with the same $\pi h_{11/2}^1~\nu h_{11/2}^{-1}$ configuration were found, which differ by the occupation of positive-parity states. The energetically lower minimum has similar positive-parity s.p.\ structure as in $^{130}$Cs and $^{132}$La. Namely, the two protons added to form $^{134}$Pr take the last free substates of $g_{7/2}$, while the two protons added to form $^{136}$Pm occupy $d_{5/2}$. This can be traced in Figs.~\ref{pr_a0a_fig} and \ref{pm_a0a_fig}. However, such solutions correspond to almost oblate shapes of $\gamma=53^{\circ}-58^{\circ}$. The other minima have $\gamma=19^{\circ}-22^{\circ}$, which is triaxial, but closer to the prolate shapes\footnote{In the following, those two minima in $^{134}$Pr and $^{136}$Pm are referred to as {\it oblate} and {\it triaxial}.}. It seems that the occupied positive-parity neutron $g_{7/2}$ and $d_{5/2}$ orbitals are now strongly mixed with the $s_{1/2}$ and $d_{3/2}$ states from the same magic shell, which may also be the case for protons. For such deformations, some substates of the negative-parity $f_{7/2}$ or $h_{9/2}$ orbitals from above $N=82$ come so low that they slip under the highest $h_{11/2}$ substate, where the valence neutron hole is put\footnote{There is only one such intruder with the SkM* force, but two with SLy4. This makes the diabatic cranking calculations with the SLy4 force difficult due to multiple level crossings. In the following, only the SkM* results are shown for the {\it triaxial} minima, and only the SLy4 results for the {\it oblate} ones.}. The highest $h_{11/2}$ state can be distinguished thanks to its alignments on the short and long axes, which are exactly opposite to those of the down-coming lowest substates of $f_{7/2}$ or $h_{9/2}$. In Figs.~\ref{pr_b0a_fig} and \ref{pm_b0a_fig}, the highest $h_{11/2}$ substate has numbers 20--$\pm$. Exact deformations in the obtained minima are displayed in Table~\ref{crifre_tab}.

\begin{table}[h]
\caption[Single-particle configurations for the HF solutions in the $N=75$ isotones.]{Parity-signature and parity single-particle configurations (see Section~\ref{hfodd_sec}) for protons ($\pi$) and neutrons ($\nu$) used to obtain the solutions in the $N=75$ isotones at zero cranking frequency.}
\label{chicon_tab}
\begin{center}
\begin{tabular}{ll|c|cc|cc}
\hline\hline
               &      &       & \multicolumn{2}{c|}{Parity-Signature}             & \multicolumn{2}{c}{Parity} \\
               &      &       & Ref.          & Exc.                              & Ref.    & Exc. \\
\hline
$^{130}$Cs     & SLy4 & $\pi$ & (14,14,14,13) &  				  & (28,27) & \\
               & SkM* & $\nu$ & (19,19,19,18) &  				  & (38,37) & \\
\hline
$^{132}$La     & SLy4 & $\pi$ & (15,15,14,13) &			        	  & (30,27) & \\
               & SkM* & $\nu$ & (19,19,19,18) &				          & (38,37) & \\
\hline
$^{134}$Pr     & SLy4 & $\pi$ & (16,16,14,13) &  				  & (32,27) & \\
{\it oblate}   &      & $\nu$ & (19,19,19,18) &  				  & (38,37) & \\
\hline
$^{134}$Pr     & SkM* & $\pi$ & (16,16,14,13) &  				  & (32,27) & \\
{\it triaxial} &      & $\nu$ & (19,19,19,18) & (19--\,+\,$\rightarrow$\,20--\,+) & (38,37) & (37--\,$\rightarrow$\,39--) \\
\hline
$^{136}$Pm     & SLy4 & $\pi$ & (17,17,14,13) &  				  & (34,27) & \\
{\it oblate}   &      & $\nu$ & (19,19,19,18) &  				  & (38,37) & \\
\hline
$^{136}$Pm     & SkM* & $\pi$ & (17,17,14,13) &  				  & (34,27) & \\
{\it triaxial} &      & $\nu$ & (19,19,19,18) & (19--\,+\,$\rightarrow$\,20--\,+) & (38,37) & (37--\,$\rightarrow$\,39--) \\
\end{tabular}
\end{center}
\end{table}

\section{PAC calculations}
\label{pac_sec}

Before going to the full HF TAC calculations in the $N=75$ isotones, it is instructive to have a closer look at the PAC results, already discussed in Section~\ref{pacrth_sec}. They contain an important information on both the valence nucleons and on the collective core. The results in this Section lead to the formulation of the classical model, that will serve as a guideline for a consecutive HF study of the chiral rotation.

\begin{figure}
\begin{center}
\includegraphics{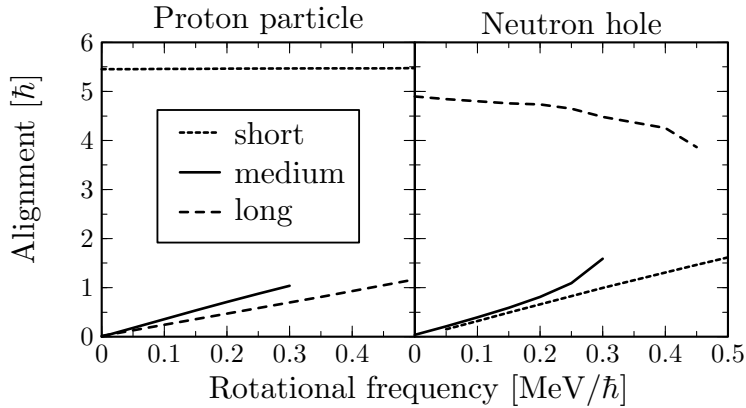}
\end{center}
\caption[Alignments of the valence $h_{11/2}$ nucleons in $^{132}$La from HF PAC.]{Single-particle alignments of the valence $h_{11/2}$ proton particle and neutron hole in $^{132}$La, obtained from one-dimensional cranking about the short, medium and long axis. The Figure shows the Hartree-Fock results for the SLy4 force and the $G$ time-odd fields included.}
\label{spalig_fig}
\end{figure}

Below, the study of the valence nucleons is carried out for the PAC calculations in $^{132}$La, performed with the SLy4 force and the $G$ time-odd fields. The conclusions derived in this sample case are supported for other nuclei and forces by the full TAC results; see Sections~\ref{plaban_sec} and \ref{chiban_sec}. Figure~\ref{spalig_fig} shows the s.p.\ alignments of the valence particle and hole, obtained from one-dimensional cranking about the three principal axes. The plot confirms that, at $\omega=0$, the particle and hole do indeed orient their spins, $\vec{j}^p$ and $\vec{j}^h$, on the short and long axis, respectively. The response of $\vec{j}^p$ and $\vec{j}^h$ to rotation is gradual and rather weak, meaning that the s.p.\ wave functions are strongly constrained by deformation (deformation-alignment). These results can be summarized as
\begin{equation}
\label{parhol_eqn}
\vec{j}^p\simeq s_s\vec{i}_s+\delta\cJ^p\vec{\omega}~, \qquad
\vec{j}^h\simeq s_l\vec{i}_l+\delta\cJ^h\vec{\omega}~,
\end{equation}
where $\vec{i}_s$ and $\vec{i}_l$ denote the unit vectors along the short and long axis, respectively. Therefore, to a reasonable approximation, the odd particle and hole can be treated like gyroscopes of spins $s_s$ and $s_l$, rigidly fixed along the short and long axes, while the small coefficients $\delta\cJ^p$ and $\delta\cJ^h$ can be incorporated into the total collective inertia tensor, $\cJ$.

\begin{figure}
\begin{center}
\includegraphics{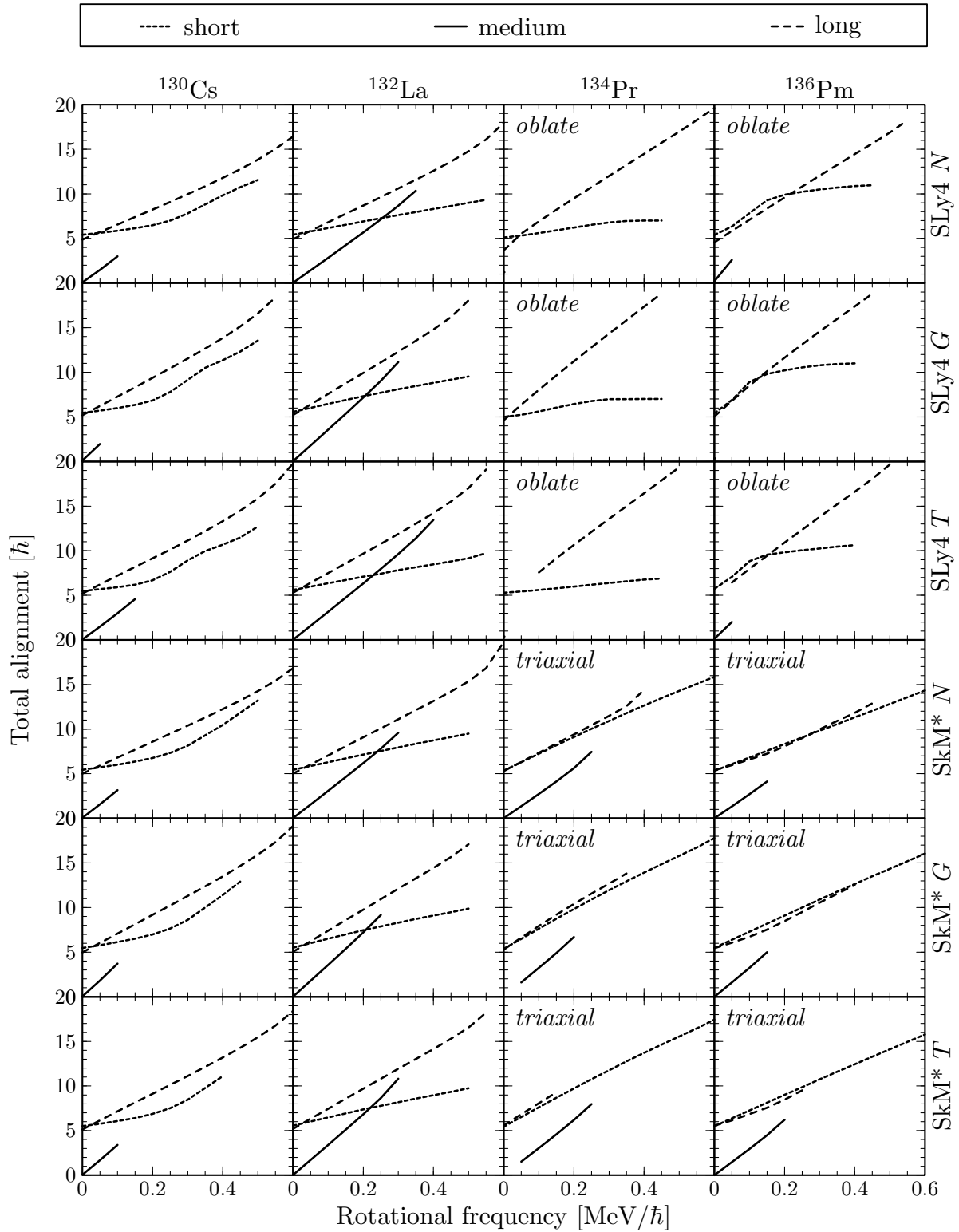}
\end{center}
\caption[Total alignments in $^{130}$Cs, $^{132}$La, $^{134}$Pr, $^{136}$Pm from HF PAC.]{Total alignments obtained from one-dimensional cranking about the short, medium and long axis in $^{130}$Cs, $^{132}$La, $^{134}$Pr, and $^{136}$Pm. Hartree-Fock results with the SLy4 and SkM* forces are shown for the $N$, $G$, $T$ time-odd fields included.}
\label{alignm_fig}
\end{figure}

The same PAC calculations provide the total alignments, $J_s$, $J_m$, $J_l$, that are plotted in Fig.~\ref{alignm_fig} for all the considered nuclei and forces. The upbend in the $J_s$ curve for $^{130}$Cs is caused by the change of slope of the last occupied positive-parity proton s.p.\ state due to its interaction with the first empty one. These are the levels 14++ and 15++ in the top-left panel of Fig.~\ref{cs_a0a_fig}, respectively. In $^{132}$La, the 14++ and 15++ states are both occupied, and no upbend occurs. Similarly, the upbend in $J_l$ for the triaxial minimum in $^{136}$Pm is because of the interaction between the neutron hole state 18++ and the particle state 20++, see the bottom-right panel in Fig.~\ref{pm_b0a_fig}. The upbend in $J_s$ for the nearly oblate solution in $^{136}$Pm originates from the smooth crossing of the 17++ and 18++ proton states, as shown in the top-left part of Fig.~\ref{pm_a0a_fig}. This crossing has no effect in $^{134}$Pr, because there are two protons less there, and the concerned states are both empty. In the nearly oblate minima in $^{134}$Pr and $^{136}$Pm, the medium axis is not well distinguished from the long one, and an attempt to crank around it immediately leads to solutions associated with the long axis.

One can see that all the observed bends in the total alignment plots are due to smooth level crossings near the Fermi surface. Otherwise, the $J_i(\omega_i)$ dependence is, to a good approximation, linear, like for the rigid rotation. The corresponding slopes give the collective total moments of inertia, $\cJ_s$, $\cJ_m$, $\cJ_l$, with respect to the short, medium, and long axis, respectively. They already contain the small contributions, $\delta\cJ^p$ and $\delta\cJ^h$, from the valence particle and hole\footnote{Since the coefficients $\cJ_i$, $i=s,m,l$, are slopes of the $J_i(\omega_i)$ dependence, they should be thought of more as the second, or dynamical moments of inertia. In classical mechanics, they would have the meaning of components of the inertia tensor.}. At zero frequency, cranking around the medium axis gives a vanishing angular momentum, while those around the other two axes give non-zero values, equal to the initial alignments, $s_s$ and $s_l$, of the odd particle and hole.

The microscopic PAC results presented so far suggest that the considered system can be modeled by two gyroscopes of spins $s_s$ and $s_l$ rigidly fixed along the short and long axes of a triaxial rigid rotor characterized by the inertia tensor, $\cJ$, whose diagonal components are equal to $\cJ_s$, $\cJ_m$, $\cJ_l$. It is instructive to solve the associated problem of motion in the classical framework, which is done in the next Section. It was mentioned in Section~\ref{symmet_sec} that for no time-odd fields included in the present HF calculations, the mean potentials for the non-rotating states have the $D_{2h}^T$ symmetry, which implies that the inertia tensor must be diagonal in the principal-axis frame of the mass distribution. With the time-odd fields included, perturbative TAC tests were performed, in which to the self-consistent non-rotating solution cranking frequency was applied in different directions in order to check the response of the mean angular momentum vector. It was found that the off-diagonal components of $\cJ$ are negligible. Therefore, the inertia tensor is taken diagonal.

\section{Classical model and the critical frequency}
\label{clasic_sec}

It was argued in the previous Section that a simple model of two gyroscopes coupled to a triaxial rigid body is appropriate for the description of rotation in the considered nuclei. Below, that model is solved in the classical framework. It possesses several analogies to the HF TAC method in its foundations. Yet first of all, it leads to a crucial conclusion that chiral rotation can only exist above some critical value of the rotational frequency or spin. A simple analytical formula allows for estimating the critical frequency on the basis of the standard PAC results.

\begin{figure}
\begin{center}
\includegraphics{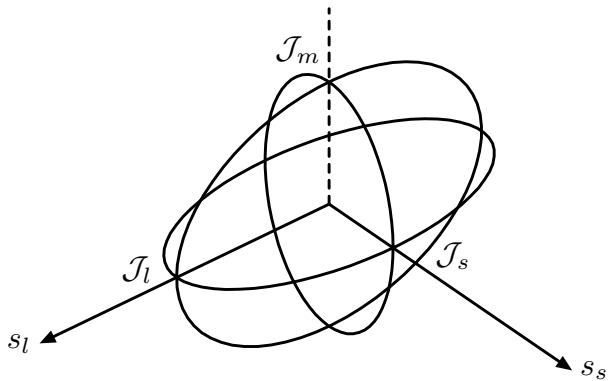}
\end{center}
\caption[The classical model for the chiral rotation.]{The classical model of chiral rotation. Two gyroscopes of spins $s_s$ and $s_l$ rigidly fixed along the short and long axes of a triaxial rigid body characterized by three moments of inertia, $\cJ_{s}$, $\cJ_{m}$, $\cJ_{l}$, associated with the short ($s$), medium ($m$) and long ($l$) axis.}
\label{clasic_fig}
\end{figure}

The model system is sketched in Fig.~\ref{clasic_fig}. A triaxial rigid body models the collective core. Its inertia tensor, $\cJ$, is diagonal with respect to the short, medium and long axes, and the corresponding components, $\cJ_{s}$, $\cJ_{m}$, $\cJ_{l}$, also include the small contributions from the valence particle and hole; see Eq.~(\ref{parhol_eqn}). The particle and hole are represented by two ideal gyroscopes of spins $s_s$ and $s_l$, rigidly fixed along the short and long axis of the core, respectively.

The notion of the ideal gyroscope requires explanation. Consider an axially symmetric rigid body with moments of inertia, $\cJ_{\parallel}$ and $\cJ_{\perp}$, with respect to the symmetry axis and an axis perpendicular to it, respectively. If such a body spins at angular frequency $\Omega$ around the symmetry axis, its kinetic energy is $T_0=\cJ_{\parallel}\Omega^2/2$. Simultaneously, if the body rotates with arbitrarily oriented frequency, $\vec{\omega}$, the kinetic energy becomes $T_{\omega}=(\cJ_{\perp}\omega^{2}\sin^{2}\vartheta+\cJ_{\parallel}(\omega\cos\vartheta+\Omega)^{2})/2$, where $\vartheta$ is the angle between $\vec{\omega}$ and the symmetry axis. The ideal gyroscope is the limiting case when $\cJ_{\parallel}$ and $\cJ_{\perp}$ go to zero and $\Omega$ increases so that $\Omega \cJ_{\parallel}$ remain constant and equal to a given value of $s$. Obviously the total spin is then equal to $s$ and points along the symmetry axis, while $T_{\omega}$ becomes infinite. However, the difference $T_{\omega}-T_0=(\cJ_{\perp}\sin^2\vartheta+\cJ_{\parallel}\cos^2\vartheta)\Omega^2/2+\cJ_{\parallel}\Omega\omega\cos\vartheta
$ has a finite limit,
\begin{equation}
T=\omega s\cos\vartheta=\vec{\omega}\vec{s}~.
\end{equation}
This difference can be considered as the kinetic energy of the ideal gyroscope measured with respect to the case with no rotation, that is with $\vec{\omega}=0$.

Coming back to the discussed classical model, the total angular momentum, $\vec{J}$, of the system reads
\begin{equation}
\vec{J}=\cJ\vec{\omega}+\vec{s}~,
\end{equation}
where $\vec{s}$ is the vector sum of spins of the the two gyroscopes. In the absence of potential interactions, the Lagrangian of the system is equal to the total kinetic energy, and is given by the formula
\begin{equation}
\label{LEkin_eqn}
L=E_{kin}=\frac{1}{2}\vec{\omega}\hat{\cJ}\vec{\omega}+\vec{\omega}\vec{s}~.
\end{equation}
Taking the laboratory components of $\vec{\omega}$ as generalized velocities, it is easy to check that the generalized momenta are equal to the laboratory components of $\vec{J}$. This fact allows us to write the Legendre transformation \cite{Gol53a} and to obtain the Hamiltonian of the system,
\begin{equation}
\label{HwIL_eqn}
H=\vec{\omega}\vec{J}-L=\frac{1}{2}\vec{\omega}\cJ\vec{\omega}=\frac{R_s^2}{2\cJ_s}+\frac{R_m^2}{2\cJ_m}+\frac{R_l^2}{2\cJ_l}~,
\end{equation}
where $\vec{R}=\vec{J}-\vec{s}$ is the angular momentum of the rigid body alone. Note that $H$ is analogous to the rotational part (\ref{prm_eqn}) of the PRM Hamiltonian. Since the Lagrangian (\ref{LEkin_eqn}) does not depend explicite on time, Hamiltonian (\ref{HwIL_eqn}) is a constant of motion, and is identified with the total energy, $E$, of the system. Consider now a particular type of the Routh function \cite{Gol53a} (Routhian), $H'$, namely such that no variables undergo the Legendre transformation. In such a case, $H'=-L$, and by rewriting the Routhian in terms of the Hamiltonian one obtains
\begin{equation}
\label{RHwI_eqn}
H'=H-\vec{\omega}\vec{J}~.
\end{equation}
This quantity is similar to Routhian (\ref{HHwJ_eqn}) appearing in the quantum cranking model.

Equations of motion for the model can be derived in the following way. The time derivatives, $\partial_{t}$ and $\partial^{\omega}_{t}$, of any vector $\vec{J}$, taken in the laboratory frame and in a frame rotating at angular frequency $\vec{\omega}$, respectively, are related by the formula $\partial_{t}\vec{J}=\partial^{\omega}_{t}\vec{J}+\vec{\omega}\times\vec{J}$ \cite{Gol53a}. But $\partial_{t}\vec{J}=0$, if $\vec{J}$ represents the angular momentum, which is conserved in the laboratory frame. Thus, one obtains the Euler equations \cite{Gol53a} for the time evolution of the angular-momentum vector in the body-fixed frame,
\begin{equation}
\label{dwtI_eqn}
\partial^{\omega}_{t}\vec{J}=-\vec{\omega}\times\vec{J}~.
\end{equation}
Thouless and Valatin \cite{Tho62a} showed that analogous equations govern the evolution of the mean angular momentum in the Time-Dependent HF theory.

Apart from the equations of motion, a convenient way to analyze rotation of the considered system is through phase portraits, which are intersections of the angular-momentum sphere with the energy ellipsoid. Dimitrov {\it et al.} carried out such an analysis \cite{Dim02a} for a slightly simplified version of the model. Separatrices defining the planar and chiral rotational regimes as well as the chiral vibrational regime (see Section~\ref{chiral_sec}) are clearly visible in this approach. However, the mean-field cranking approximation can only account for the so-called {\it uniform} rotations, in which the mean angular-momentum vector is constant in the intrinsic frame of the nucleus, $\partial^{\omega}_{t}\vec{J}=0$. Because of that, also the present study within the classical model is limited to such uniform motions. The Euler equations (\ref{dwtI_eqn}) now take the form $\vec{\omega}\times\vec{J}=0$, and require that $\vec{\omega}$ and $\vec{J}$ be parallel. The same condition holds for the HF solutions and is known as the Kerman-Onishi theorem; see Section~\ref{tac_sec} and \cite{Ker81a}. These equations can be easily solved for the considered model. However, to show further analogies with the HF method, here the variational principle is employed to find uniform solutions.

According to the Hamilton's principle, motion of a mechanical system can be found by making the action integral, $\int L~dt$, stationary. Here, real uniform rotations are sought, and all of them obviously belong to a wider class of trial motions with $\vec{\omega}$ constant in the intrinsic frame. Within this class, Lagrangian (\ref{LEkin_eqn}) does not depend on time. Therefore, extremizing the action for the given $\omega$ reduces to finding extrema of the Lagrangian as a function of the intrinsic-frame components of $\vec{\omega}$. Since $H'=-L$, the Routhian (\ref{RHwI_eqn}) can be equally used for this purpose. This serves us as a bridge between the classical model and quantum cranking theory, where an analogous Routhian is minimized in the space of the trial wavefunctions.

Extrema of $H'$ with respect to the intrinsic-frame components of $\vec{\omega}$ at a given length of $\omega$ can be found by using a Lagrange multiplier, $\mu$, for $\omega^{2}$. Setting to zero the derivatives of the quantity
\begin{equation}
H'+\frac{1}{2}\mu\omega^2=\frac{1}{2}\left[(\mu-\cJ_s)\omega_s^2+(\mu-\cJ_m)\omega_m^2+(\mu-\cJ_l)\omega_l^2\right]-(\omega_s s_s+\omega_l s_l)
\end{equation}
with respect to $\omega_{s}$, $\omega_{m}$, $\omega_{l}$, one obtains
\begin{eqnarray}
\omega_s            & = & s_s/(\mu-\cJ_s)~, \label{wsssmJs} \\
\omega_m(\mu-\cJ_m) & = & 0~,               \label{wmmJm0} \\
\omega_l            & = & s_l/(\mu-\cJ_l)~. \label{wlslmJl}
\end{eqnarray}
Equation (\ref{wmmJm0}) gives either $\omega_{m}=0$ or $\mu={\cJ}_{m}$, leading to two distinct classes of solutions.

\begin{figure}
\begin{center}
\includegraphics{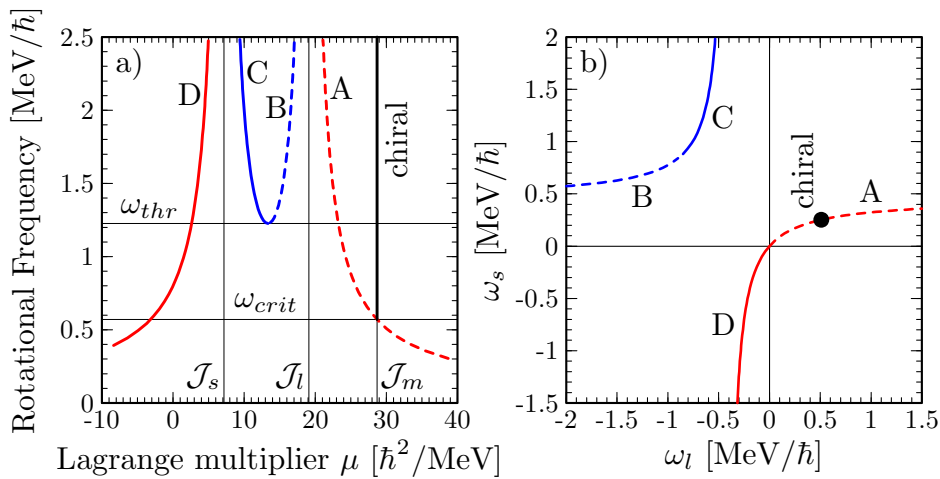}
\end{center}
\caption[The $\omega(\mu)$ dependence for the classical model and the classical planar and chiral solutions in the $\vec{\omega}$ space.]{a) Rotational frequency $\omega(\mu)$ for the four planar bands (marked as A, B, C, D) and for the chiral doublet (the vertical line) obtained in the classical model. b) Intrinsic-frame trajectory of the rotational frequency vector along those bands. The chiral solution corresponds to a straight line perpendicular to the figure plane, and intersecting it at the marked point. That perpendicular direction represents $\omega_m$.}
\label{wmwswl_fig}
\end{figure}

\begin{figure}
\begin{center}
\includegraphics{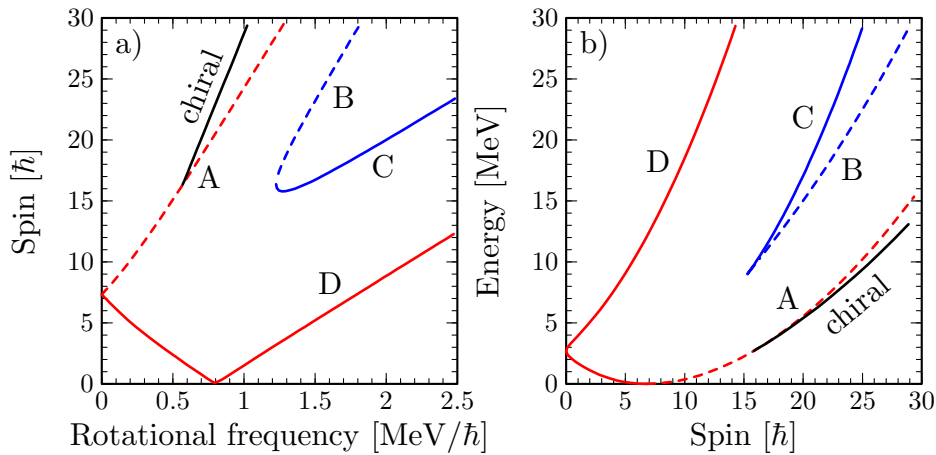}
\end{center}
\caption[Spins and energies for the planar and chiral bands in the classical model.]{a) Angular momenta $J(\omega)$ and b) Energies $E(I)$ for the four planar bands and the chiral doublet obtained in the classical model.}
\label{JwEI_fig}
\end{figure}

If $\omega_{m}=0$ then both $\vec{\omega}$ and $\vec{J}$ lie in the $s$-$l$ plane. This gives planar solutions, for which the chiral symmetry is not broken. All values of $\mu$ are allowed, and the Lagrange multiplier must be determined from the length of $\vec{\omega}$ calculated in the obvious way from (\ref{wsssmJs}) and (\ref{wlslmJl}). Figure~\ref{wmwswl_fig}a shows $\omega$ versus $\mu$ for sample model parameters, extracted from the HF PAC solutions with the SLy4 force with no time-odd fields, and listed in Table~\ref{crifre_tab}. The solutions marked as A and D exist for all values of $\omega$, while above some threshold frequency, $\omega_{thr}$, two more solutions appear, B and C. This threshold frequency can be determined by finding the minimum of $\omega$ in function of $\mu$, and reads
\begin{equation}
\omega_{thr}=\frac{\left(s_s^{2/3}+s_l^{2/3}\right)^{3/2}}{|\cJ_l-\cJ_s|}~.
\end{equation}
The value of $\omega_{thr}$ coming from the present HF calculations is rather high, higher than $1\,\mathrm{MeV}/\hbar$. Since bands B and C are situated far above the yrast line (see Fig.~\ref{JwEI_fig}) they will not be subject of further analysis.

For $\mu$=$\cJ_m$, all values of $\omega_m$ are allowed, while components in the $s$-$l$ plane are fixed at $\omega_s=s_s/(\cJ_m-\cJ_s)$ and $\omega_l=s_l/(\cJ_m-\cJ_l)$. Consequently, the angular momentum has non-zero components along all three axes, and the chiral symmetry is broken. For each value of $\omega$, there are two cases differing by the sign of $\omega_{m}$, and thus giving the chiral doublet. The fact that $\omega_{s}$ and $\omega_{l}$ are constant leads to the principal conclusion that chiral solutions cannot exist for $\omega$ smaller than the critical frequency
\begin{equation}
\label{crifre_eqn}
\omega_{crit}=\left[\left(\frac{s_s}{\cJ_m-\cJ_s}\right)^2
                   +\left(\frac{s_l}{\cJ_m-\cJ_l}\right)^2\right]^{1/2}~.
\end{equation}
At that frequency, and with $\omega_m$=0, the chiral solution coincides with the planar band A. 

In the $\vec{\omega}$ space, the four planar solutions form a hyperbola in the $s$-$l$ plane, while the chiral doublet corresponds to a straight line perpendicular to that plane. These curves are shown in Fig.~\ref{wmwswl_fig}b. Figure~\ref{JwEI_fig}a gives the angular momentum in function of rotational frequency for all the presented bands. With increasing $\omega$, the second moment of inertia, $\mathrm{d}J/\mathrm{d}\omega$, asymptotically approaches $\cJ_l$ for bands A and B, and $\cJ_s$ for bands C and D. For the chiral band, $J$ is exactly proportional to $\omega$ with the coefficient $\cJ_m$. Thus, the critical spin, $J_{crit}$, corresponding to the critical frequency
(\ref{crifre_eqn}) reads
\begin{equation}
J_{crit}=\cJ_m\omega_{crit}~.
\end{equation}
Figure~\ref{JwEI_fig}b summarizes the energies in function of spin. At low angular momenta, the yrast line coincides with the planar band D. Then it continues along the planar solution A. Since the moment of inertia $\cJ_m$ is the largest, beyond the critical frequency the chiral solution becomes yrast, thereby yielding good prospects for experimental observation.

Since the works of Volterra of 1898 \cite{Vol98a}, similar classical particle-rotor models have been considered in several contexts different than the chiral rotation. They do not always assume a rigid coupling of gyroscopes to the core, but introduce a potential interaction between them. Bohr and Mottelson \cite{Boh80a}, and later also Kamchatnov \cite{Kam90a} examined one gyroscope interacting with an axially-symmetric core, to study high-K bands. As far as the magnetic rotation is concerned, Macchiavelli {\it et al.} \cite{Mac99a} considered two gyroscopes coupled to a spherical core to investigate the competition between the collective rotation and shears mechanism; see Section~\ref{magnet_sec}. That model was generalized by Pasternak \cite{Pas04a} to the case of axially deformed core.

\section{Classical estimates of the critical frequency}
\label{crifre_sec}

A useful feature of the classical model presented in the previous Section is that all its parameters can be easily extracted from the standard PAC calculations (see Section~\ref{pac_sec}), that can be performed by many existing mean-field codes. Then, Eq.~(\ref{crifre_eqn}) provides an estimate of the critical frequency, and gives insight into where its value comes from. Below, the critical frequency and spin are estimated on the basis of the HF PAC results of Section~\ref{pac_sec}, and also from the PAC calculations with pairing, performed within the Total Routhian Surface (TRS) method.

The model parameters, $\cJ_s$, $\cJ_m$, $\cJ_l$, and $s_s$, $s_l$, were obtained from a least-square fit of straight lines, $J_i=\cJ_i\omega_i+s_i$, $i=s,m,l$, to the calculated points shown in Fig.~\ref{alignm_fig}. Parameter $s_m$ was set to zero. Whenever there was a bend in the calculated $J_i(\omega_i)$ dependence (see Section~\ref{pac_sec}), the line was fitted in the rotational frequency range below the bend. The fitted moments of inertia and initial alignments, together with the resulting values of $\omega_{crit}^{clas}$ and $I_{crit}^{clas}$(\footnote{In the present work, the symbol $\vec{J}$ denotes the average angular-momentum vector calculated from a mean-field solution or the classical angular-momentum vector. The symbol $I$ is used for the spin quantum number, as it is measured in experiment. If $I$ is approximated from mean-field or classical results, the relation $I(I+1)=\vec{J}^2$ is used.}), are listed in Table~\ref{crifre_tab}.

As expected from the irrotational-flow model \cite{Boh75a}, the moment of inertia with respect to the medium axis is the largest. Note, however, that the $\cJ_s:\cJ_m:\cJ_l$ ratio is far from that predicted by the irrotational-flow formula (\ref{irrflo_eqn}) used for the calculated triaxiality $\gamma$. The inclusion of the time-odd fields $G$ (see Section~\ref{timodd_sec}) increases all the moments of inertia, but particularly $\cJ_m$, with respect to the case, $N$, with no time-odd fields. This causes a decrease in $\omega_{crit}^{clas}$, but the corresponding $I_{crit}^{clas}$ does not change much, because $\cJ_m$ is larger. Switching on the fields $T$ results in values of $\cJ_m$ between those obtained for the cases $N$ and $G$. The critical frequency always becomes higher than for the $G$ fields, and the resulting critical spin is always the highest among all the examined sets of time-odd fields. These variations of $I_{crit}^{clas}$ are of the order of a few spin units.

For the chiral bands in $^{132}$La, the full HF TAC solutions were found (see Section~\ref{chiban_sec}), and the obtained values of the critical frequency and spin, $\omega_{crit}^{HF}$ and $I_{crit}^{HF}$, are also listed in Table~\ref{crifre_tab}. They are slightly higher than the classical estimates, $\omega_{crit}^{clas}$ and $I_{crit}^{clas}$, but the agreement can be considered good, given the simplicity of the classical model. When the time-odd fields are included, values of $\omega_{crit}^{HF}$ and $I_{crit}^{HF}$ vary similarly to those of $\omega_{crit}^{clas}$ and $I_{crit}^{clas}$.

To examine the possible role of pairing correlations, PAC calculations within the TRS approach were performed for $^{132}$La; see Appendix~\ref{detals_app} for details. Deformation of $\beta\approx0.20$ and $\gamma\approx25^{\circ}$ was obtained from energy minimization. In the cranking calculations, TRS effectively interchanges values of $\cJ_s$ and $\cJ_l$, and enlarges $\cJ_m$ roughly twice with respect to the HF results. On the basis of the irrotational-flow model, these changes in the moments of inertia can be partly understood as due to the change in triaxiality; the value of $\gamma$ obtained from TRS is closer to $30^\circ$ and on the other side of $30^\circ$ than the HF results. As governed by Eq.~(\ref{crifre_eqn}), the increase in $\cJ_m$ significantly lowers the values of $\omega^{clas}_{crit}$ and $I^{clas}_{crit}$, see Table~\ref{crifre_tab}. However, inclusion of pairing in the full HF TAC calculations (Hartree-Fock-Bogolyubov method) is necessary to verify this trend.

Values of the critical spin obtained from the HF calculations are rather high as compared to the spin range in which the supposed chiral bands are observed in nuclei under study. A detailed comparison with experiment is given in Section~\ref{chiexp_sec}.

\begin{table}
\caption[Deformations, parameters of the classical model, and critical frequencies and spins for the $N=75$ isotopes.]{The $\beta$ and $\gamma$ deformations, parameters of the classical model, $\cJ_{s,m,l}$ $[\hbar^2/\mbox{MeV}]$, $s_{s,l}$ $[\hbar]$, classical estimates for the critical frequency and spin, $\omega_{crit}^{clas}$ $[\mathrm{MeV}/\hbar]$ and $I_{crit}^{clas}$ $[\hbar]$, and full HF TAC results for those quantities, $\omega_{crit}^{HF}$ $[\mathrm{MeV}/\hbar]$ and $I_{crit}^{HF}$ $[\hbar]$, for the $N=75$ isotopes. HF results with the SLy4 and SkM* forces are listed for the $N$, $G$, $T$ time-odd fields included. For $^{132}$La, results from TRS calculations are given, too.}
\label{crifre_tab}
\begin{center}
\begin{tabular}{lc|cc|ccc|cc|cccc}
\hline\hline
$^{130}$Cs     &     & $\beta$ & $\gamma$ & $\cJ_s$ & $\cJ_m$ & $\cJ_l$ & $s_s$ & $s_l$ & $\omega^{clas}_{crit}$ & $I^{clas}_{crit}$ & $\omega^{HF}_{crit}$ & $I^{HF}_{crit}$ \\
\hline
SLy4           & $N$ & 0.24    & 49	     & 4.81    & 29.2    & 17.0    & 5.41  & 4.86  & 0.46		       & 12.8		   &			  &		    \\
               & $G$ & 0.24    & 49       & 5.50    & 37.1    & 21.3    & 5.45  & 5.16  & 0.37		       & 13.2		   &			  &	      \\
               & $T$ & 0.24    & 49	     & 4.16    & 29.4    & 19.9    & 5.49  & 5.20  & 0.59		       & 16.8		   &			  &	      \\
\hline
SkM*           & $N$ & 0.23    & 47	     & 5.86    & 31.3    & 17.9    & 5.43  & 5.01  & 0.43		       & 13.0		   &			  &	      \\
               & $G$ & 0.23    & 47	     & 6.55    & 36.7    & 21.1    & 5.47  & 4.97  & 0.37		       & 13.0		   &			  &	      \\
               & $T$ & 0.23    & 47       & 5.69    & 33.4    & 20.0    & 5.49  & 5.14  & 0.43		       & 13.9		   &			  &	      \\
\hline\hline
$^{132}$La     &   & $\beta$ & $\gamma$ & $\cJ_s$ & $\cJ_m$ & $\cJ_l$ & $s_s$ & $s_l$ & $\omega^{clas}_{crit}$ & $I^{clas}_{crit}$ & $\omega^{HF}_{crit}$ & $I^{HF}_{crit}$ \\
\hline
SLy4           & $N$ & 0.26    & 46       & 7.18    & 28.7    & 19.1    & 5.44  & 4.90  & 0.57		       & 15.9		   & 0.68		  & 18.8 \\
               & $G$ & 0.26    & 46       & 8.45    & 36.0    & 23.7    & 5.60  & 5.21  & 0.47		       & 16.4		   & 0.60		  & 20.3 \\
               & $T$ & 0.26    & 46       & 7.12    & 31.7    & 22.2    & 5.64  & 5.26  & 0.60		       & 18.5		   &			  &	 \\
\hline
SkM*           & $N$ & 0.25    & 45       & 8.19    & 30.8    & 20.3    & 5.47  & 5.03  & 0.54		       & 16.0		   & 0.62		  & 17.8 \\
               & $G$ & 0.25    & 45       & 8.81    & 35.9    & 23.5    & 5.60  & 5.06  & 0.46		       & 15.9		   & 0.54		  & 17.8 \\
               & $T$ & 0.25    & 45       & 8.37    & 34.0    & 22.4    & 5.63  & 5.21  & 0.50		       & 16.5		   & 0.58		  & 18.5 \\
\hline
TRS            &     & 0.20    & 25       & 15.7    & 65.6    & 9.33    & 6.33  & 4.23  & 0.15                   & 9.2               &                      &      \\
\hline\hline
$^{134}$Pr     &   & $\beta$ & $\gamma$ & $\cJ_s$ & $\cJ_m$ & $\cJ_l$ & $s_s$ & $s_l$ & $\omega^{clas}_{crit}$ & $I^{clas}_{crit}$ & $\omega^{HF}_{crit}$ & $I^{HF}_{crit}$ \\
\hline
SLy4           & $N$ & 0.26    & 58       & 6.11    &         & 25.4    & 5.00  & 4.36  &                        &                   &                      &                 \\
{\it oblate}   & $G$ & 0.26    & 58       & 7.21    &         & 31.3    & 4.92  & 4.84  &                        &                   &                      &                 \\
               & $T$ & 0.26    & 56       & 3.68    &         & 29.8    & 5.26  & 4.60  &                        &                   &                      &                 \\
\hline
SkM*           & $N$ & 0.23    & 22       & 18.5    & 28.1    & 20.7    & 5.38  & 5.28  & 0.91                   & 25.0              &                      &                 \\
{\it triaxial} & $G$ & 0.23    & 22       & 21.4    & 32.6    & 24.3    & 5.46  & 5.44  & 0.82                   & 26.1              &                      &                 \\
               & $T$ & 0.23    & 22       & 20.7    & 30.8    & 24.8    & 5.52  & 5.57  & 1.08                   & 32.7              &                      &                 \\
\hline\hline
$^{136}$Pm     &     & $\beta$ & $\gamma$ & $\cJ_s$ & $\cJ_m$ & $\cJ_l$ & $s_s$ & $s_l$ & $\omega^{clas}_{crit}$ & $I^{clas}_{crit}$ & $\omega^{HF}_{crit}$ & $I^{HF}_{crit}$ \\
\hline
SLy4           & $N$ & 0.25    & 53       &	  &	    & 24.7    &       & 4.62  & 		       &		   &			  &		    \\
{\it oblate}   & $G$ & 0.25    & 53       &	     &         & 30.3	 &	 & 5.38  &			  &		      & 		     &  	       \\
               & $T$ & 0.25    & 52       &	  &	    & 29.0    &       & 5.05  & 		       &		   &			  &		    \\
\hline
SkM*           & $N$ & 0.22    & 19       & 15.0    & 27.5    & 12.4    & 5.32  & 5.38  & 0.56		       & 14.8		   &			  &		    \\
{\it triaxial} & $G$ & 0.22    & 19       & 17.9    & 33.2    & 13.0	& 5.50  & 5.42  & 0.45  		 & 14.4 	     &  		    &		      \\
               & $T$ & 0.22    & 19       & 17.4    & 29.8    & 13.4    & 5.51  & 5.53  & 0.56		       & 16.1		   &			  &		    \\
\end{tabular}
\end{center}
\end{table}

\section{Planar HF bands}
\label{plaban_sec}

The next step in the HF analysis of the $\pi h_{11/2}^1~\nu h_{11/2}^{-1}$ configuration in the $N=75$ isotones was to obtain planar solutions, much similar to the shears solutions in $^{142}$Gd; see Section~\ref{shemin_sec}. They serve us as starting points in the search for the chiral bands, as described in the next two Sections.

In the $N=75$ isotones, planar solutions were found by applying a tilted cranking vector to converged non-rotating states. Since in the triaxial minima in $^{134}$Pr and $^{136}$Pm, the neutron $h_{11/2}$ states interleave with other negative-parity levels (see Section~\ref{chimin_sec}), it was very difficult to follow the proper configuration diabatically, and the planar bands in those minima were obtained only for very low angular frequencies. Figures~\ref{la_00c_fig}--\ref{la_0Tc_fig} give sample s.p.\ Routhians for the planar bands in $^{132}$La.

\begin{figure}[h]
\begin{center}
\includegraphics{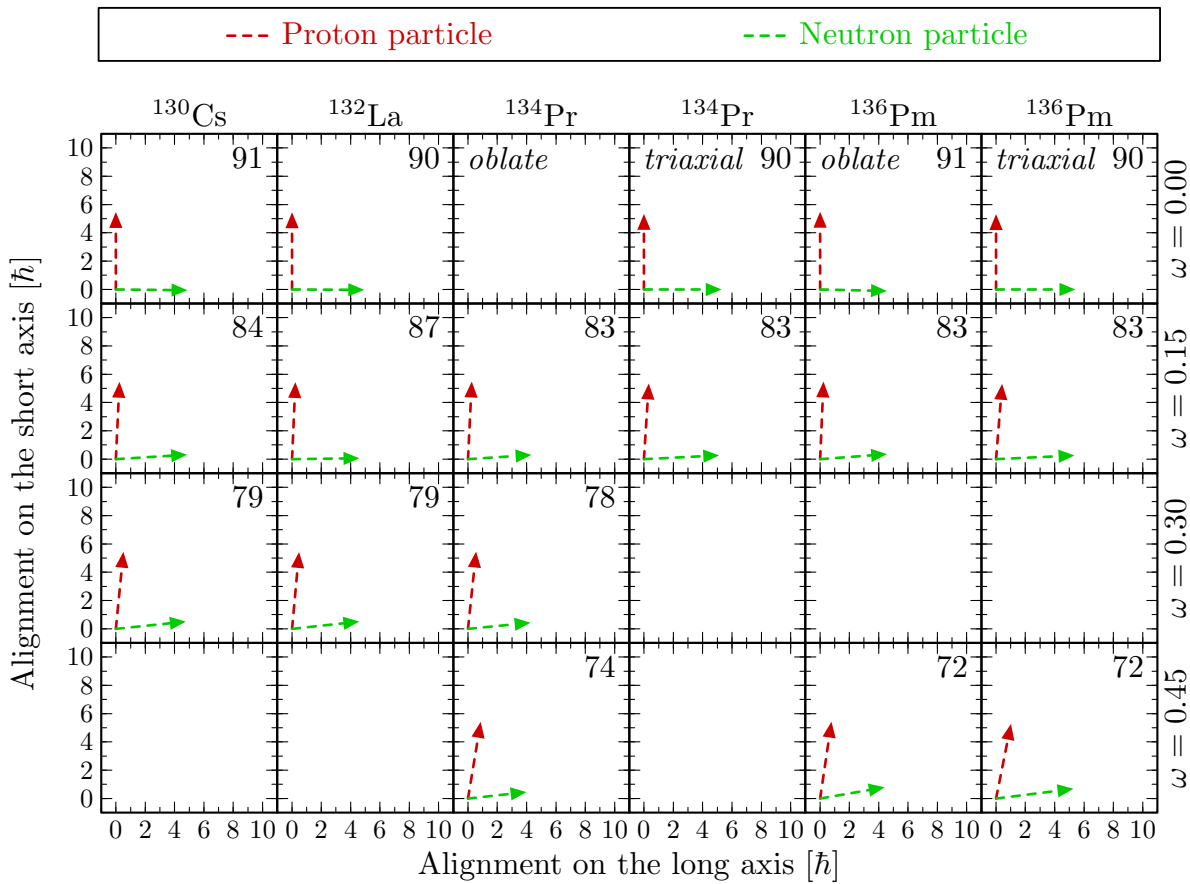}
\end{center}
\caption[Spin vectors of the valence $h_{11/2}$ nucleons along the HF planar bands in the $N=75$ isotones.]{Angular momentum vectors of the valence $h_{11/2}$ proton particle and neutron hole along the HF planar bands found in the $N=75$ isotones. Results with no time-odd fields are shown; the {\it triaxial} solutions in $^{134}$Pr and $^{136}$Pm were calculated with the SkM* force, while the remaining results displayed were obtained with SLy4. The angle between the two spin vectors is given in each panel. Rotational frequency is given in $\mathrm{MeV}/\hbar$. The scale of the plots and the values of $\omega$ are identical to those in Fig.~\ref{shemec_fig}, for the shears band in $^{142}$Gd.}
\label{plaspi_fig}
\end{figure}

Figure~\ref{plaspi_fig} shows the angular momenta of the valence $h_{11/2}$ proton particle and neutron hole for the planar bands in the studied $N=75$ isotones. Results with no time-odd fields are shown, as an example, because they were found in the largest frequency range, but inclusion of the $G$ or $T$ fields very little alters the alignments, like it was already demonstrated for the shears band in $^{142}$Gd; see Fig.~\ref{shemec_fig}. One can see that, indeed, the individual spins of the valence nucleons are rather tightly aligned along the short and long axis, as it was expected from the PAC calculations in Section~\ref{pac_sec}. Interestingly, there is almost no difference between the triaxial and oblate minima in $^{134}$Pr and $^{136}$Pm as far as this point is concerned. Neither do the four $N=75$ isotones differ much in this regard. Note, however, that the alignments in the $N=75$ isotones are visibly stiffer than for the shears band in $^{142}$Gd (see, e.g., the angles between the blades given in Figs.~\ref{plaspi_fig} and \ref{shemec_fig}). This is because the deformation in the considered $N=75$ is larger than that in $^{142}$Gd.

Contrary to the shears band in $^{142}$Gd (see Chapter~\ref{magnet_cha}), the deformation is almost constant along the planar bands in the $N=75$ isotones. The absolute deformation $\beta$ drops by not more than $0.02$ in the frequency range, where the planar solutions were found, and the triaxiality varies by about $3^\circ$, at most. This is presumably connected to the fact that there are only two valence nucleons, compared to four in $^{142}$Gd, and they are unable to polarize the core so much.

\begin{figure}
\begin{center}
\includegraphics{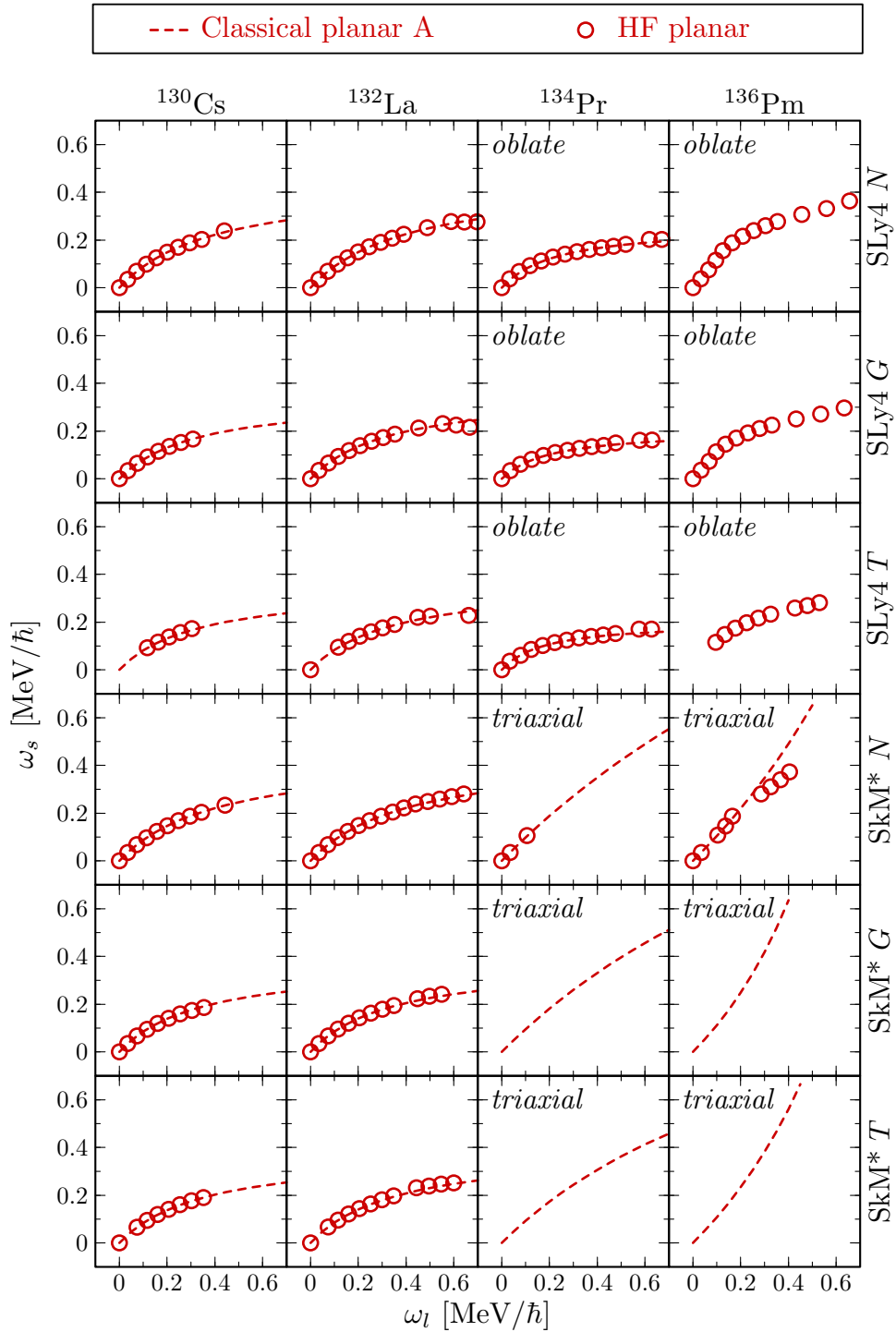}
\end{center}
\caption[HF and classical trajectory of $\vec{\omega}$ along the planar bands in the $N=75$ isotones.]{Intrinsic-frame trajectory of the angular frequency vector along the HF planar bands in the $N=75$ isotones, compared to the classical predictions. HF results with the SLy4 and SkM* forces are shown for the $N$, $G$, $T$ time-odd fields included.}
\label{plaome_fig}
\end{figure}

The HF planar solutions discussed in this Section start at zero rotational frequency (convergence problems with the $G$ and $T$ fields withstanding), and the individual angular momenta of the valence particle and hole have positive projections on the $\vec{\omega}$ vector. Thus, these solutions have all the characteristics of the planar band $A$ of the classical model; see Section~\ref{clasic_sec}. Indeed, the trajectory of $\vec{\omega}$ along the HF bands almost exactly follows the classical hyperbola, as illustrated in Fig.~\ref{plaome_fig}. Also the calculated energy as function of spin agree very well with the classical prediction, see Figs.~\ref{csener_fig}--\ref{ppener_fig}.

\section{Search for the chiral HF solutions}

Planar HF solutions were easily obtained by applying a small cranking frequency to the non-rotating state (at least with no time-odd fields). For chiral bands, analogous task was more difficult, because these bands start at a finite frequency, which in the present case is not lower than $\approx0.5\,\mathrm{MeV}/\hbar$. Several level crossings may occur between $\omega=0$ and such a high frequency, and it is a priori difficult to identify the required s.p.\ configuration. To follow a given configuration diabatically, it is desirable to find a continuous path linking the chiral solution to the non-rotating state. A hint on how to proceed comes from the classical model prediction that the chiral band sprouts from the planar solution (at the point corresponding to the critical frequency). One can thus restart iterations from the converged planar band, rather than from the non-rotating state. Such a way of proceeding was adopted in this work, to find the solutions described in the next Section. However, the author also found chiral solutions by performing a perturbative search along the planar band. That method gives some understanding of why chiral solutions do not appear in some cases, and therefore merits a note in this Section.

Specifically, to each converged point of the planar band, a small additional component, $\omega_m$, of the angular frequency along the medium axis was added. The resulting Routhian was diagonalized only once, and the same parity s.p.\ configuration was required as for the planar state. Then, it was checked whether in the resulting (non-selfconsistent) state the angular momentum and rotational frequency vectors were parallel or not. The Kerman-Onishi theorem \cite{Ker81a} requires that their parallelism is a necessary condition for self-consistency; see Section~\ref{tac_sec}. Presuming that in nuclei that are stiff against deformation changes, the direction of $\vec{J}$ is the only degree of freedom, the Kerman-Onishi condition is also sufficient. In other words, if $\vec{J}$ is parallel to $\vec{\omega}$ in the considered state after one diagonalization, then it is very probable that further iterations may lead to a converged chiral solution. Indeed, in practice it was always the case, and never a chiral solution was obtained, in spite of several attempts, if that simple test gave negative result.

\begin{figure}
\begin{center}
\includegraphics{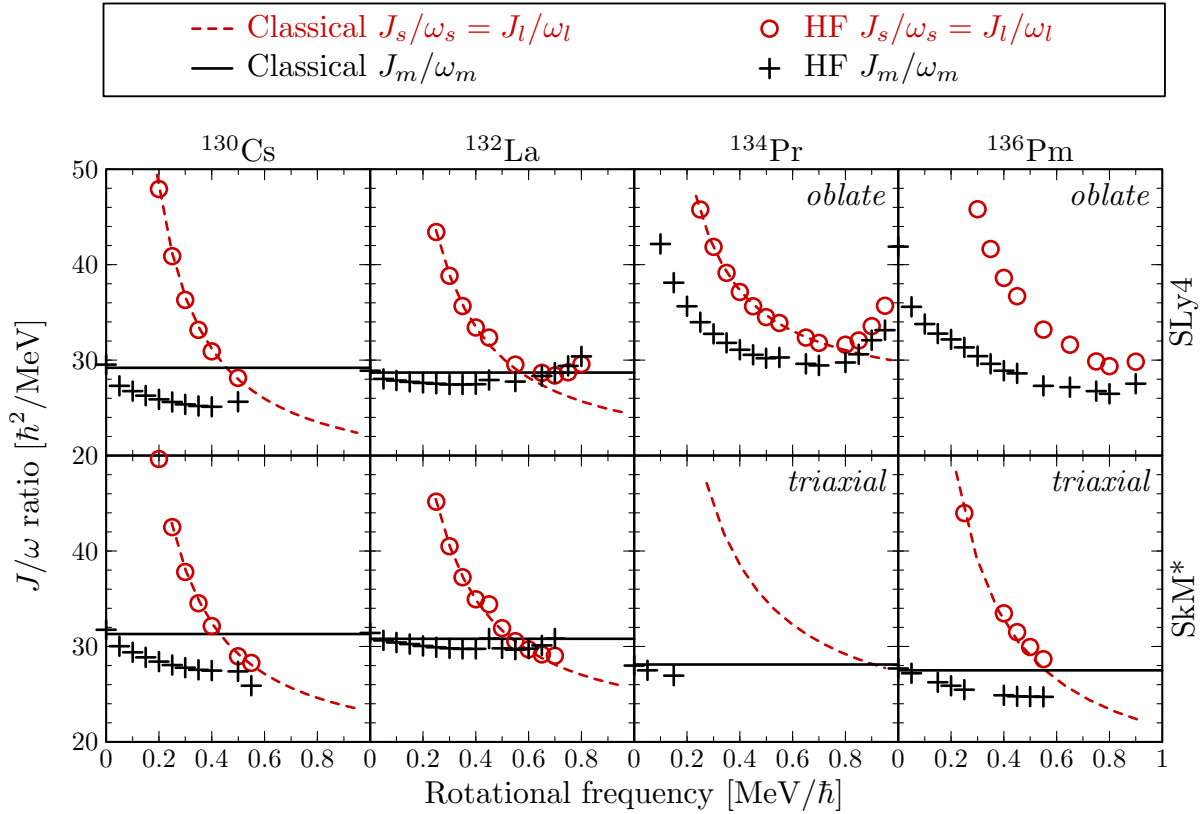}
\end{center}
\caption[Perturbative search for HF chiral solutions.]{The HF and classical-model values of the $J_m/\omega_m$ and $J_s/\omega_s\approx J_l/\omega_l$ ratios for the perturbative search for the HF chiral solutions along the planar bands in the $N=75$ isotones (see text). HF results for the SLy4 and SkM* forces with no time-odd fields are shown.}
\label{search_fig}
\end{figure}

The condition for $\vec{J}$ and $\vec{\omega}$ being parallel can be written in the form
\begin{equation}
\label{IsImIl_eqn}
\frac{J_s}{\omega_s}=\frac{J_m}{\omega_m}=\frac{J_l}{\omega_l}~.
\end{equation}
Note that the $J_s/\omega_s$ and $J_l/\omega_l$ ratios must be very close to each other in the non-selfconsistent state, because the Kerman-Onishi condition is fulfilled for the self-consistent planar state. Therefore, the test consists in checking for each point of the planar band if $J_m/\omega_m$ is equal to $J_s/\omega_s\approx J_l/\omega_l$. In fact, this is only reliable if the time-odd fields are switched off, because in their presence, the $J_m/\omega_m$ ratio calculated perturbatively is significantly smaller than the self-consistent result would be\footnote{This can be seen, e.g., from the comparison of non-selfconsistent and self-consistent PAC results. The reason is that the relevant components of the time-odd fields only become active after self-consistency is achieved for a non-zero cranking frequency.}. The test was made with $\omega_m=0.05\,\mathrm{MeV}/\hbar$, and it has been verified that changing this value up to $\omega_m\approx0.10\,\mathrm{MeV}/\hbar$ does not influence the results. The discussed ratios, calculated for all the HF planar bands found in the $N=75$ isotones, are plotted in Fig.~\ref{search_fig}, together with the classical-model predictions.

It can be a priori expected that chiral solutions do not appear in the oblate minima in $^{134}$Pr and $^{136}$Pm, because of insufficient triaxiality. Indeed, the calculated values of the $J_m/\omega_m$ and $J_s/\omega_s\approx J_l/\omega_l$ ratios exhibit a complicated behavior, and do not become equal one to another at any point. In $^{130}$Cs, as well as in the triaxial minimum in $^{136}$Pm, the two ratios clearly approach each other. It seems that the only reason why they do not attain equality is that the planar bands were not found up to sufficiently high frequencies, because of level crossings. Note, however, that the moment of inertia associated with the medium axis, $J_m/\omega_m$, significantly drops with angular frequency, which takes $J_m/\omega_m$ away from $J_s/\omega_s\approx J_l/\omega_l$, and defers their equalization to higher frequencies. This does not occur in $^{132}$La, where the ratios do become equal, and this happens near the point expected from the classical model. As described in the next Section, self-consistent chiral solutions were indeed found in this case.

\section{Self-consistent chiral Skyrme-HF solutions}
\label{chiban_sec}

Apart from the perturbative test described in the previous Section, self-consistent chiral solutions were sought in the following way. To the mean field of each converged point of the planar band, a cranking term was added with the rotational frequency of the same magnitude as for the planar solution, but with equal components on all the three intrinsic axes. Of course, other choices for $\vec{\omega}$ are possible; the important point is to break the chiral symmetry at the starting point by applying a non-zero $\omega_m$, to render the calculations symmetry-unrestricted; see Section~\ref{tac_sec}. From the Routhian constructed in such a way, regular HF iterations were restarted with the same parity s.p.\ configuration as for the planar state.

In $^{130}$Cs, $^{134}$Pr, and $^{136}$Pm, as well as for low rotational frequencies in $^{132}$La, the iterations converged to planar solutions that differed from the planar starting points uniquely by orientation in space (due to different direction of the applied $\vec{\omega}$). This confirms in the frame of the Skyrme-HF method that, for the concerned configuration, the chiral symmetry is not dynamically broken at low frequencies, as predicted by the classical model. In $^{132}$La, for frequencies high enough, converged solutions were obtained with the angular momentum having non-zero components on all the three intrinsic axes, and thus violating the chiral symmetry, $\hat{R}^T$. These are the first fully self-consistent results that confirm the spontaneous breaking of this symmetry in rotating atomic nuclei.

\begin{figure}[h]
\begin{center}
\includegraphics{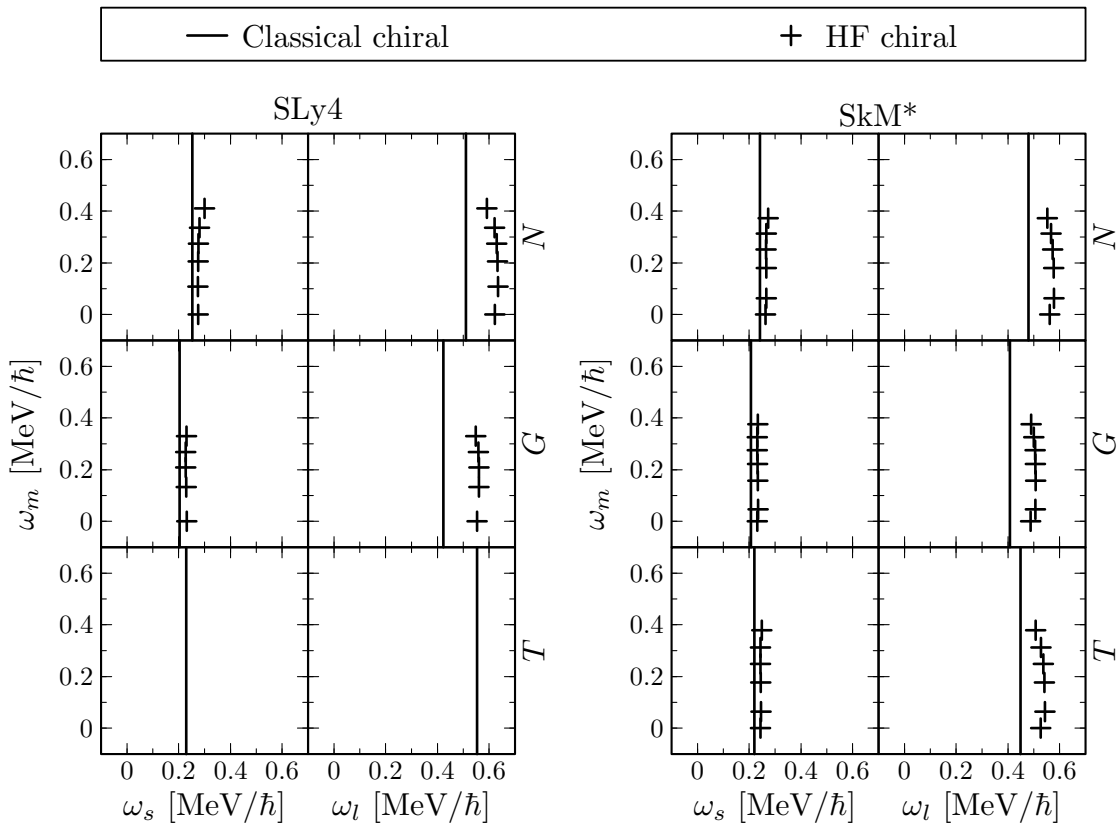}
\end{center}
\caption[Trajectory of $\vec{\omega}$ along the HF chiral bands in $^{132}$La, compared to the classical predictions.]{Trajectory of the angular frequency vector along the HF chiral bands in $^{132}$La, compared to the classical predictions. Projections onto the $s$-$m$ and $l$-$m$ planes of the intrinsic frame are shown. The Figure gives the HF results with the SLy4 and SkM* forces for the $N$, $G$, $T$ time-odd fields included.}
\label{chiome_fig}
\end{figure}

The found fragments of chiral bands were then used as starting points to obtain chiral solutions for lower and higher angular frequencies. Calculations were performed with a frequency step of $0.02\,\mathrm{MeV}/\hbar$. At a certain value of decreasing $\omega$, the planar orientation of the angular momentum was regained in the intrinsic frame, and the solution merged into the previously obtained planar band\footnote{With the exception of the SLy4 case with the $T$ time-odd fields. Although a chiral solution has been found for this force, it could not be connected to the planar band, and it is not sure whether it is built on the $\pi h_{11/2}^1~\nu h_{11/2}^{-1}$ configuration. In the following, results for this band are not discussed.}. This confirms the classical prediction that the planar and chiral bands have a common point. In a natural way, that junction value of $\omega$ can be regarded as the Skyrme-HF result for the critical frequency, and is denoted in the following as $\omega_{crit}^{HF}$. From Figs.~\ref{la_00c_fig} to \ref{la_0Tc_fig}, it can be seen that the chiral and planar s.p.\ Routhians do indeed coincide at $\omega=\omega_{crit}^{HF}$. Values of $\omega_{crit}^{HF}$ are collected in Table~\ref{crifre_tab} and are discussed in the next Section. For highest frequencies, chiral solutions were obtained up to a certain value of $\omega$, and all attempts to go higher caused iterations to fall into another, although also chiral, minimum. This is probably due to multiple smooth crossings of occupied and empty levels, particularly in neutrons; see the s.p.\ Routhians in Figs.~\ref{la_00c_fig} to \ref{la_0Tc_fig}. That other minimum is not build on the $\pi h_{11/2}^1~\nu h_{11/2}^{-1}$ configuration, and will not be subject of further studies here. Although the chiral solutions have been found in a rather narrow $\omega$ interval, of about $0.1\,\mathrm{MeV}/\hbar$, the accompanying increase in $\omega_m$ is significant, from zero to about $0.4\,\mathrm{MeV}/\hbar$; see Fig.~\ref{chiome_fig}.

Similarly to the planar solutions described in the previous Section, there is almost no deformation change along the HF chiral bands in $^{132}$La. This yields good prospects for the validity of the classical descriptions of those bands; see Section~\ref{clasic_sec}. Figure~\ref{chiome_fig} gives the intrinsic-frame trajectories of the angular frequency vector along the HF chiral solutions. In each case, the HF trajectory is almost a straight line, parallel to the medium axis, as predicted by the classical model. The only difference is that $\omega_{crit}^{HF}$ is a bit higher than $\omega_{crit}^{clas}$, and the line is shifted along the planar band to higher frequencies.

\begin{figure}
\begin{center}
\includegraphics{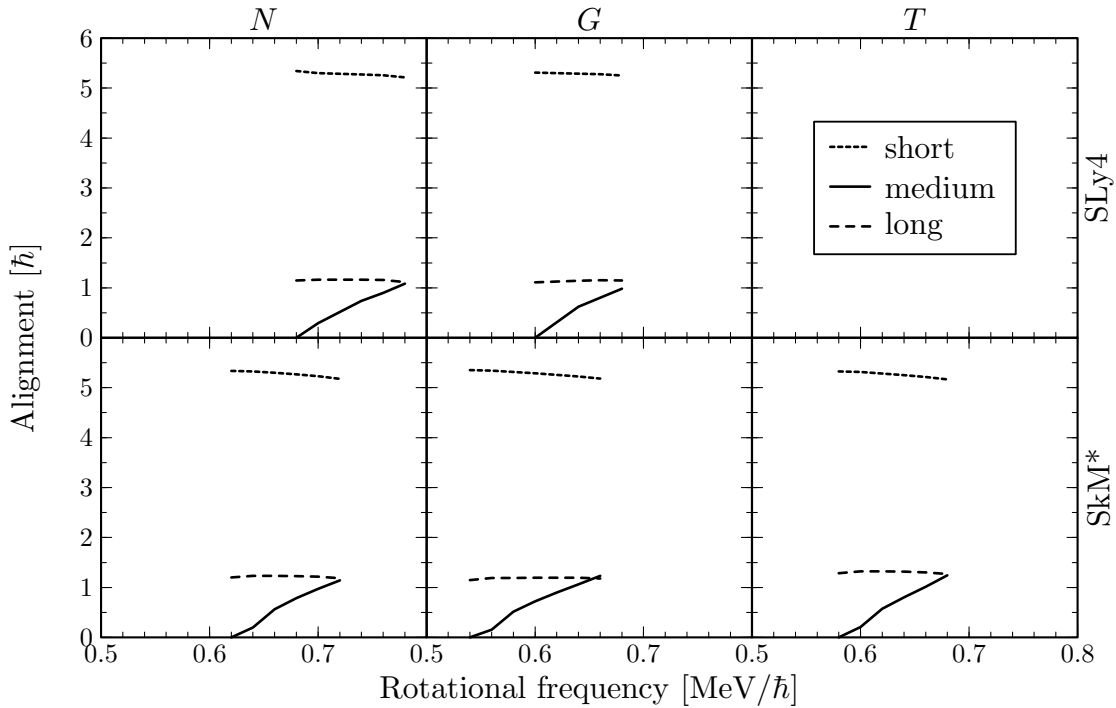}
\end{center}
\caption[Alignments of the $h_{11/2}$ proton on the principal axes in the HF chiral solution in $^{132}$La.]{Alignments of the valence $h_{11/2}$ proton particle on the short, medium and long intrinsic axis along the HF chiral band in $^{132}$La. HF results with the SLy4 and SkM* forces are shown for the $N$, $G$, $T$ time-odd fields included.}
\label{chiali_fig}
\end{figure}

In the vicinity of $\omega_{crit}^{HF}$, the $h_{11/2}$ Routhian associated with the valence neutron hole enters a region of high density of negative-parity levels interacting among themselves. Therefore, it is doubtful whether the considered hole can still be identified with a single s.p.\ state. However, the $h_{11/2}$ proton particle is still well separated. Figure~\ref{chiali_fig} gives its spin alignments on the three intrinsic axes along the chiral band. The plot confirms the stiff character of those alignments. Like in the case of $\vec{\omega}$, only the component on the medium axis changes, the other two remaining nearly constant.

\section{Comparison with experiment}
\label{chiexp_sec}

\begin{figure}
\begin{center}
\includegraphics{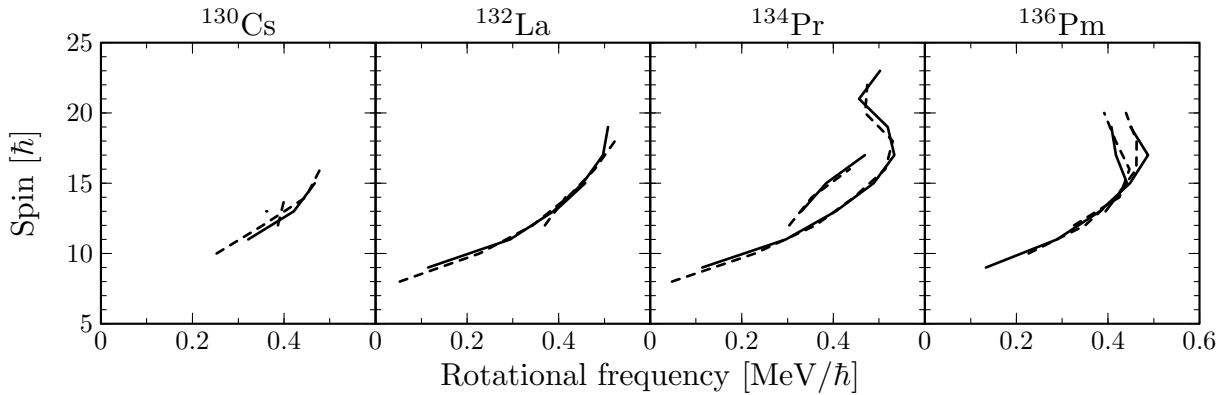}
\end{center}
\caption[Experimental $I(\omega)$ dependence for the chiral bands in $N=75$ isotones.]{Experimental values of spin versus rotational frequency calculated as $\omega(I)=[E(I+1)-E(I-1)]/2$ for the candidate chiral bands in the four $N=75$ isotones. Solid and dashed lines connect the points corresponding to odd and even spins, respectively.}
\label{differ_fig}
\end{figure}

Unfortunately, mean-field results for the chiral rotation are not directly comparable with experimental data, as those of the PRM. This is because the pure mean-field approach does not take into account the interaction between the left- and right-handed solutions, which are exactly degenerate; see Section~\ref{chiral_sec}. Thus, splitting between the chiral partners cannot be reproduced, nor their exact energies. It is argued in the literature that the mean-field chiral solution can be viewed as a kind of average of the two partner bands, and one can thus compare the mean trends. One can also speculate about the value of the critical frequency. It should correspond to the kink in the $I(\omega)$ dependence, as predicted by the PRM and by the classical model; see Figs.~\ref{prm_fig} and \ref{JwEI_fig}, respectively. However, there seems to be no distinct kinks in the experimental bands in the $N=75$ isotones; see Fig.~\ref{differ_fig}.

\begin{figure}
\begin{center}
\includegraphics{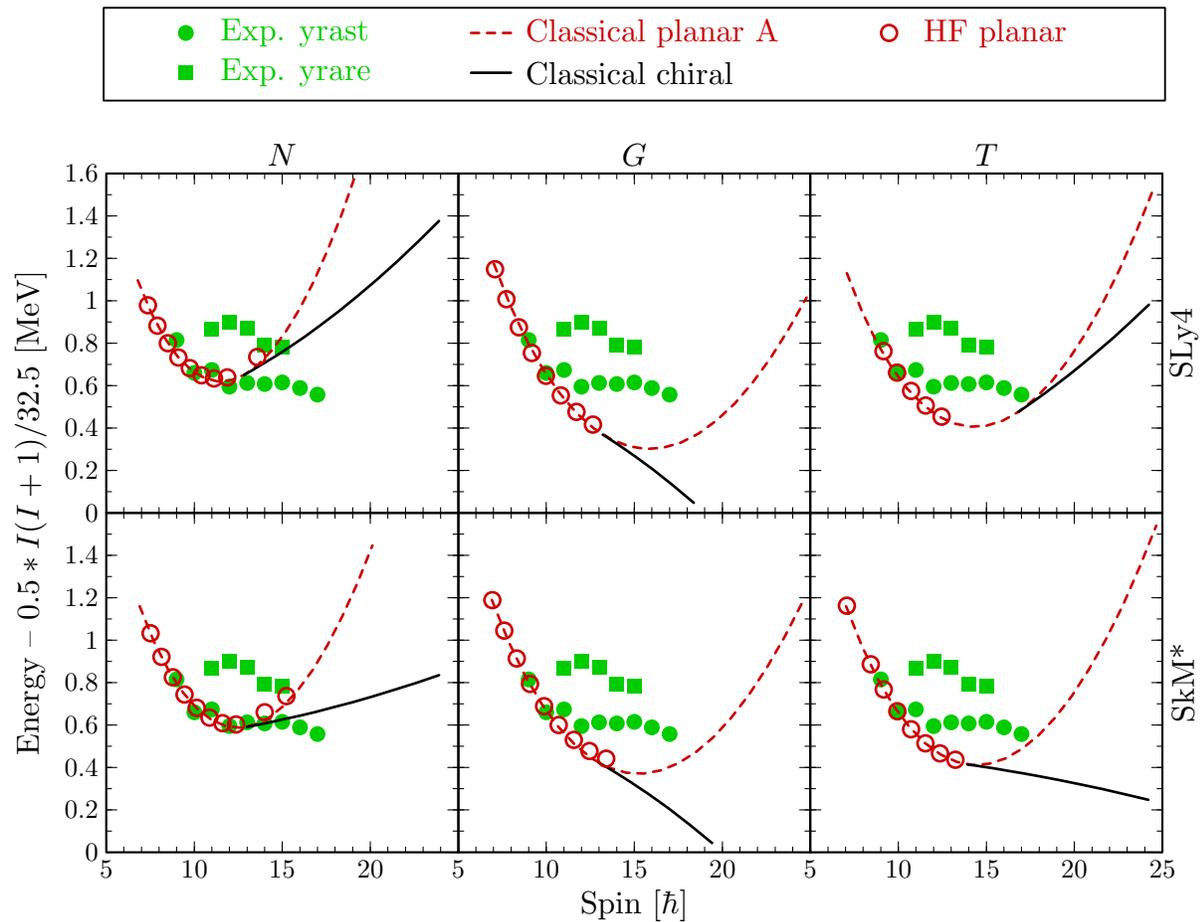}
\end{center}
\caption[HF, classical and experimental energies for the chiral bands in $^{130}$Cs.]{Energies from the HF TAC calculations and from the classical model, compared with the experimental data on the candidate chiral partners in $^{130}$Cs. HF results with the SLy4 and SkM* forces are shown for the $N$, $G$, $T$ time-odd fields included.}
\label{csener_fig}
\end{figure}

\begin{figure}
\begin{center}
\includegraphics{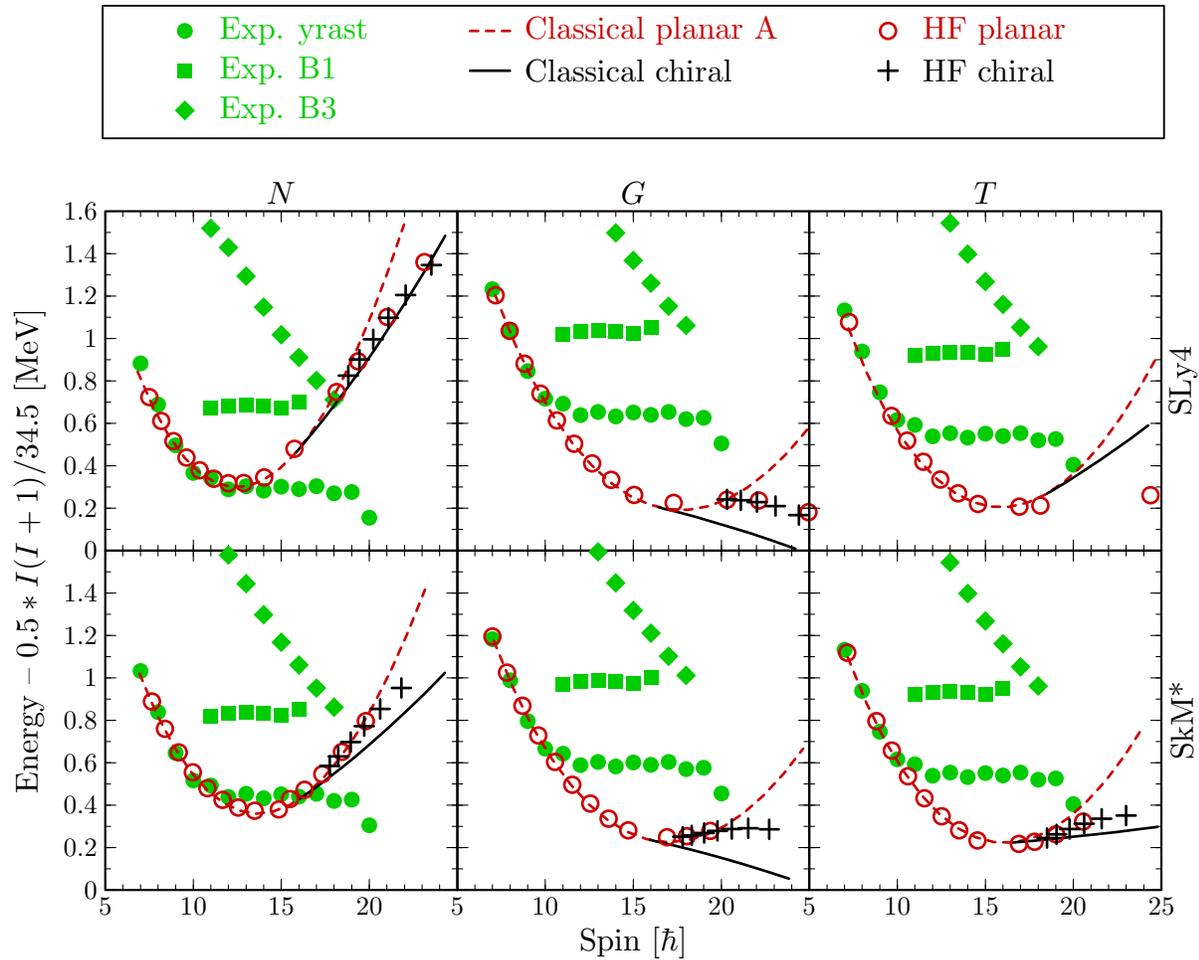}
\end{center}
\caption[HF, classical and experimental energies for the chiral bands in $^{132}$La.]{Energies from the HF TAC calculations and from the classical model, compared with the experimental data in $^{132}$La. The band B1 is the previously known candidate chiral partner \cite{Sta02a}, while B3 is the recently discovered third band \cite{Gro04a}. HF results with the SLy4 and SkM* forces are shown for the $N$, $G$, $T$ time-odd fields included.}
\label{laener_fig}
\end{figure}

\begin{figure}
\begin{center}
\includegraphics{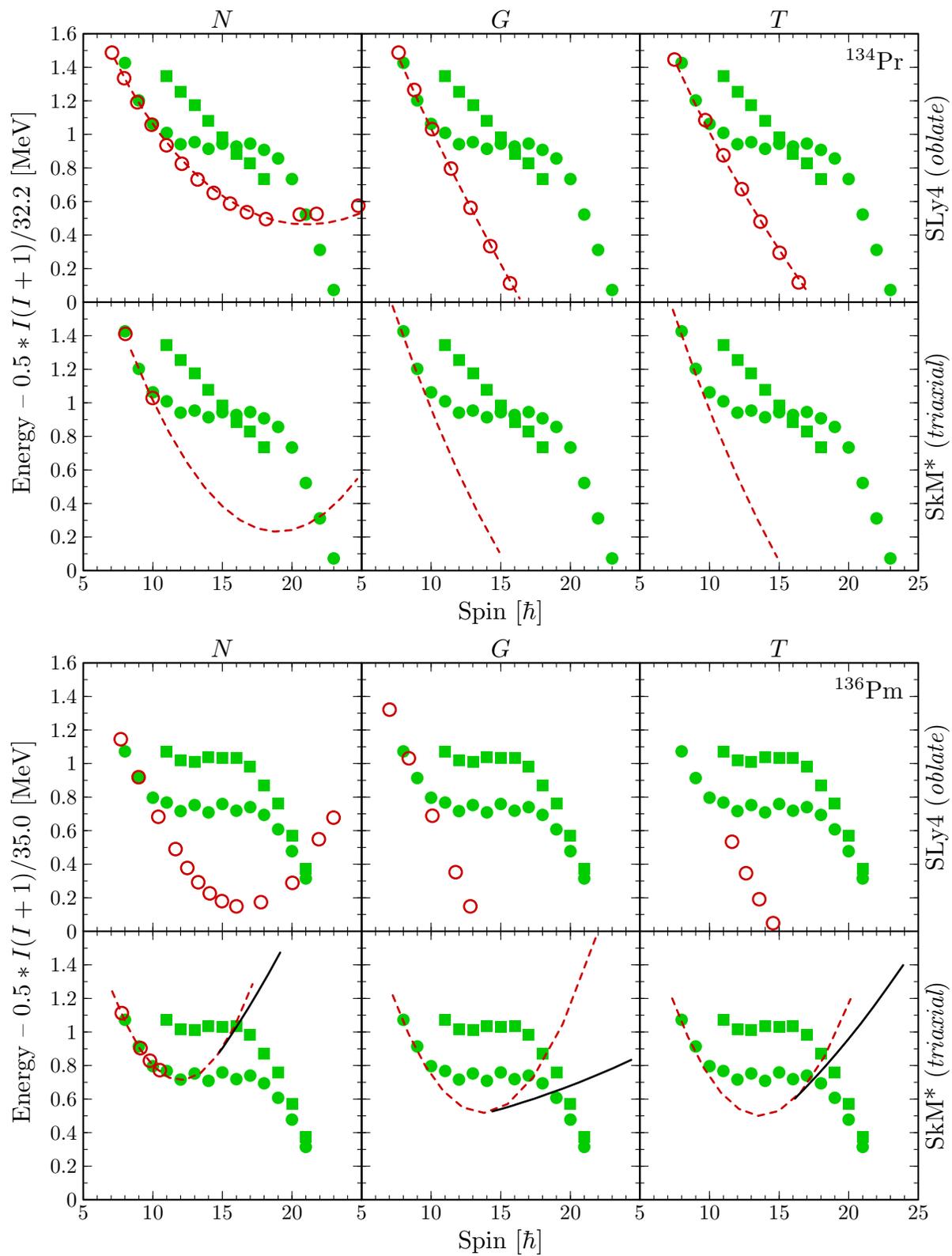}
\end{center}
\caption[HF, classical and experimental energies for the chiral bands in $^{134}$Pr, $^{136}$Pm.]{Same as Fig.~\ref{csener_fig}, but for the bands in $^{134}$Pr and $^{136}$Pm.}
\label{ppener_fig}
\end{figure}

Figures~\ref{csener_fig}--\ref{ppener_fig} give a comparison of calculated and experimental energies. At low spins, where the supposed chiral partners have not been observed, the level energies are reasonably reproduced by the HF planar solutions. It may mean, on the one hand, that the yrast band is not yet perturbed by interactions with the chiral partner, and, on the other hand, that the HF moments of inertia are correct. Roughly at the spin where the chiral partners commence to be visible, the HF planar solutions cease following the yrast bands, which significantly change their behavior. One may argue, therefore, that this is due to the interaction between the left- and right-handed minima.

In $^{132}$La, the HF results for the critical spin, $I_{crit}^{HF}=15.9-18.5\hbar$ (see Table~\ref{crifre_tab} or Fig.~\ref{laener_fig}) are rather high as compared to the spin range, in which the two candidate chiral partners are observed. Yet, the classical estimate,  $I_{crit}^{clas}=9.2\hbar$, evaluated for the TRS PAC results with pairing (see Section~\ref{crifre_sec}) is already below that range. It means that the inclusion of pairing in the calculations may be important for the correct interpretation of the data. But it also suggests that the actual critical spin is somewhere between the HF and TRS results, that is roughly in the middle of the spin range of the chiral partners. The HF-PAC-based values of $I_{crit}^{clas}$ calculated for $^{130}$Cs and $^{136}$Pm also lie at half-spin-range of the partner bands; see Figs.~\ref{csener_fig} and \ref{ppener_fig}. The closeness of the side bands to the critical spin may imply that those structures represent the transition from planar to chiral rotation, as discussed in Section~\ref{chiral_sec}. Although this transition is abrupt in the cranking model, and even in the PRM as indicated by the sharp kink in the $I(\omega)$ dependence in Fig.~\ref{prm_fig}b, it may be rather smooth and complex in real nuclei. This is an interesting topic for study including techniques beyond the mean field.

\chapter{Conclusions and outlook}
\label{conout_cha}

To conclude, a numerical tool has been constructed that allows for Skyrme-Hartree-Fock (Skyrme-HF) Tilted-Axis Cranking (TAC) calculations with no symmetries imposed. Thus, it can be used for investigating the spontaneous breaking of various symmetries in rotating nuclei. As a first application, the phenomena of the magnetic and chiral rotation were studied. For their cranking description, both two-dimensional (planar) and three-dimensional (chiral) TAC solutions were obtained, which is a robust test of the method. The present results provide one of the first fully self-consistent proofs for the possibility of nuclear rotation about an axis tilted with respect to the principal axes. They constitute the first fully self-consistent description of the chiral rotation, and were preceded only by a single Relativistic Mean Field calculation \cite{Mad00a}, as far as the magnetic rotation is concerned.

Before going to the full HF TAC calculations, properties of the $h_{11/2}$ valence nucleons were examined within the PAC frame and on the basis of pure symmetry considerations. It was found that the valence particles and holes align their angular momenta along the short and long axis of the nucleus, respectively. It was argued that their spins are bound to those axes rather tightly, and possibly more tightly in triaxial nuclei than in axially-symmetric ones. By considering a very simple model it was concluded that pairing correlations may soften those alignments.

For the study of the shears bands, the $\pi h_{11/2}^2~\nu h_{11/2}^{-2}$ configuration in $^{142}$Gd was chosen. A substantially triaxial solution was obtained. The HF results corroborate that an important portion of the angular momentum is generated by the shears mechanism, as expected from earlier investigations. It was found that the aligning valence nucleons polarize the core quite strongly, and the deformation decreases with spin. However, the rotating core also generates angular momentum, and actually wins the competition with the shears mechanism. This may be due to overestimated deformation or the lack of pairing in the calculations. It is also possible that triaxiality acts against the shears mechanism by stiffening the alignments of the valence nucleons; such an effect has not been considered so far. The too big collective moment of inertia causes that agreement with experiment is not satisfactory.

The chiral solutions were sought in four $N=75$ isotones, $^{130}$Cs, $^{132}$La, $^{134}$Pr, $^{136}$Pm, with the $\pi h_{11/2}^1~\nu h_{11/2}^{-1}$ configuration. Strongly triaxial solutions were found, but nearly oblate minima proved to be lower in energy for $^{134}$Pr and $^{136}$Pm. One-dimensional cranking about the three axes premised that the alignments of the valence nucleons are very stiff and that the core responds to rotation linearly, like a rigid body. On the basis of these results, a classical model was constructed, in which the chiral rotor is represented by two gyroscopes coupled to a triaxial rigid body. The solutions to that model agree surprisingly well with the full HF TAC results, and elucidate the salient features of the chiral rotation. The HF chiral solutions were found only in $^{132}$La. It was established that the chiral rotation appears only above a certain critical value of the angular frequency; a conclusion that had been known before as a purely numerical result. In the present work, the origin of the critical frequency is explained in an illustrative way within the classical model, that also gives an analytical estimate for that quantity. The HF chiral solutions correctly reproduce the mean $E(I)$ behavior of the observed doublet bands in $^{132}$La. The calculated critical frequency is rather high as compared to the spin range in which the chiral partner bands are observed. PAC calculations with pairing suggested that pairing correlations may reduce its value. It is also possible that the bands in the $N=75$ isotones represent the transition from planar to chiral rotation.

One purpose of the present study was to investigate the influence of the Skyrme time-odd fields on the magnetic and chiral results. The inclusion of those fields changes mainly the moments of inertia, as expected, and causes convergence problems, mainly at low cranking frequencies. It does not alter, however, the qualitative features of the solutions.

Granted, the pairing correlations do influence the shears-rotor competition in the magnetic bands and the critical frequency in the chiral rotation. Their inclusion in the HF TAC calculations in the frame of the Hartree-Fock-Bogolyubov method would be very desirable. The pure mean-field approach is not capable of estimating the splitting between the chiral partner bands, because it does not take into account the interaction between the left- and right-handed solutions. Inclusion of that interaction requires techniques beyond the mean field, preferably the restoration of broken chiral symmetry. To describe the smooth transition from planar to chiral rotation, it should be the projection before variation.

\appendix

\chapter{Invariance of the time-even density}
\label{eveden_app}

This Appendix gives the proof that unitary mixing of states in a Kramers pair does not alter the time-even part of the density matrix, even if only one particle is put on such a pair. This lemma was used in Section~\ref{cond2t_sec}.

The density matrix, $\rho$, of an arbitrary Slater determinant, $|\Psi\rangle$, is defined as
\begin{equation}
\rho_{\alpha\beta}=\langle\Psi|a_\beta^+a_\alpha|\Psi\rangle~.
\end{equation}
It can always be expressed as a sum of contributions from the single-particle (s.p.) states, $|\nu_i\rangle$, of which $|\Psi\rangle$ is constructed,
\begin{equation}
\label{rabsi_eqn}
\rho_{\alpha\beta}=\sum_i\langle 0|a_{\nu_i}a_\beta^+a_\alpha a_{\nu_i}^+|0\rangle~.
\end{equation}
The density $\rho$ can be uniquely decomposed into a time-even part, $\rho^{(+)}$, and a time-odd part, $\rho^{(-)}$,
\begin{equation}
\rho=\rho^{(+)}+\rho^{(-)}~, \qquad
\hat{T}^+\rho^{(+)}\hat{T}=+\rho^{(+)}~, \qquad
\hat{T}^+\rho^{(-)}\hat{T}=-\rho^{(-)}~,
\end{equation}
in the following way:
\begin{equation}
\label{rp12rbr}
\rho^{(+)}=\frac{1}{2}(\rho+\bar\rho)~, \qquad \rho^{(-)}=\frac{1}{2}(\rho-\bar\rho)~,
\end{equation}
where $\bar{\rho}$ is the time-reversed density, i.e. the density of the state $\hat{T}|\Psi\rangle$,
\begin{equation}
\bar\rho=\hat{T}^+\rho\hat{T}~, \qquad \bar\rho_{\alpha\beta}=(\langle\Psi|\hat{T}^+)a_\beta^+a_\alpha(\hat{T}|\Psi\rangle)~.
\end{equation}
Since the time-reversal operator, $\hat{T}$, is anti-unitary and anti-hermitian when acting on s.p.\ states, one can always choose the s.p.\ basis so that it consist of pairs of states, ($|\mu\rangle$,~$|\bar\mu\rangle$), mutually reversed in time up to phase factors $s_\mu$, $s_{\bar\mu}$ (Kramers pairs),
\begin{equation}
\label{kramer_eqn}
\hat{T}|\mu\rangle=s_\mu|\bar\mu\rangle~, \qquad
\hat{T}|\bar\mu\rangle=s_{\bar\mu}|\mu\rangle~, \qquad
s_{\bar\mu}=-s_\mu~.
\end{equation}
In such a basis, the time-reversed density matrix, $\bar\rho$, is related to the original density, $\rho$, by a simple formula,
\begin{equation}
\label{brabsasbrab}
\bar\rho_{\alpha\beta}=s_\alpha s_\beta^*\rho_{\alpha\beta}^*~.
\end{equation}
See, e.g., \cite{Dob87a} for detailed derivations.

If the s.p.\ Hamiltonian, $\hat{h}$, is even under transformation through the time-reversal,
\begin{equation}
\hat{T}^+\hat{h}\hat{T}=+\hat{h}~,
\end{equation}
then its eigenstates can be chosen so that they form a basis like above. The states $|\mu\rangle$ and $|\bar\mu\rangle$ have equal energies (Kramers degeneracy). Suppose now that of such eigenstates one constructs a Slater determinant, $|\Psi\rangle$, and in at least one Kramers pair only one state is occupied, say $|\mu\rangle$. This may happen if the number of particles is odd or if there are particle-hole excitations. According to (\ref{rabsi_eqn}), the contribution to the density matrix, $\rho$, from the particle on $|\mu\rangle$ is
\begin{equation}
\label{Drab_eqn}
\Delta\rho_{\alpha\beta}=\langle0|a_\mu a_\beta^+a_\alpha a_\mu^+|0\rangle=\delta_{\alpha\mu}\delta_{\beta\mu}~,
\end{equation}
where there is no summation over $\mu$, and $\rho$ has been expressed in the basis of the same s.p.\ states, of which $|\Psi\rangle$ is built. By using (\ref{rp12rbr}) and (\ref{brabsasbrab}) one finds that such a single particle contributes both to the time-even and to the time-odd density,
\begin{equation}
\label{DrpDrm_eqn}
\Delta\rho^{(+)}_{\alpha\beta}=\frac{1}{2}(\delta_{\alpha\mu}\delta_{\beta\mu}+\delta_{\alpha\bar\mu}\delta_{\beta\bar\mu})~, \qquad \Delta\rho^{(-)}_{\alpha\beta}=\frac{1}{2}(\delta_{\alpha\mu}\delta_{\beta\mu}-\delta_{\alpha\bar\mu}\delta_{\beta\bar\mu})~.
\end{equation}
Since the states $|\mu\rangle,|\bar\mu\rangle$ are degenerate, all their unitary combinations are equally good eigenstates of $\hat{h}$ and can serve for constructing Slater determinants. The theorem used in Section~\ref{cond2t_sec} states that if the particle is put not on $|\mu\rangle$, but on a state $|\mu'\rangle$,
\begin{equation}
|\mu'\rangle=p|\mu\rangle+q|\bar\mu\rangle~, \qquad p^*p+q^*q=1~,
\end{equation}
then its contribution to the time-even density does not change. Indeed, its new contribution, $\Delta\rho'$, to the total density equals
\begin{eqnarray}
\Delta\rho'_{\alpha\beta} & = & \langle0|(a_\mu p^*+a_{\bar\mu} q^*)a_\beta^+a_\alpha(pa_\mu^++qa_{\bar\mu}^+)|0\rangle \\
                          & = & p^*p\delta_{\alpha\mu}\delta_{\beta\mu}+p^*q\delta_{\alpha\bar\mu}\delta_{\beta\mu}+q^*p\delta_{\alpha\mu}\delta_{\beta\bar\mu}+q^*q\delta_{\alpha\bar\mu}\delta_{\beta\bar\mu}~,
\end{eqnarray}
where the same basis has been used as in (\ref{Drab_eqn}). Again, from (\ref{rp12rbr}) and (\ref{brabsasbrab}) one calculates directly that $\Delta\rho'^{(+)}=\Delta\rho^{(+)}$, and the time-odd term, $\Delta\rho^{(-)}$, changes. Note yet, from (\ref{DrpDrm_eqn}), that if both states are occupied in a Kramers pair, then their contributions to the time-odd density cancel.

\chapter{Alignment and decoupling vectors}
\label{polvec_app}

For a $D_2^T$-symmetric single-particle (s.p.) Hamiltonian, the alignment and decoupling vectors of eigenstates forming a Kramers pair are expressed in terms of three real parameters. Thus, the Equations (\ref{JxMx00_eqn}, \ref{Jy0My0_eqn}, \ref{Jz00Mz_eqn}) of Section~\ref{cond2t_sec} are derived.

Consider a Kramers pair, ($|\mu\rangle$,~$|\bar\mu\rangle$), as defined by (\ref{kramer_eqn}). For the state $|\mu\rangle$ one can define the real {\it alignment vector}, $\vec{J}^\mu$, and the complex {\it decoupling vector}, $\vec{D}^\mu$,
\begin{equation}
\vec{J}^\mu=\langle\mu|\hat{\vec{J}}|\mu\rangle~, \qquad \vec{D}^\mu=\langle\mu|\hat{\vec{J}}|\bar\mu\rangle~.
\end{equation}
Although the decoupling vector changes its phase when $|\mu\rangle,|\bar\mu\rangle$ change their phases, the relative phases of its components do not depend on the phase convention. Since the angular-momentum operator is odd under the time reversal, it can easily be verified that
\begin{equation}
\label{tirejd_eqn}
\vec{J}^{\bar\mu}=\langle\bar\mu|\hat{\vec{J}}|\bar\mu\rangle=-\vec{J}^\mu~, \qquad \vec{D}^{\bar\mu}=\langle\bar\mu|\hat{\vec{J}}|\mu\rangle=\vec{D}^{\mu~*}~.
\end{equation}

Here, we are concerned by properties of a single Kramers pair formed by eigenstates of a s.p.\ Hamiltonian, $\hat{h}$, that is even under the $D_2^T$ group. Therefore, it is possible to choose the states of the pair as eigenstates of either of the three signatures, $\hat{R}_i$, where $i=x,y,z$, but only one at a time, because the signature operators do not commute among themselves. This results, respectively,  in three formally different pairs, ($|\mu_i\rangle$,~$|\bar\mu_i\rangle$), that correspond to just three different bases in the same two-dimensional eigenspace of $\hat{h}$. Here, it is also convened that the states $|\mu_i\rangle$ have eigenvalues of $-i$ under the action of $\hat{R}_i$, while the eigenvalues of $|\bar\mu_i\rangle$ are $+i$.

The fact that $|\mu_i\rangle$ and $|\bar\mu_i\rangle$ are eigenstates of $\hat{R}_i$, together with the transformation rules of the components, $\hat{J}_j$, of the angular momentum operator under the three signatures,
\begin{equation}
\hat{R}_i^+\hat{J}_j\hat{R}_i=\left\{
\begin{array}{ll}
+\hat{J}_j & j=i \\
-\hat{J}_j & j\neq i \\
\end{array}\right.~,
\end{equation}
induces limitations on the components, $J^{\mu_i}_j$ and $D^{\mu_i}_j$, of the alignment and decoupling vectors. Namely,
\begin{equation}
\label{JmiDmi}
J^{\mu_i}_j=\left\{
\begin{array}{ll}
\mbox{non-zero} & j=i \\
0               & j\neq i \\
\end{array}\right.~, \qquad
D^{\mu_i}_j=\left\{
\begin{array}{ll}
0               & j=i \\
\mbox{non-zero} & j\neq i \\
\end{array}\right.~.
\end{equation}
In other words, $\vec{J}^{\mu_i}$ is confined to the axis $i$ and $\vec{D}^{\mu_i}$ to the plane perpendicular to that axis.

In order to establish connections between the alignment and decoupling vectors associated with the three pairs, ($|\mu_i\rangle$,~$|\bar\mu_i\rangle$), one needs to express the states of one pair as linear combinations of those in the remaining two. Below, states $|\mu_x\rangle$, $|\bar\mu_x\rangle$ and $|\mu_y\rangle$, $|\bar\mu_y\rangle$ are expressed as linear combinations of $|\mu_z\rangle$, $|\bar\mu_z\rangle$ by diagonalization of $\hat{R}_x$ and $\hat{R}_y$ in the basis of $|\mu_z\rangle$, $|\bar\mu_z\rangle$, after having found the corresponding two-by-two matrices. Taking into account that $\hat{R}_x$ and $\hat{R}_y$ are odd under $\hat{R}_z$,
\begin{equation}
\hat{R}_z^+\hat{R}_x\hat{R}_z=-\hat{R}_x~, \qquad \hat{R}_z^+\hat{R}_y\hat{R}_z=-\hat{R}_y~,
\end{equation}
one arrives at a conclusion that the diagonal matrix elements of $\hat{R}_x$ and $\hat{R}_y$ in the $|z\rangle$, $|\bar z\rangle$ basis vanish. The off-diagonal ones depend on the phases of $|z\rangle$, $|\bar z\rangle$, but are restricted by the fact that the signatures are unitary anti-hermitian operators. Moreover, the off-diagonal elements of $\hat{R}_x$ and $\hat{R}_y$ are connected by the equality
\begin{equation}
\langle z|\hat{R}_x|\bar z\rangle=i\langle z|\hat{R}_y|\bar z\rangle~,
\end{equation}
which is due to the multiplication rule
\begin{equation}
\hat{R}_x=\hat{R}_y\hat{R}_z~.
\end{equation}
All this means that, with a proper choice of phases in $|\mu_z\rangle$, $|\bar\mu_z\rangle$, one can always make that the corresponding matrices of $\hat{R}_i$ be related to the Pauli matrices\footnote{This assures compatibility of notation with the case of a particle possessing only the intrinsic spin and no space degrees of freedom, where the general definition, $\hat{R}_i=\exp(-i\pi\hat{J}_i)$, leads explicite to $\hat{R}_i=-i\hat{\sigma}_i$. For such a particle, the states of spin $|\uparrow\rangle$ and $|\downarrow\rangle$ have the $z$-signature signature eigenvalues of $-i$ and $+i$, respectively, and play the role of $|\mu_z\rangle$ and $|\bar\mu_z\rangle$.}, $\hat{\sigma}_i$,
\begin{equation}
\hat{R}_i=-i\hat{\sigma}_i~.
\end{equation}
By diagonalization of $\hat{R}_x$ and $\hat{R}_y$ one obtains, with an arbitrary choice of phases:
\begin{equation}
\label{x_eqn}
|\mu_x\rangle=\sqrt{\frac{-i}{2}}(|\mu_z\rangle+|\bar\mu_z\rangle)~, \qquad
|\bar\mu_x\rangle=-\sqrt{\frac{i}{2}}(|\mu_z\rangle-|\bar\mu_z\rangle)~,
\end{equation}
\begin{equation}
\label{y_eqn}
|\mu_y\rangle=\sqrt{\frac{i}{2}}(|\mu_z\rangle+i|\bar\mu_z\rangle)~, \qquad
|\bar\mu_y\rangle=\sqrt{\frac{-i}{2}}(i|\mu_z\rangle+|\bar\mu_z\rangle)~.
\end{equation}
These formulae, together with relations (\ref{tirejd_eqn}), allow to express $\vec{J}^{\mu_x}$, $\vec{D}^{\mu_x}$ and $\vec{J}^{\mu_y}$, $\vec{D}^{\mu_y}$ in terms of $\vec{J}^{\mu_z}$, $\vec{D}^{\mu_z}$. Taking into account the constraining conditions (\ref{JmiDmi}) on the left-hand side, one obtains the following set of equations:
\begin{equation}
(J^{\mu_x}_x,0,0)=(\re D^{\mu_z}_x,\re D^{\mu_z}_y,0)~, \quad (0,D^{\mu_x}_y,D^{\mu_x}_z)=(-\im D^{\mu_z}_x,-\im D^{\mu_z}_y,-iJ^{\mu_z}_z)~,
\end{equation}
\begin{equation}
(0,J^{\mu_y}_y,0)=(-\im D^{\mu_z}_x,-\im D^{\mu_z}_y,0)~, \quad (D^{\mu_y}_x,0,D^{\mu_y}_z)=(-i\re D^{\mu_z}_x,-i\re D^{\mu_z}_y,J^{\mu_z}_z)~.
\end{equation}
A glance at them convinces us that all of the quantities $J^{\mu_i}_j$, $D^{\mu_i}_j$ express through the three "diagonal" components, $J^{\mu_i}_i$. One eventually arrives at:
\begin{equation}
\vec{J}^{\mu_x}=(J^{\mu_x}_x,0,0)~, \qquad \vec{D}^{\mu_x}=(0,J^{\mu_y}_y,-iJ^{\mu_z}_z)~,
\end{equation}
\begin{equation}
\vec{J}^{\mu_y}=(0,J^{\mu_y}_y,0)~, \qquad \vec{D}^{\mu_y}=(-iJ^{\mu_x}_x,0,J^{\mu_z}_z)~,
\end{equation}
\begin{equation}
\vec{J}^{\mu_z}=(0,0,J^{\mu_z}_z)~, \qquad \vec{D}^{\mu_z}=(J^{\mu_x}_x,-iJ^{\mu_y}_y,0)~.
\end{equation}

If the Hamiltonian, $\hat{h}$, has no extra symmetries, then $J^{\mu_i}_i$ can assume arbitrary values. If $\hat{h}$ is axially symmetric, say with respect to the $z$ axis, then the states $|\mu_z\rangle$, $|\bar\mu_z\rangle$ are eigenstates of $\hat{J}_z$, which leads to quantization of $J^{\mu_z}_z$. In fact, $J^{\mu_z}_z=+1/2,-3/2,...$, because $\hat{R}_z=\exp(-i\pi\hat{J}_z)$, while in the adopted convention $\hat{R}_z|\mu_z\rangle=-i|\mu_z\rangle$. For states $|\mu_x\rangle$ and $|\mu_y\rangle$, defined by (\ref{x_eqn}) and (\ref{y_eqn}), one easily finds
\begin{equation}
\label{JxxJyy_eqn}
J^{\mu_x}_x=\frac{1}{2}\mathrm{Re}\langle\mu_z|\hat{J}_++\hat{J}_-|\bar\mu_z\rangle~, \qquad
J^{\mu_y}_y=\frac{1}{2}\mathrm{Re}\langle\mu_z|\hat{J}_+-\hat{J}_-|\bar\mu_z\rangle~,
\end{equation}
where $\hat{J}_+=\hat{J}_x+i\hat{J}_y$ and $\hat{J}_-=\hat{J}_x-i\hat{J}_y$ are the ladder operators, that increment and decrement the magnetic quantum number, $J_z$, of an eigenstate, $|J_z\rangle$, of $\hat{J}_z$,
\begin{equation}
\hat{J}_+|J_z\rangle\sim|J_z+1\rangle~, \qquad
\hat{J}_-|J_z\rangle\sim|J_z-1\rangle~.
\end{equation}
One can see, therefore, that the matrix elements in (\ref{JxxJyy_eqn}) can be non-zero only if $|\mu_z\rangle$ and $|\bar\mu_z\rangle$ differ in $J_z$ by one, that is if $J^{\mu_z}_z=1/2$. In such a case, $\langle\mu_z|\hat{J}_-|\bar\mu_z\rangle=0$, and $J^{\mu_x}_x=J^{\mu_y}_y$. These results can be summarized as
\begin{equation}
(J^{\mu_x}_x,J^{\mu_y}_y,J^{\mu_z}_z)=\left\{
\begin{array}{lll}
(J^{\mu_\perp}_\perp,J^{\mu_\perp}_\perp,J^{\mu_\parallel}_\parallel) & \mathrm{for} & J^{\mu_\parallel}_\parallel=1/2 \\
(0,0,J^{\mu_\parallel}_\parallel)                                     & \mathrm{for} & J^{\mu_\parallel}_\parallel=3/2,~5/2,~... \\
\end{array}\right.
\end{equation}
The parameter $J^{\mu_\perp}_\perp$ is not restricted by the above kinematic conditions, and is usually called {\it decoupling parameter}.

\chapter{Technical details of the calculations}
\label{detals_app}

This Appendix summarizes all technical details of the present Hartree-Fock (HF) calculations, and of the auxiliary calculations performed in Chapters~\ref{magnet_cha} and \ref{chiral_cha} within the Total Routhian Surface (TRS) approach.

All the HF calculations presented in this work were performed with the help of the code \pr{HFODD}, version (v2.05c), which is described in Section~\ref{hfodd_sec}. Twelve spherical shells of the Harmonic Oscillator were taken as the single-particle (s.p.) basis. It has been verified for the one-dimensional cranking in $^{132}$La that varying the number of shells from 10 up to 16 changes the quantities important for the present study, like deformation, alignments, moments of inertia, by less than 1\%. The calculations were performed with two Skyrme parameter sets, SLy4 \cite{Cha97a} and SkM* \cite{Bar82a}. Attempts were undertaken to obtain each solution with three sets of the HF time-odd fields \cite{Dob95a,Dob97a} included, as listed in Table~\ref{timodd_tab} and explained in Chapter~\ref{timodd_sec}. Those sets are denoted in the text as $N$, $G$, $T$. Since the treatment of the pairing correlations is not yet implemented in \pr{HFODD} in the Tilted-Axis Cranking (TAC) mode, pairing was not included. In all Principal-Axis Cranking (PAC) calculations, the $\hat{P}$, $\hat{R}_y$, $\hat{S}^T_x$, $\hat{S}^T_z$ symmetries (and their products) were imposed, while only $\hat{P}$ was kept in the TAC case. The respective parity-signature and parity configurations (see Section~\ref{hfodd_sec}) for all the considered solutions are given in Tabs.~\ref{shecon_tab} and \ref{chicon_tab}. The configurations were followed diabatically in each band, either by choosing the occupied state by hand, whenever a crossing was about to occur, or by using the diabatic-blocking techniques described in Section~\ref{hfodd_sec}. In all planar solutions, the direction of $\vec{\omega}$ was chosen so that it had equal components on two axes of the program frame (see Section~\ref{tac_sec}) and zero component on the third axis. In the chiral bands, $\vec{\omega}$ had equal projections on all the three program axes. In each case, the direction of $\vec{\omega}$ was kept fixed, and only its magnitude was changed. The axial and planar bands were calculated with an $\omega$ step of $0.05\,\mathrm{MeV}/\hbar$, while the step of $0.02\,\mathrm{MeV}/\hbar$ was used for the chiral solutions. With no time-odd fields included, the iterations for each value of $\omega$ were restarted from the previously converged solution at the prior value of $\omega$. Then, each point converged with no time-odd fields served as a starting point for calculations with the time-odd fields included. As explained in Section~\ref{hfodd_sec}, the convergence rate is sometimes very low in TAC calculations. To deal with this difficulty, all solutions were first obtained at the level of convergence defined by the \pr{HFODD} flag EPSITE=0.000001; see the \pr{HFODD} manual in \cite{Dob00a}. Afterwards, additional iterations were repeated in portions until the functional energy ceased changing from one iteration to another by more than $1\,\mathrm{eV}$. This criterion could be fulfilled after hundreds of iterations at higher frequencies, $\omega>0.4\,\mathrm{MeV}/\hbar$, and after thousands of iterations for lower frequencies, $\omega<0.4\,\mathrm{MeV}/\hbar$. This process could have been much faster, had the technique of resetting $\vec{\omega}$ parallel to $\vec{J}$ been known earlier; see Section~\ref{hfodd_sec}. The constraint on $\vec{\omega}\times\vec{J}$ was not used, because it turned out to be very ineffective; see Section~\ref{hfodd_sec}.

To estimate the possible influence of the pairing correlations, auxiliary calculations within the TRS method were performed in the present work. In Chapter~\ref{magnet_cha}, the energy surface for the ground state in $^{142}$Gd was examined, and one-dimensional cranking for $^{132}$La was considered in Chapter~\ref{chiral_cha}. The calculations were done by using the TRS software developed by Satu\l{}a and Wyss \cite{Sat94a}. The method employs the phenomenological Woods-Saxon potential and a correction for the liquid-drop energy. Self-consistency is assured in the pairing channel, and the strength of the monopole pairing is adjusted to experimental data following the average gap method. The quadrupole pairing is also included, with the coupling constant chosen so that the pairing field possess the local galilean invariance. Approximate projection onto good particle number is implemented within the Lipkin-Nogami method. The code does not allow for TAC calculations. See \cite{Sat94a} for all details and further references.

\chapter{Single-particle Routhians}
\label{routhi_app}

In this Appendix, all Figures showing the Hartree-Fock (HF) single-particle (s.p.) Routhians are collected.

The Routhians marked as Axial in the Figures' captions come from one-dimensional cranking, with imposed parity and signature. The Routhians denoted Planar and Chiral were obtained from Tilted-Axis Cranking (TAC) calculations, with only the parity imposed. In both kinds of Routhians, solid lines are used for positive-parity levels, and dashed lines for negative-parity levels. In the case of Principal-Axis Cranking (PAC), levels of positive signature ($+i$) are plotted in black, and levels of negative signature ($-i$) in gray. Each s.p.\ level is labeled with its total ordinal number and its ordinal number in the given symmetry block. For the PAC or TAC Routhians, the second number refers to the parity-signature or parity blocks, respectively. For example, the state marked as 62 14--+ is the 62nd state among all and the 14th of negative parity and positive signature. The last item in each label indicates whether the given state is occupied (1) or not (0). This information is not provided for states far from the Fermi energy. For clarity, the Routhians bearing the $h_{11/2}$ nucleons are marked with points, put at the values of the cranking frequency, at which the HF solutions were obtained. The proton particles and the neutron holes are marked with full and open circles, respectively. The symbols $N$, $G$, $T$, used in the captions, refer to calculations with various Skyrme time-odd fields included; see Section~\ref{timodd_sec}. The labels {\it oblate} and {\it triaxial}, used for $^{134}$Pr and $^{136}$Pm, denote the nearly oblate and triaxial minima found in those nuclei, see Section~\ref{chimin_sec}. Figures~\ref{la_00c_fig}--\ref{la_0Tc_fig} show the Routhians for the planar and chiral bands in $^{132}$La in the same plots. A vertical line is drawn at the value of the HF critical frequency, $\omega_{crit}^{HF}$; see Section~\ref{chiban_sec}. The Routhians left to that line correspond to the planar band, and the Routhians to the right correspond to the chiral band.

\newpage

\begin{figure}
\begin{center}
\includegraphics{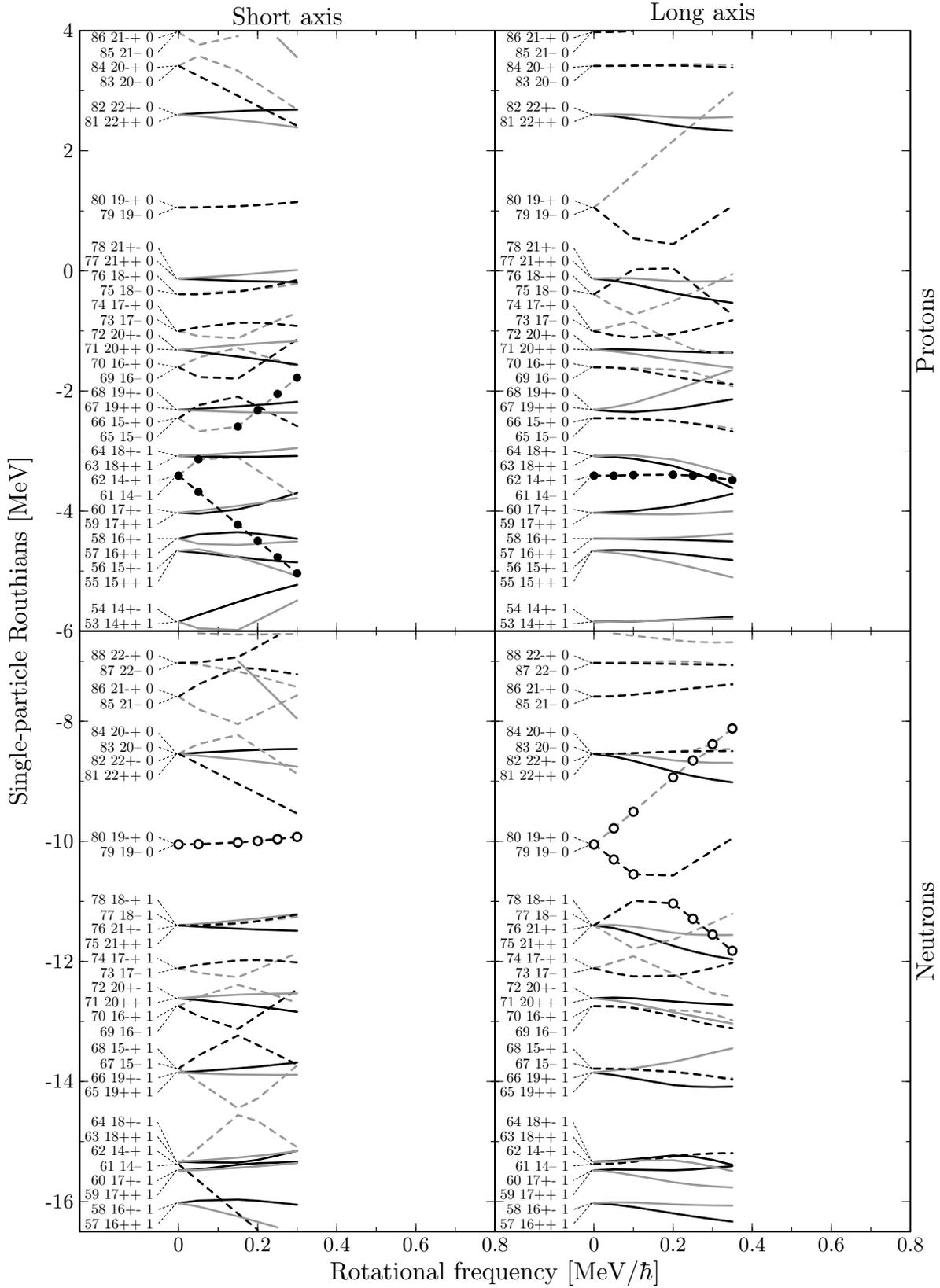}
\end{center}
\caption{Axial Routhians in $^{142}$Gd for SLy4 force with $N$ time-odd fields.}
\label{gd_a0a_fig}
\end{figure}

\begin{figure}
\begin{center}
\includegraphics{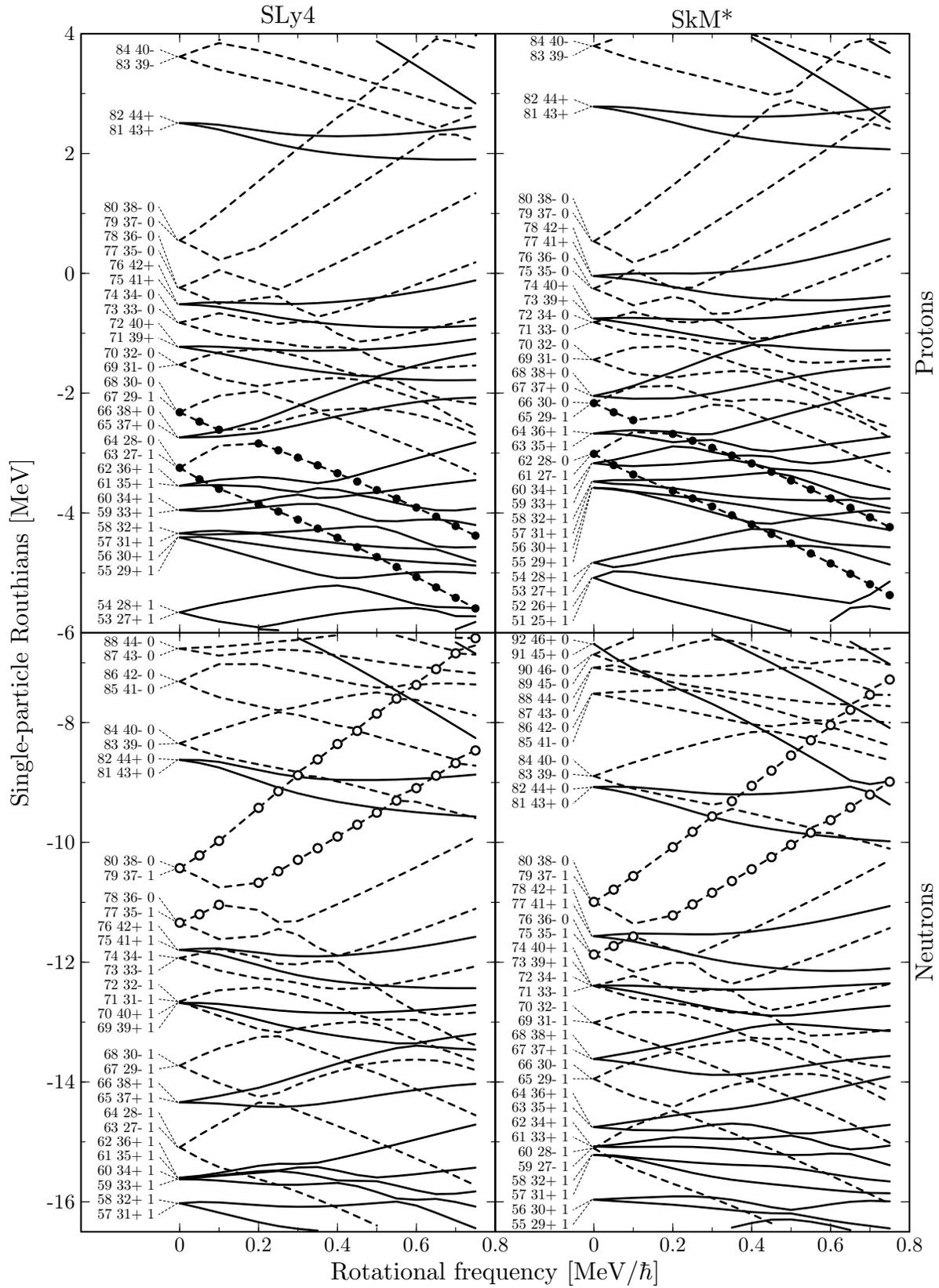}
\end{center}
\caption{Planar Routhians in $^{142}$Gd with $N$ time-odd fields.}
\label{gd_00p_fig}
\end{figure}

\begin{figure}
\begin{center}
\includegraphics{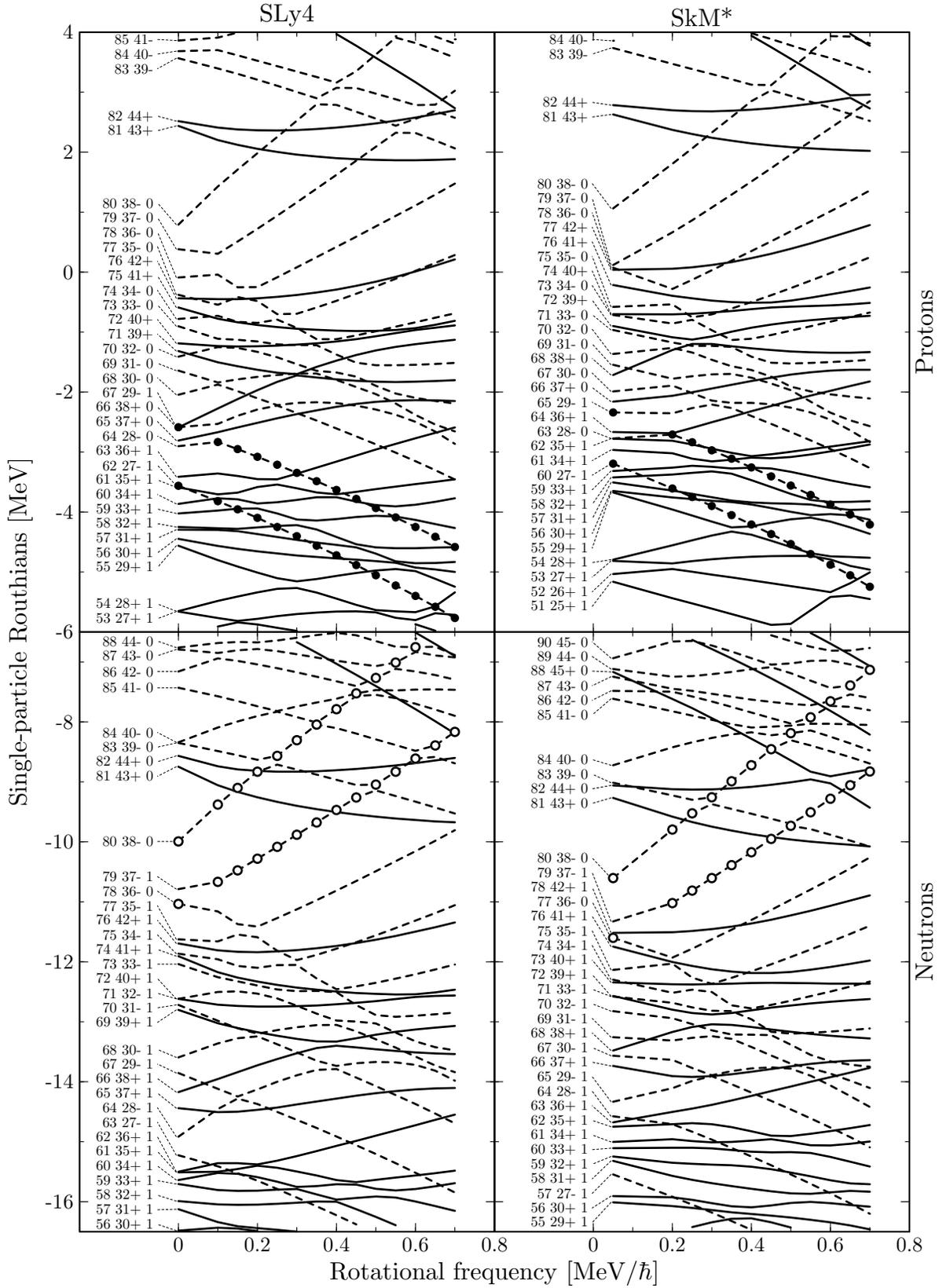}
\end{center}
\caption{Planar Routhians in $^{142}$Gd with $G$ time-odd fields.}
\label{gd_0Gp_fig}
\end{figure}

\begin{figure}
\begin{center}
\includegraphics{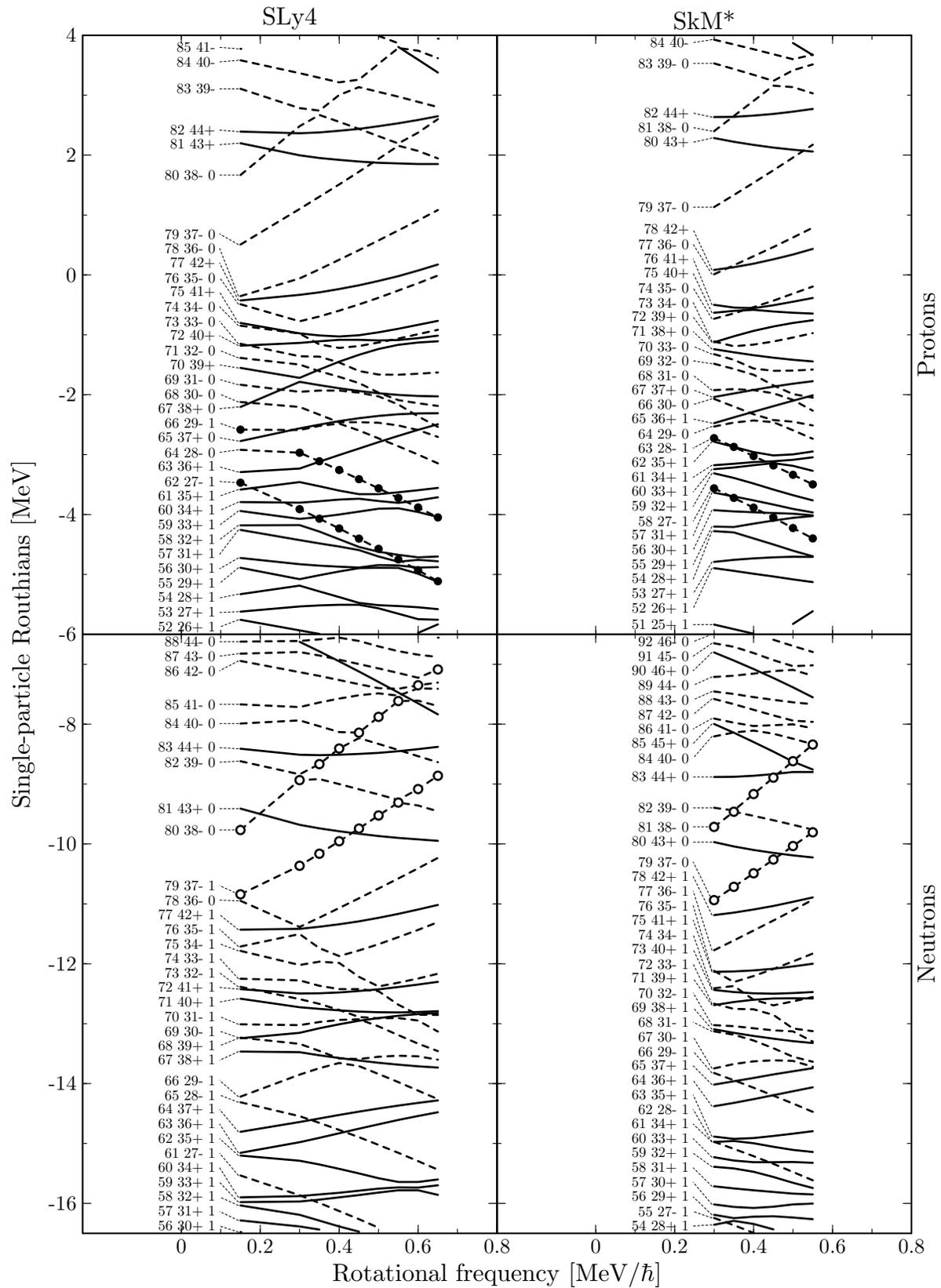}
\end{center}
\caption{Planar Routhians in $^{142}$Gd with $T$ time-odd fields.}
\label{gd_0Tp_fig}
\end{figure}

\begin{figure}
\begin{center}
\includegraphics{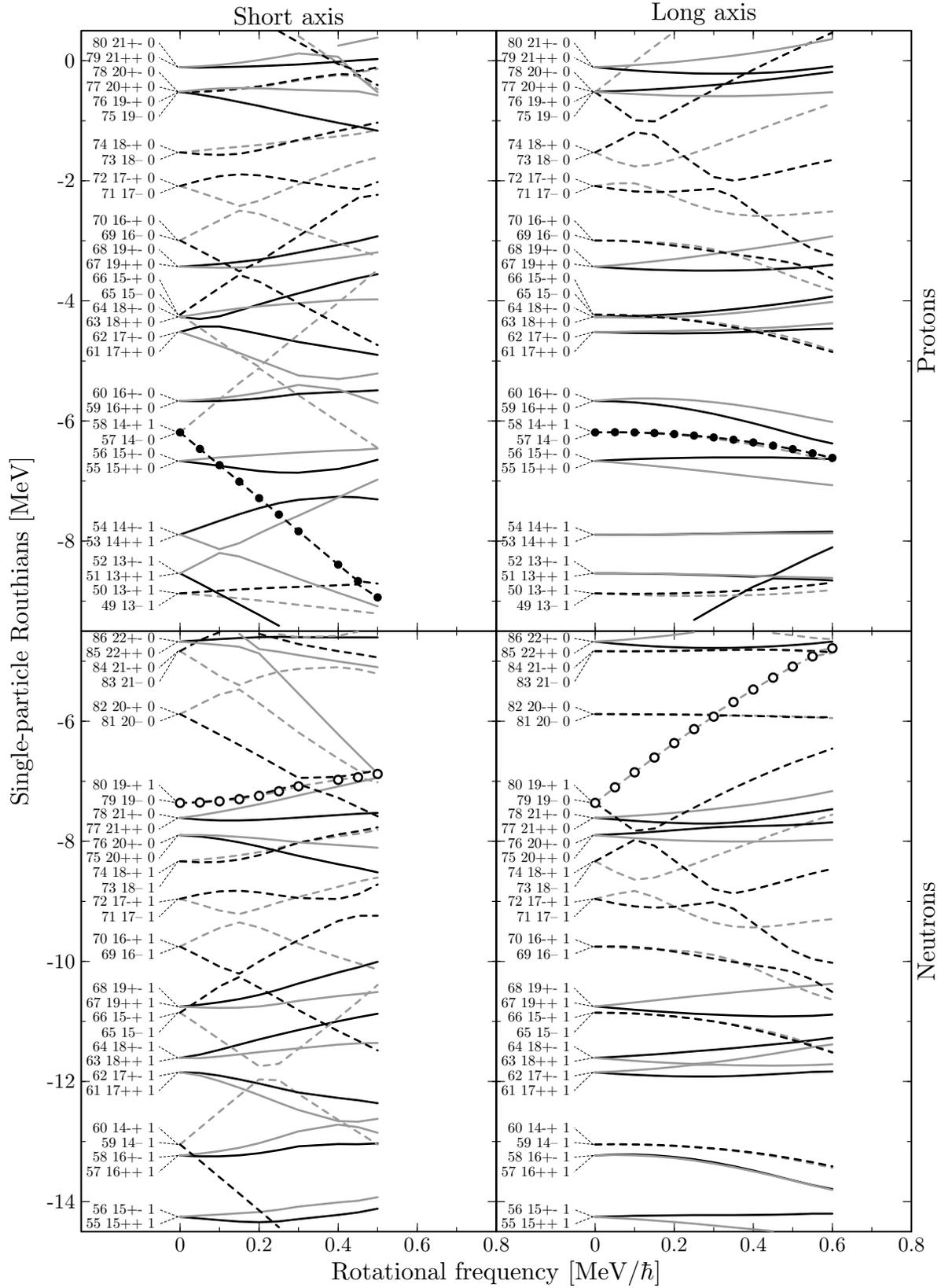}
\end{center}
\caption{Axial Routhians in $^{130}$Cs for SLy4 force with $N$ time-odd fields.}
\label{cs_a0a_fig}
\end{figure}

\begin{figure}
\begin{center}
\includegraphics[angle=90]{la_pro}
\end{center}
\caption{Proton Axial Routhians in $^{132}$La for SLy4 force with $N$ time-odd fields.}
\label{la_pro_fig}
\end{figure}

\begin{figure}
\begin{center}
\includegraphics[angle=90]{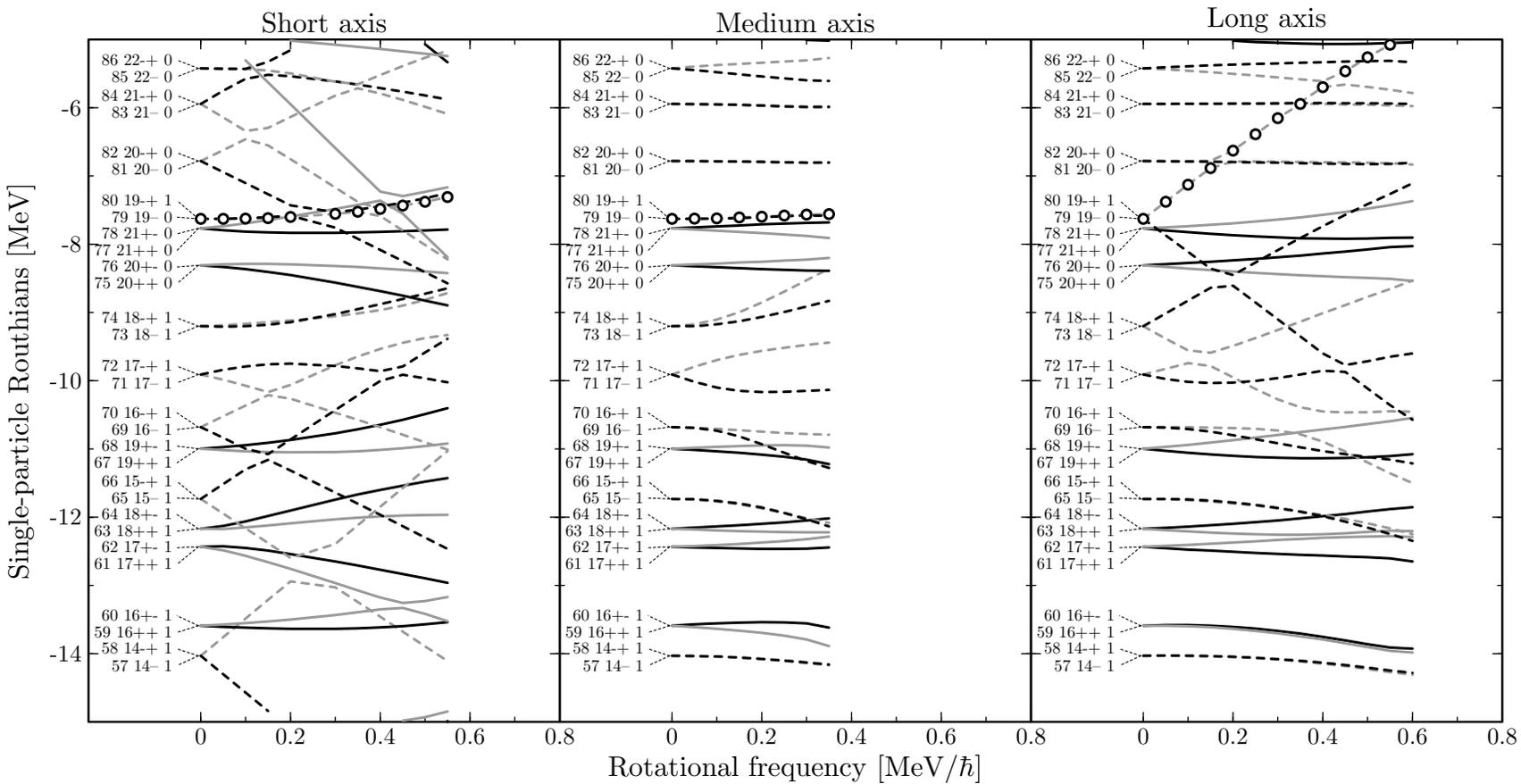}
\end{center}
\caption{Neutron Axial Routhians in $^{132}$La for SLy4 force with $N$ time-odd fields.}
\label{la_neu_fig}
\end{figure}

\begin{figure}
\begin{center}
\includegraphics{pr_a0a}
\end{center}
\caption{Axial Routhians in {\it oblate} $^{134}$Pr for SLy4 force with $N$ time-odd fields.}
\label{pr_a0a_fig}
\end{figure}

\begin{figure}
\begin{center}
\includegraphics{pr_b0a}
\end{center}
\caption{Axial Routhians in {\it triaxial} $^{134}$Pr for SkM* force with $N$ time-odd fields.}
\label{pr_b0a_fig}
\end{figure}

\begin{figure}
\begin{center}
\includegraphics{pm_a0a}
\end{center}
\caption{Axial Routhians in {\it oblate} $^{136}$Pm for SLy4 force with $N$ time-odd fields.}
\label{pm_a0a_fig}
\end{figure}

\begin{figure}
\begin{center}
\includegraphics{pm_b0a}
\end{center}
\caption{Axial Routhians in {\it triaxial} $^{136}$Pm for SkM* force with $N$ time-odd fields.}
\label{pm_b0a_fig}
\end{figure}

\begin{figure}
\begin{center}
\includegraphics{la_00c}
\end{center}
\caption{Planar and Chiral Routhians in $^{132}$La with $N$ time-odd fields.}
\label{la_00c_fig}
\end{figure}

\begin{figure}
\begin{center}
\includegraphics{la_0Gc}
\end{center}
\caption{Planar and Chiral Routhians in $^{132}$La with $G$ time-odd fields.}
\label{la_0Gc_fig}
\end{figure}

\begin{figure}
\begin{center}
\includegraphics{la_0Tc}
\end{center}
\caption{Planar and Chiral Routhians in $^{132}$La with $T$ time-odd fields.}
\label{la_0Tc_fig}
\end{figure}

\end{document}